\pdfoutput=1

\documentclass{aa}

\usepackage[T1]{fontenc} 
\usepackage{amssymb}
\usepackage{bbm}
\usepackage{txfonts}
\usepackage{psfrag}
\usepackage{graphicx}
\usepackage{natbib}
\usepackage{if then}
\usepackage{longtable}
\usepackage{ulem}
\usepackage{color}
\usepackage[dvipsnames]{xcolor}
\usepackage{verbatim}
\usepackage{algorithm,algorithmicx,algpseudocode}
\usepackage{tikzscale}
\usepackage{filecontents}
\usepackage{amsmath}
\usepackage{bm}
\usepackage{multirow}
\usepackage{todonotes}
\usepackage[breaklinks=true]{hyperref}
\usepackage{booktabs}
\usepackage{colortbl}
\usepackage{xfrac}

\def\del#1{{}}

\newcommand{\vect}[1]{{\mathbf {#1} }}
\newcommand{\mat}[1]{{#1}}

\newcommand{\borg}{{\tt BORG}}

\newcommand{\ktmpp}{K_\text{2M++}}

\newcommand{\Mpch}{$h^{-1}$~Mpc~}

\renewcommand{\deg}{\ensuremath{{}^\circ}}

\newcommand{\tmpp}{{2M++}}

\hypersetup{
  colorlinks   = true, 
  urlcolor     = RoyalBlue, 
  linkcolor    = RoyalBlue, 
  citecolor   = MidnightBlue
}
\definecolor{airforceblue}{rgb}{0.36, 0.54, 0.66}
 \definecolor{azure}{rgb}{0.0, 0.5, 1.0}

\begin{document}
\bibpunct{(}{)}{;}{a}{}{,}

\titlerunning{Physical Bayesian modelling of the non-linear matter distribution}

\title{Physical Bayesian modelling of the non-linear matter distribution: new insights into the Nearby Universe}

\author{ J. Jasche\inst{1,2} \and G. Lavaux\inst{3}}
\institute{The Oskar Klein Centre, Department of Physics, Stockholm University, Albanova University Center, SE 106 91 Stockholm, Sweden \and Excellence Cluster Universe, Technische Universit\"at M\"unchen, Boltzmannstrasse 2, 85748 Garching, Germany
\and CNRS \& Sorbonne Universit\'{e}, UMR7095, Institut d'Astrophysique de Paris, F-75014, Paris, France
}

\date{submitted to A\&A  25.06.2018}

\label{firstpage}

\abstract{
 Accurate analyses of present and next-generation cosmological galaxy surveys require new ways to handle effects of non-linear gravitational structure formation processes in data. To address these needs we present an extension of our previously developed algorithm for Bayesian Origin Reconstruction from Galaxies to analyse matter clustering at non-linear scales in observations. This is achieved by incorporating a numerical particle mesh model of gravitational structure formation into our Bayesian inference framework.
The algorithm simultaneously infers the three-dimensional primordial matter fluctuations from which present non-linear observations formed and provides reconstructions of velocity fields and structure formation histories. The physical forward modelling approach automatically accounts for the non-Gaussian features in gravitationally evolved matter density fields and addresses the redshift space distortion problem associated with peculiar motions of observed galaxies. Our algorithm employs a hierarchical Bayes approach to jointly account for various observational effects, such as unknown galaxy biases, selection effects, and observational noise. Corresponding parameters of the data model are marginalized out via a sophisticated Markov Chain Monte Carlo  approach relying on a combination of a multiple block sampling framework and an efficient implementation of a Hamiltonian Monte Carlo sampler.  We demonstrate the performance of the method by applying it to the 2M++ galaxy compilation, tracing the matter distribution of the Nearby Universe. We show accurate and detailed inferences of the three-dimensional non-linear dark matter distribution of the Nearby Universe. As exemplified in the case of the Coma cluster, our method provides complementary mass estimates that are compatible with those obtained from weak lensing and X-ray observations. For the first time, we also present a reconstruction of the vorticity of the non-linear velocity field from observations. In summary, our method provides plausible and very detailed inferences of the dark matter and velocity fields of our cosmic neighbourhood.
}
\keywords{Methods: data analysis, Cosmology: large-scale structure, Methods: statistical, Cosmology: observations, Galaxies: statistics
}


\maketitle
\flushbottom

\section{Introduction}

The goal of modern cosmology is the investigation of the dynamics of the
Universe and the formation of structures to determine the underlying
gravitational world model. Especially observations of the cosmic microwave background
(CMB), as provided by ESA's Planck satellite mission, have contributed to firmly
establishing the $\Lambda$CDM framework as the standard model of cosmology
\citep{2016A&A...594A..13P}. This model reconciles the homogeneous expansion
dynamics of the Universe with the generation and evolution of cosmic structures.
In particular, the present dynamical evolution of our Universe is believed to be
governed by dark energy and dark matter, constituting about $95\%$  of its
total energy budget. Although they are required to explain the formation of all
observable structures within the standard picture of Einstein's gravity,
dark matter and dark energy so far elude direct observations and they have not yet been
identified as particles or fields within more fundamental theories
\citep[see e.g.][]{2017arXiv170101840F}.

Making progress in understanding the cosmological phenomenology requires both taking ever more data and developing increasingly accurate and precise data analyses methods. This is particularly important when
attempting to identify those subtle signals that could hint us towards the true
nature of the physics driving the dynamical evolution of our Universe.

\begin{figure*}
        \centering
        \includegraphics[width=\hsize]{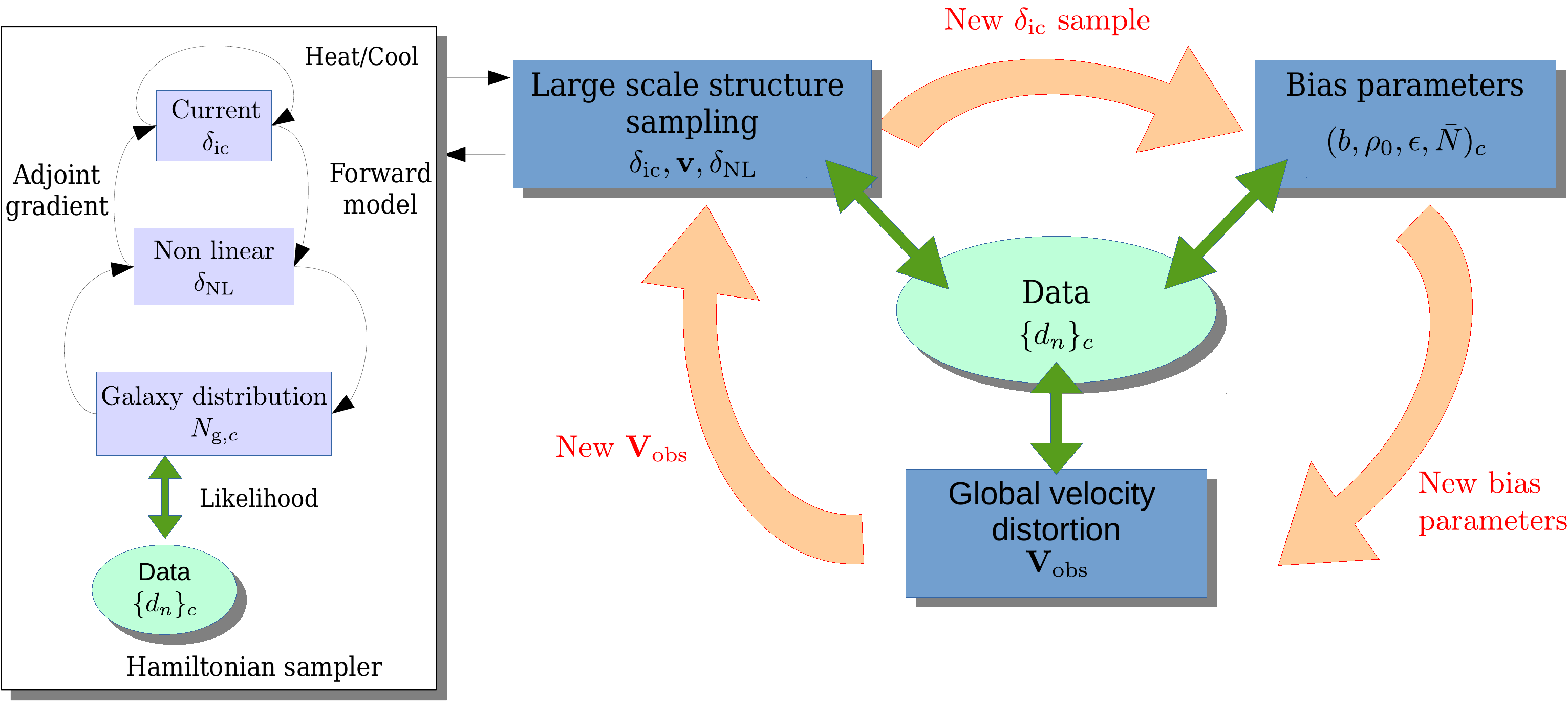}
    \caption{
      Flow chart depicting the multi-step iterative block sampling procedure. In the first step, a three-dimensional density field will be realized conditional on the galaxy observations (top left corner). In subsequent steps observer velocity, bias parameters and the normalization parameters for the galaxy distribution are sampled conditional on respective previous samples. Iteration of this process yields samples from the full joint posterior distribution generated by the \borg{} algorithm. \label{fig:flowchart}
    }
\end{figure*}

In recent times the field of cosmology has evolved from focusing on studies of
the homogeneous expansion dynamics, with supernov\ae{} of type Ia \citep{Perlmutter1999,Riess2016} and the
observation of small density perturbations in the linear regime with CMB experiments \citep{Mather1990,Smoot1992,Spergel2003,2016A&A...594A..13P}, to observations of linearly evolving structures in galaxy redshift surveys \citep[see e.g.][]{TEGMARK_2004,2007MNRAS.381.1053P,2016MNRAS.460.4210G}.
The natural next step consists of analysing non-linear cosmic structures in observations. In particular, about 80\% of the total cosmological signal provided by next-generation cosmological instruments, such as the Euclid satellite or the Large Synoptic Survey Telescope (LSST), will be generated by the cosmic matter distribution at non-linear scales \citep{2009arXiv0912.0201L,2011arXiv1110.3193L,2017arXiv170104469S}.

Accessing non-linear scales in observations also promises to extract additional cosmological information. As regularly mentioned in the literature \citep[e.g.][]{2012ApJ...754..109L,2016PhRvD..93h3510M}, the number of observable modes at smaller non-linear scales is much larger than that at larger scales, which is intrinsically limited by the size of the observable Hubble volume. In addition inference of large scale fluctuations is affected most by survey geometries and selection effects which can be quite complex \citep[see e.g.][]{Davis1991,Peacock2001}.

However,  cosmological information at non-linear scales is locked in very complex higher order statistics and cannot be accessed entirely by only measuring simple two-point statistics \citep{2016PhRvD..93h3510M,2017arXiv170104469S}.
As a consequence novel data analysis methods need to study non-linearly evolving structures to make the most of coming cosmological data sets \citep{2017arXiv170104469S}.
This requires developing novel and complex data models capable of accounting for intrinsic stochastic and systematic uncertainties of the data but also for the properties of non-linear
gravitational structure formation responsible for the non-Gaussian features in observations of the non-linear cosmic Large Scale Structure (LSS).

In many aspects, this requires to go beyond state-of-the-art in data analysis, currently relying mostly on linear data models including a linear perturbative description of observed cosmic structures
\citep{HOFFMAN1994,1994ApJ...423L..93L,LAHAV1994,1995ApJ...449..446Z,1995MNRAS.272..885F,1997MNRAS.287..425W,1999ApJ...520..413Z,ERDOGDU2004,KITAURA2008MNRAS,KITAURA2009MNRAS,JASCHESPEC2010,JaschePspec2013,2013A&A...549A.111E,2015MNRAS.447.1204J,2015A&A...583A..61G}. There has also been a considerable effort in going beyond linear data models to better capture the non-Gaussian nature of the observed galaxy distribution via Bayesian log-normal Poisson modelling \citep{KITAURA2010MNRAS,JASCHE2010HADESMETHOD,JASCHE2010HADESDATA}.

In addition, to account for non-linear structure formation processes, we have proposed to perform Bayesian analyses of galaxy observations with full three-dimensional and physical models of gravitational structure formation \citep{JASCHEBORG2012,JLW15,2016MNRAS.455.3169L}.
By exploiting physical models of the in-homogeneous evolution of cosmic matter, our approach allows for  inferring spatially distributed density and velocity fields and quantifying corresponding uncertainties, via an efficient Markov Chain Monte Carlo (MCMC) approach.

Incorporating a physical model of gravitational structure formation into the Bayesian inference approach turns the task of analysing observed non-linear cosmic structures into a statistical initial conditions problem. More specifically, we aim at inferring plausible three-dimensional initial density fields from which presently observed non-linear structures formed. In this fashion, our approach establishes an immediate link between observed present cosmic structures and their primordial initial conditions from which they formed via non-linear gravitational evolution.

It must be mentioned that naive inversion of the flow of time in corresponding physical structure formation models, to obtain initial conditions from non-linear density fields, is generally not possible due to the ill-posedness of the inverse problem \citep{NUSSER1992,Crocce2006}.

In this context ill-posedness is a statement on the existence of a range of feasible inference solutions that are consistent with noisy and incomplete observations, generally defying a unique model reconstruction. More specifically, in the context of the cosmic large scale structures, ill-posedness results from several instances. In particular, we usually deal with incomplete and noisy data but also dissipative processes, coarse-graining effects or incomplete access to the dark matter phase-space distribution. The combination of these effects eliminates information on the dark matter phase space distribution and prevents unique recovery of information on cosmic initial conditions via Liouville's theorem for Hamiltonian dynamics \citep{liouville1838,nla.cat-vn881490}.

However, detailed information on the reason for ill-posedness is not required to address the problem via statistical inference. As already discussed in our previous works, we address the issue of ill-posedness by performing thorough Bayesian inference via physical forward modelling within sophisticated Hamiltonian Monte Carlo sampling approach \citep{JASCHEBORG2012,JLW15,2016MNRAS.455.3169L}. This MCMC approach correctly explores the space of feasible solutions for the large-scale structure inference problem, which are compatible with noisy and incomplete observations. More specifically our approach infers a set of plausible three-dimensional primordial density fields from which structures in present observations formed. Since our algorithm tries out feasible solutions purely via forward simulations, it is not affected by the problems of traditional inverse modelling, as summarized above.

Our approach also shares many beneficial properties with proposed ad-hoc BAO reconstruction methods, which have been demonstrated to increase the detectability of the BAO peaks from three to four sigma  \citep[see e.g.][]{2009PhRvD..80l3501N,2012MNRAS.427.2146X,2015PhRvD..92l3522S,2018PhRvD..97b3505S}. By now several groups have proposed approaches to incorporate physical models into data analysis frameworks \citep{Nusser2000,BRENIER2003,Lavaux10,JASCHEBORG2012,Doumler2013,Wang2013,Kitaura13,Schmittfull2017,Seljak2017}

While previous approaches relied on perturbative descriptions of cosmic large-scale structure in this work we go beyond such limitations by incorporating fully non-linear and non-perturbative
computer models of structure formation into our previously proposed algorithm for Bayesian Origin Reconstruction from Galaxies (\borg{}).
More specifically we seek to fit a gravitational particle mesh (PM) model in its entirety to galaxy observations of the Nearby Universe. In contrast to contemporary analyses, limited to studying the lowest order moments of the density field (e.g. power- and bi-spectra), physical modelling of the entire three-dimensional matter
distribution in observations permits us to implicitly access the entire hierarchy of higher order poly-spectra by directly fitting the filamentary three-dimensional distribution of matter in the Universe.

A particularly important advantage of our approach is that it does not only provide single point estimates, such as mean or mode, but it characterizes the corresponding posterior distribution in terms of MCMC samples and thus allows for a thorough uncertainty quantification (UQ) \citep[such as][]{JASCHEBORG2012,JLW15}. Previously such approaches have been considered computationally too prohibitive for numerical $N$-body models of structure formation. This work introduces our implementation of a non-linear large-scale structure inference framework, on the basis of the latest advances in Bayesian methodology and sampling algorithms. This permits us to apply sophisticated MCMC techniques to the title problem at scales previously inaccessible to cosmological data analysis.

Analysing cosmological surveys subject to noise and systematics is generally challenging and requiring the data model to handle a variety of nuisances. In order to address this issue we turned our \borg{} algorithm into a modular statistical programming engine that exploits hierarchical Bayes and block sampling techniques to flexibly build data models for different data sets. Different building blocks of the data model can be added to the Markov Chain and their respective parameters will be jointly inferred within the multiple block sampling approach as visualized in figure~\ref{fig:flowchart}.

The present work also aims at applying our techniques to infer a coherent and consistent physical model of the three-dimensional large-scale dark matter distribution, its dynamics and formation histories in the Nearby Universe.
This will be achieved by applying the \borg{} algorithm to the \tmpp{} galaxy sample \citep[][]{LH11}.

These results will provide us with detailed and accurate maps of the expected dark matter distribution in the Nearby Universe and will permit us to measure the masses of prominent cosmic structures.
Specifically for the case of the Coma cluster, we will demonstrate that we can obtain mass estimates that are compatible with gold-standard weak lensing measurements.
We further seek to determine dynamical properties of cosmic structures and test their potential to impact cosmological measurements in the Nearby Universe via effects of peculiar velocities.

The manuscript is structured as follows. In Section~\ref{sec:methodology} we describe the methodology and the modifications to the \borg{} algorithm.
Section~\ref{sec:data_model} provides a detailed overview of the data model required to compare predictions of the structure formation model with observed galaxy surveys.
The main part of this work focuses on the application of our algorithm to data of the \tmpp{} compilation. The corresponding description of setting up these analysis run and the employed data is given in Section~\ref{sec:data_application}. Section~\ref{sec:inference_results} highlights some of our inference results. In particular we showcase results on galaxy biases (Section~\ref{sec:galaxy_biases}), the inferred three-dimensional density field at the initial conditions and in the present epoch (Section~\ref{sec:result_3d_density}), the formation history of the Supergalactic plane (\ref{sec:form_hist}), the estimated mass and corresponding mass profile of the Coma cluster (Section~\ref{sec:mass_reconstruction}), the velocity field of the Local Universe (Section~\ref{sec:inf_vel_field}) and its possible impact on Hubble constant measurements in the Nearby Universe (Section~\ref{sec:impact_hubble}). Finally, in Section~\ref{sec:Summary_an_Conclusion}, we conclude the paper and discuss future developments.

\section{Bayesian inference with the \borg{} algorithm }

\label{sec:methodology}
This section provides an overview of our previously developed Bayesian inference framework including the modifications as introduced in this work.

\subsection{The  \borg{} algorithm }
\label{sec:borg_model}
The presented project builds upon our previously developed algorithm for Bayesian Origin Reconstruction from Galaxies (\borg{}), aiming at the analysis of three-dimensional cosmic matter distribution at linear and non-linear scales of structure formation in galaxy surveys \citep[see e.g.][]{JASCHEBORG2012,JLW15,2016MNRAS.455.3169L}. More explicitly the \borg{} algorithm fits three-dimensional models of gravitational structure formation to data.

Interestingly, introducing a physical model of gravitational structure growth immediately into the inference process turns the task of analysing the present non-linear galaxy distribution into a statistical initial conditions problem. More specifically the \borg{} algorithm seeks to infer the cosmic initial conditions from which present three-dimensional structures in the distribution of galaxies have formed via non-linear gravitational mass aggregation.

The \borg{} algorithm explores a cosmic LSS posterior distribution consisting of a Gaussian prior for the initial density field at a initial cosmic scale factor of $a=10^{-3}$ and a Poissonian model of galaxy
formation at a scale factor $a=1$, while initial density fields are related to the present galaxy distribution via a second order Lagrangian perturbation theory (2LPT) or a full particle mesh, as described in this work, model of gravitational structure formation \citep[for details see][]{JASCHEBORG2012}. By exploiting non-linear structure growth models the \borg{} algorithm naturally accounts for the filamentary structure of the cosmic web typically associated with higher-order statistics induced by non-linear gravitational processes.
As described in our previous works the posterior distribution also accounts for systematic and stochastic uncertainties, such as survey geometries, selection effects, unknown noise and galaxy biases as well as foreground contaminations \citep[see e.g.][]{JASCHEBORG2012,JLW15,2016MNRAS.455.3169L,2017A&A...606A..37J}.

The resultant procedure is numerically highly non-trivial since it requires to explore the very high-dimensional and non-linear space of possible solutions to the initial conditions problem within a fully probabilistic approach.
Typically, these spaces comprise $10^6$ to $10^7$ parameters, corresponding to amplitudes of the primordial density at different volume elements of a regular mesh in Lagrangian space for grids between $128^3$ and $256^3$ elements. Numerically efficient exploration of this highly non-Gaussian and non-linear posterior distribution is achieved via an efficient implementation of a Hamiltonian Markov Chain Monte Carlo sampling algorithm \citep[see][for details]{DUANE1987,2012arXiv1206.1901N,JASCHEBORG2012}.

It is important to remark that our inference process requires at no point the inversion of the flow of time in the dynamical physics model. The analysis solely depends on forward model evaluations, alleviating many of the problems encountered in previous approaches to the inference of initial conditions, such as spurious decaying mode amplification \citep[see e.g.][]{NUSSER1992,Crocce2006}. Specifically \citep{Crocce2006} nicely demonstrate that inference of initial conditions is a fundamentally ill-posed problem. Recovering information on the initial conditions becomes harder and increasingly uncertain towards smaller scales, generally preventing unique backward in time integration of the final density field.  Rather than inferring the initial conditions by backward in time integration, our approach builds a fully probabilistic non-linear filter using the dynamical forward model as a prior. This prior singles out physically reasonable LSS states from the space of all possible solutions to the statistical initial conditions problem. However, they do not strictly limit the space of initial conditions that must be searched to match observations. If for some reason unlikely events are required to explain observational data, the algorithm explores prior regions that are {\it a priori} unlikely. This allows for the potential characterization of primordial non-Gaussian signals in the recovered initial conditions for example.

\begin{figure*}
\includegraphics[width=\hsize]{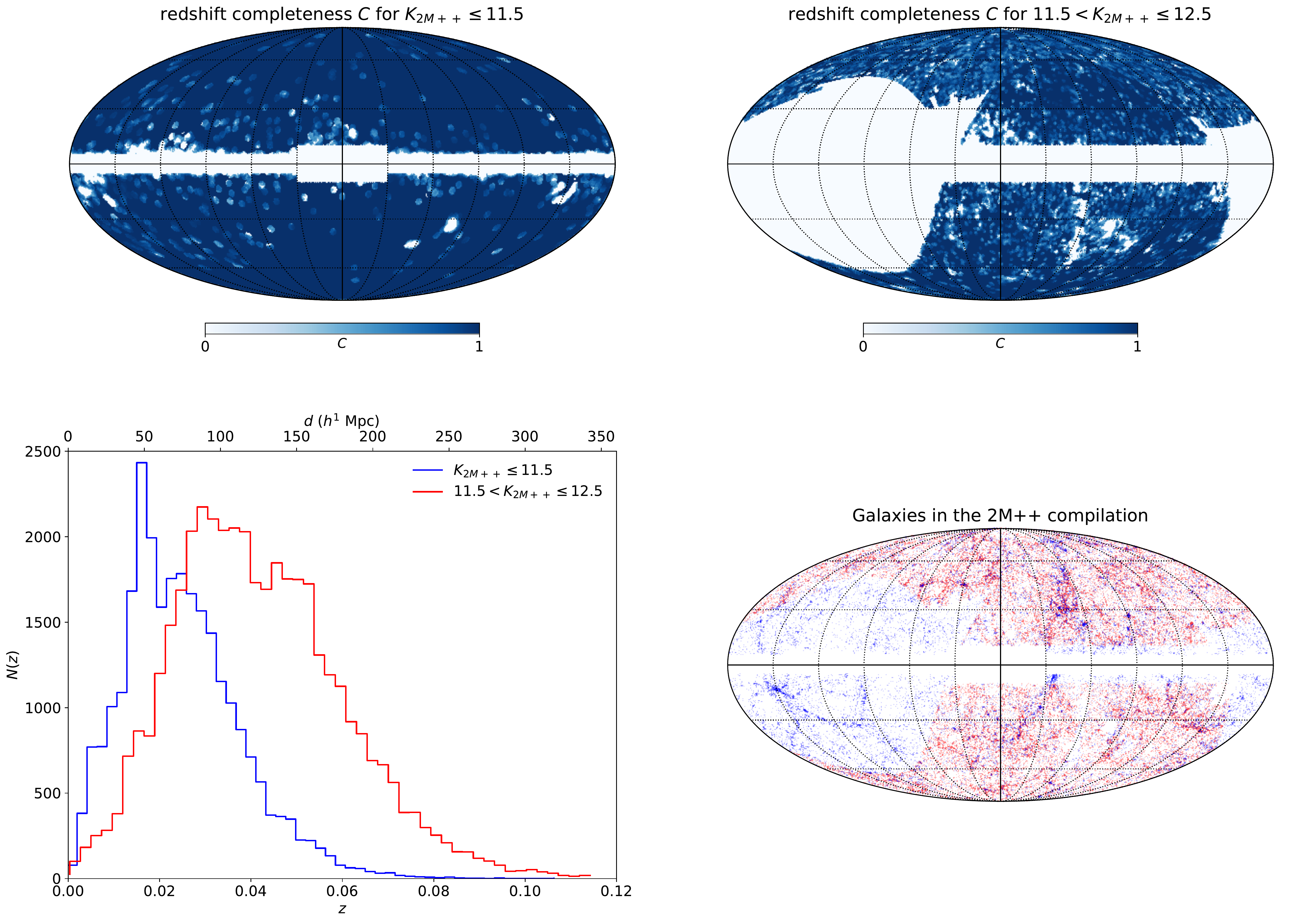}
\caption{\label{fig:data_description} This figure illustrates the 2M++ data and its selection properties. Top left panel: the sky completeness at $K_\mathrm{2M++} \leq 11.5$, derived as the number of observed redshifts versus the number of targets in the 2MASS photometric sample. Top right panel: The same quantity is shown but for apparent magnitudes $11.5 < K_\mathrm{2M++} \leq 12.5$. Bottom left panel: Number count of galaxies in thin radial shells for the two different magnitude cuts shown in the top row. We see that the catalogue covers a volume up to a redshift $z\sim 0.06-0.08$. Bottom right panel: Sky projection of the positions of the galaxies of the 2M++ catalogue. The local large-scale structures are clearly visible.}
\end{figure*}

Since the \borg{} algorithm provides an approximation to non-linear large-scale dynamics, it automatically provides information on the dynamical evolution of the large-scale matter distribution. In particular, it explores the space of plausible dynamical structure formation \textit{histories} compatible with both data and model. Also note, that the \borg{} algorithm naturally infers initial density fields at their Lagrangian coordinates, while final density fields are recovered at corresponding final Eulerian coordinates. Therefore the algorithm accounts for the displacement of matter in the course of structure formation.

As results the algorithm provides measurements of the three dimensional density field but also performs full four-dimensional state inference and recovers the dynamic formation history and velocity fields of the cosmic LSS.

Some examples of secondary projects derived from these results aimed at studying dark matter voids in the galaxy distribution  \citep{Leclercq2014A}, the phase-space distribution of matter in the SDSS survey \citep{2017JCAP...06..049L}, properties of the population of Active Galactic Nuclei (AGN) \citep{2018A&A...612A..31P} as well as gravitational screening mechanisms \citep{2018MNRAS.474.3152D,2018arXiv180207206D} and cosmic magnetic fields \citep{Hutschenreuter2018}.

\subsection{Hamiltonian Monte Carlo sampling}
\label{sec:hamiltonian_sampling}
Large-scale Bayesian inverse problems, as described in this work, belong to the most challenging tasks in the field of modern cosmological data analysis. This is mostly due to the numerical complexity of the physical model to test with data but even more so due to the high dimensional nature of the inference task itself. The combination of numerically expensive model evaluations and the curse of dimension typically renders large-scale Bayesian inverse problems numerically impractical \citep{bellman1961adaptive}.

A particular interesting algorithm to circumvent some of the problems associated to the curse of dimensionality is the Hamiltonian Monte Carlo (HMC) algorithm. Its numerical and statistical efficiency originates from the fact that it exploits techniques developed to follow classical dynamical particle motion in potentials. This approach provides deterministic proposals to the Metropolis-Hastings algorithm that can be accepted with very high probability \citep[][]{DUANE1987,NEAL1993,NEAL1996}.

The HMC can be used to generate random realizations of a set of parameters \(\{x_i\}\) of size $N$ from any target distribution  \({\Pi}(\{x_i\})\) by interpreting its negative logarithm as a potential for classical particle motion given as:
\begin{equation}
\label{eq:Potential}
\Psi(\{x_i\})=-\ln {\Pi}(\{x_i\}) \, .
\end{equation}
Introducing additional sets of auxiliary quantities, referred to as 'momenta' $\{p_i\}$ and a 'mass matrix' $\mat{M}$, it is possible to define a Hamiltonian function analogous to classical mechanics:
\begin{equation}
\label{eq:Hamiltonian}
H(\{x_i\},\{p_i\}) = \frac{1}{2} \sum_{i,j} p_i\,M_{i,j}^{-1}\,p_j +\Psi(\{x_i\}) \, .
\end{equation}
It is important to remark that the joint distribution for parameters \(\{x_i\}\) and \(\{p_i\}\) can then be obtained via exponentiating the Hamiltonian given in equation (\ref{eq:Hamiltonian}):
\begin{equation}
\label{eq:joint_TARGET_DISTRIBUTION}
\Pi(\{x_i\},\{p_i\}) \propto \mathrm{e}^{-H} ={\Pi}(\{x_i\})\,\exp\left(-\frac{1}{2}\,\sum_{i,j}\,p_i\,M_{i,j}^{-1}\,p_j\right)\, .
\end{equation}
As can be seen the joint distribution in equation (\ref{eq:joint_TARGET_DISTRIBUTION}) factorizes in a product of our target distribution  \({\Pi}(\{x_i\})\) and a Gaussian distribution in the momenta \(\{p_i\}\). This demonstrates that the two sets of variables \(\{p_i\}\) and \(\{x_i\}\) are statistically independent and marginalization over auxiliary momenta yields the desired target distribution \({\Pi}(\{x_i\})\).

It is now possible to explore the joint parameter space of variables $\{p_m\}$ and $\{x_m\}$ by following persistent trajectories for some fixed amount of pseudo time $\tau$ according to Hamilton's equations of motion:
\begin{equation}
\label{eq:HAMILTON1}
	\frac{\mathrm{d}x_m}{\mathrm{d}t} = \frac{\partial H}{\partial p_m}\, ,
\end{equation}
and
\begin{equation}
\label{eq:HAMILTON2}
	\frac{\mathrm{d}p_m}{\mathrm{d}t} = \frac{\partial H}{\partial x_m} = - \frac{\partial \Psi(\{x_i\})}{\partial x_m}\, .
\end{equation}
In the above equation, the Hamiltonian forces are given by the gradient of the logarithmic target distribution with respect to inference parameters. Therefore, 'particles' do not move freely in this high dimensional parameter space and they tend to be attracted towards regions with higher probability.
New random realizations for the parameters $\{p'_i\}$ and $\{x'_i\}$ are then obtained by starting at the current
position in phase space characterized by the values  $\{p_i\}$ and $\{x_i\}$ and following Hamiltonian dynamics for
a certain amount of pseudo time $\tau$. The endpoint of this trajectory will then be accepted according to the standard Metropolis-Hastings acceptance rule:
\begin{equation}
	\label{eq:acceptance_rule}
	{\Pi}_A = \min\Big\{1,\exp\big[-\left(H(\{x'_i\},\{p'_i\})-H(\{x_i\},\{p_i\}\right)\big]\Big\}\, .
\end{equation}
The particular feature that renders HMC an excellent algorithm for high dimensional parameter space exploration is precisely the conservation of the Hamiltonian by the above equation of motions. Consequently the expected acceptance rate given by Equation~\eqref{eq:acceptance_rule} for the exact Hamiltonian dynamics has a value of unity.

In practice the acceptance rate may be lower due to numerical inaccuracies of the numerical integration scheme. To generate a valid Markov chain, auxiliary momenta are randomly re-drawn from a Gaussian distribution after each acceptance step and the procedure starts again.
Individual momenta $\{p_i\}$ are not stored. Discarding auxiliary momenta simply amounts to marginalization and yields the target distribution  ${\Pi}(\{x_i\})$.

In summary, two particular features of the HMC algorithm render it ideal for the exploration of high dimensional parameter spaces with complex physical models. First of all, it exploits conserved quantities such as the Hamiltonian to provide a high Metropolis-Hastings acceptance probability, hence reducing the amount of rejected model evaluations. More importantly, the HMC exploits gradient information of the target posterior distribution, preventing it from performing a blind random walk in high dimensions. This leads the algorithm follows targeted persistent trajectories to efficiently explore parameter spaces. For details on the numerical
implementation of the HMC for cosmic large-scale structure analyses, the interested reader is also encouraged to have a look at our previous work \citep{JASCHE2010HADESMETHOD,JASCHEBORG2012,JaschePspec2013}.

\subsection{Modular statistical programming via Block sampling}

A particular feature of the full Bayesian inference approach, as presented here, is the possibility to perform modular statistical programming. In particular, the \borg{} algorithm can solve any hierarchical Bayesian problem by simply adding additional components to a block sampling framework, as outlined in figure~\ref{fig:flowchart}. This block sampling approach allows for straightforwardly accounting for additional observational systematics by building more complex data models and adding corresponding parameter samplers to the block sampling framework.

In this work, we use this block sampling framework to jointly account for unknown parameters of a galaxy biasing model, as described further below, and unknown noise levels for respective galaxy samples (see figure~\ref{fig:flowchart}). Iterating this block sampling framework by conditionally drawing random realizations of parameters in sequence will then result in a correct Markov Chain that asymptotes towards the desired joint target posterior distribution \citep[e.g.][]{GEMA1984}.

\section{A data model for non-linear LSS inference}
\label{sec:data_model}

This section describes the development and implementation of a non-perturbative data model to analyse the three-dimensional cosmic LSS at non-linear scales in data.

\subsection{The general data model}
The aim of the \borg{} algorithm is to provide a full characterization of the three-dimensional cosmic large-scale structure in observations by providing a numerical representation of the associated posterior distribution via sophisticated MCMC methods. More specifically the  \borg{} algorithm provides data constrained realizations of a set of plausible three-dimensional matter density contrast amplitudes
$\{\delta_i\}$ underlying a set of observed galaxy number counts $\{N^g_i\}$ for various volume elements in the observed domain indexed by $i$. Using Bayes rule, the most general form of this posterior distribution can be expressed as:
\begin{equation}
\Pi\left(\{\delta_i\}|\{N^g_i\}\right)=\frac{\Pi\left(\{\delta_i\}\right)\, \Pi\left(\{N^g_i\}|\{\delta_i\}\right)}{\Pi\left(\{N^g_i\}\right)} \, ,
\end{equation}
where the prior distribution $\Pi\left(\{\delta_i\}\right)$ describes our a priori knowledge on the three-dimensional matter distribution in the Universe, $\Pi\left(\{N^g_i\}|\{\delta_i\}\right)$ is the likelihood describing the statistical process of obtaining a set of observations $\{N^g_i\}$ given a specific realization of the matter field $\{\delta_i\}$ and $\Pi\left(\{N^g_i\}\right)$ is the so-called evidence which normalizes the probability distribution.
We note that $\Pi\left(\{\delta_i\}\right)$ may depend on cosmological parameters and other auxiliary parameters, sometimes hyper-parameters, that we skip to represent for the moment in the notation.

As already described in our previous work, a major complication arises from the fact that the prior distribution $\Pi\left(\{\delta_i\}\right)$ for non-linear gravitationally formed density fields is not known in closed form, such as in terms of a multivariate probability density distribution \citep{JASCHEBORG2012}. State-of-the-art approaches, therefore, assume Gaussian or log-normal distributions as approximations to the prior for the matter density contrast. However, since these distributions model only the one- and two-point statistics, they fail to capture the filamentary features of the observed cosmic web that are associated with higher order statistics \citep[see e.g.][]{1980lssu.book.....P}.

Additional complexity for the analysis of next-generation deep surveys arises from the fact that observed galaxy number counts are not solely determined by underlying density amplitudes but are additionally affected by dynamic effects such as redshift space distortions or light cone effects. Naive treatment of such additional dynamic structure formation processes in data would require to also self-consistently infer the three-dimensional velocity field from data. We would need to use a joint posterior distribution for density amplitudes $ \{\delta_i\}$ and peculiar velocities $ \{\vec{v}_i\}$ given as:
\begin{equation}
\Pi\left(\{\delta_i\},\{\vec{v}_i\}|\{N^g_i\}\right)=\frac{\Pi\left(\{\delta_i\},\{\vec{v}_i\}\right)\, \Pi\left(\{N^g_i\}|\{\delta_i\},\{\vec{v}_i\}\right)}{\Pi\left(\{N^g_i\}\right)} \, .
\end{equation}
Not only does this approach aggravate the search for a suitable prior distribution $\Pi\left(\{\delta_i\},\{\vec{v}_i\}\right)$ but it also dramatically increases the amounts of parameters to be inferred with the three components of the spatial velocity field  $ \{\vec{v}_i\}$. We note that, generally, parameter space exploration becomes exponentially harder with the number of inference parameters. This fact is known as the curse of dimensions. Naive addition of a few million velocity amplitudes would therefore not be a wise decision when seeking to perform parameter space exploration.

While velocity fields at the present epoch are not uniquely related to the dark matter density field, the theory of gravitational structure formation and the cosmic microwave background yields indication that primordial matter fluctuations were almost at rest with respect to the Hubble flow in the early Universe \citep{1980lssu.book.....P}. In this picture, tiny fluctuations in the primordial peculiar velocity field derive uniquely from the field of initial density fluctuations by being proportional to the gradient of their gravitational potential \citep[see e.g.][]{1980lssu.book.....P,2002PhR...367....1B}.

Also, the primordial fluctuations field exhibits almost trivial statistical properties. In accordance with present theory and observations by the Planck satellite mission, the initial density field is an almost ideal Gaussian random field with zero mean and a covariance matrix corresponding to the post-recombination initial cosmological power-spectrum \citep[see e.g.][]{2016A&A...594A..13P}.

If we could, therefore, cast the problem of analysing the non-linear structures in the present Universe into a problem of inferring their initial conditions, we would be able to simultaneously address the problem of finding a suitable prior distribution without the need to increase the parameter space when having to deal with the present velocity field.

The required large scale structure posterior distribution would then turn into a joint distribution of the present density field $\{\delta_i\}$ and the set of primordial density fluctuation amplitudes $\{\delta^{\mathrm{IC}}_i\}$ conditional on observations of galaxy number counts $\{N^g_i\}$, given as:
\begin{align}
\Pi\left(\{\delta_i\},\{\delta^{\mathrm{IC}}_i\}|\{N^g_i\}\right)& =
   \frac{\Pi\left(\{\delta_i\},\{\delta^{\mathrm{IC}}_i\}\right)\, \Pi\left(\{N^g_i\}|\{\delta_i\},\{\delta^{\mathrm{IC}}_i\}\right)}{\Pi\left(\{N^g_i\}\right)} \nonumber \\
& =
   \frac{\Pi\left(\{\delta^{\mathrm{IC}}_i\}\right) \, \Pi\left(\{\delta_i\}|\{\delta^{\mathrm{IC}}_i\}\right)\, \Pi\left(\{N^g_i\}|\{\delta_i\}\right)}{\Pi\left(\{N^g_i\}\right)} \, , \label{eq:posterior0}
\end{align}
where $\Pi\left(\{\delta^{\mathrm{IC}}_i\}\right)$ is the prior distribution of cosmic initial fluctuations and $\Pi\left(\{\delta_i\}|\{\delta^{\mathrm{IC}}_i\}\right)$ describes the process by which the present matter distribution has been obtained from their initial conditions. We further assume conditional independence $\Pi\left(\{N^g_i\}|\{\delta_i\},\{\delta^{\mathrm{IC}}_i\}\right) = \Pi\left(\{N^g_i\}|\{\delta_i\}\right)$, that is, galaxy observations are conditionally independent of the primordial fluctuations once the final density field is given. This last assumption is not a fundamental limitation of the probabilistic model but it simplifies greatly the comparison to observations at the level considered in this work. The fundamental assumption is that galaxy formation is expected to depend only on the scalar fluctuations of the final conditions. Further extensions of the model, for which the galaxy formation would depend on the entire history of the dynamics, would be possible at additional computational costs.

The distribution $\Pi\left(\{\delta_i\}|\{\delta^{\mathrm{IC}}_i\}\right)$ describes gravitational structure formation. It encodes the processes by which the present matter fluctuations $\{\delta_i\}$ derive from the initial conditions $\{\delta^{\mathrm{IC}}_i\}$. Here we will assume that the final matter distribution derives uniquely from the initial conditions. This is, of course, the standard of cosmological modelling since cosmological simulations provide deterministic results when integrating the structure formation model. Thus we can model the final density field as a function of the initial density field:
\begin{equation}
\delta_i = G_i(\{\delta^{\mathrm{IC}}_i\}) \,,
\end{equation}
where $G_i(\{\delta^{\mathrm{IC}}_i\}))$ is our structure formation model that transforms initial conditions into final density fields. Since we assume this process to be deterministic we immediately obtain:
\begin{equation}
\Pi\left(\{\delta_i\}|\{\delta^{\mathrm{IC}}_i\}\right) = \prod_i \delta^D\left(\delta_i - G_i(\{\delta^{\mathrm{IC}}_i\})\right) \, ,
\end{equation}
where $\delta^D(x)$ denotes the Dirac delta distribution.
This yields the following large scale structure posterior distribution:
\begin{multline}
\Pi\left(\{\delta_i\},\{\delta^{\mathrm{IC}}_i\}|\{N^g_i\}\right)
= \\
\frac{\Pi\left(\{\delta^{\mathrm{IC}}_i\}\right)}{\Pi\left(\{N^g_i\}\right)} \left(\prod_j \delta^D\left(\delta_j - G_j(\{\delta^{\mathrm{IC}}_i\})\right)\right) \Pi\left(\{N^g_i\}|\{\delta_i\}\right) \, , \label{eq:posterior}
\end{multline}
Marginalization over the final density fields $\{\delta_i\}$ then yields our posterior distribution:
 \begin{equation}
\Pi\left(\{\delta^{\mathrm{IC}}_i\}|\{N^g_i\}\right)
=\frac{\Pi\left(\{\delta^{\mathrm{IC}}_i\}\right) \, \Pi\left(\{N^g_i\}|\{G_i(\{\delta^{\mathrm{IC}}_i\})\}\right)}{\Pi\left(\{N^g_i\}\right)} \, . \label{eq:final_posterior}
\end{equation}
This distribution links the present observations of the galaxy distributions $\{N^g_i\}$ to the corresponding initial conditions $\{\delta^{\mathrm{IC}}_i\}$ from which they originate via a gravitational structure formation model $\left\{G_j(\{\delta^{\mathrm{IC}}_i\})\right\}$.

Embedding a physical structure formation model into the posterior distribution to analyse three-dimensional cosmic structures in observations thus solves many outstanding questions. Most importantly we can now address issues related to structure formation dynamics, such as redshift space distortions, light cone effects and higher order statistics associated with the filamentary structure of the cosmic web. In the following, we will discuss how to perform inferences with non-linear structure formation models.

\begin{figure}
	    \centering
        \includegraphics[width=\hsize]{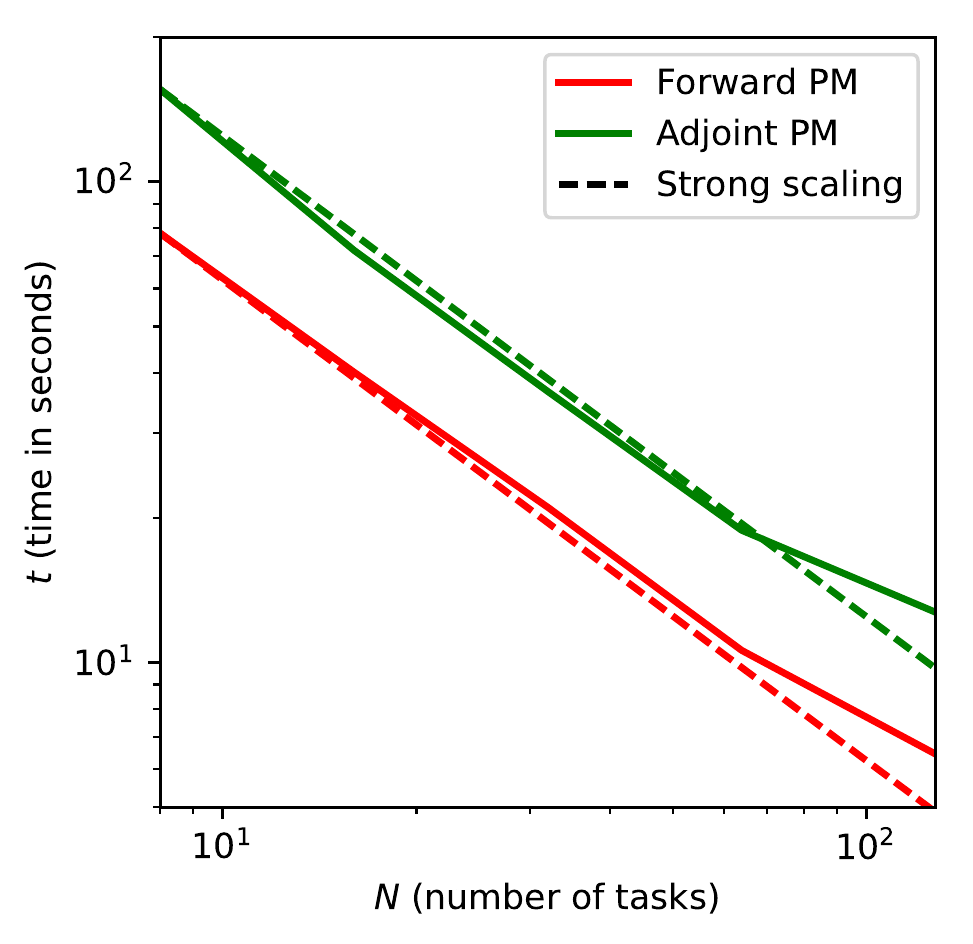}
	\caption{This plot shows the computational scaling properties of the code over MPI-tasks. The x-axis is the number of MPI tasks, each task being given 8 cores with OpenMP parallelization. The y-axis is the wall time seconds taken by the software to execute the indicated part of the algorithm. The red lines correspond to the evaluation of one time-step of the BORG-PM forward model, i.e. the N-body simulation including gravity solver. The green lines correspond to the time taken to compute the adjoint gradient of that same model. We note that the cost of the adjoint gradient takes only twice as much time as the forward model itself over the entire range. Also, the scaling is strong up to $\sim$100 cores, the break visible at the end being because of the core saturation and the use of hyper-threading on the supercomputer. \label{fig:bench}
    }
\end{figure}

\subsection{The non-linear structure formation model}
\label{sec:pm_model}
Our previous work relied on second order Lagrangian perturbation theory (LPT) to model cosmic structure formation \citep{JASCHEBORG2012,JLW15,2016MNRAS.455.3169L}. Even though LPT provides good approximations to the cosmic large-scale structure at the largest scales there are clear limits to its validity. Most notably the LPT approach relies on a convergence of series expansion. This expansion fails to accurately describe multi-streaming regions in high-density objects and cannot accurately capture the dynamic of gravitational evolution of dark matter at scales $l \lesssim 10 \mathrm{h^{-1}\,Mpc}$ \citep[see e.g.][]{Melott1995,Tassev2012}.

\begin{figure*}
  \begin{center}
      \includegraphics[width=\hsize]{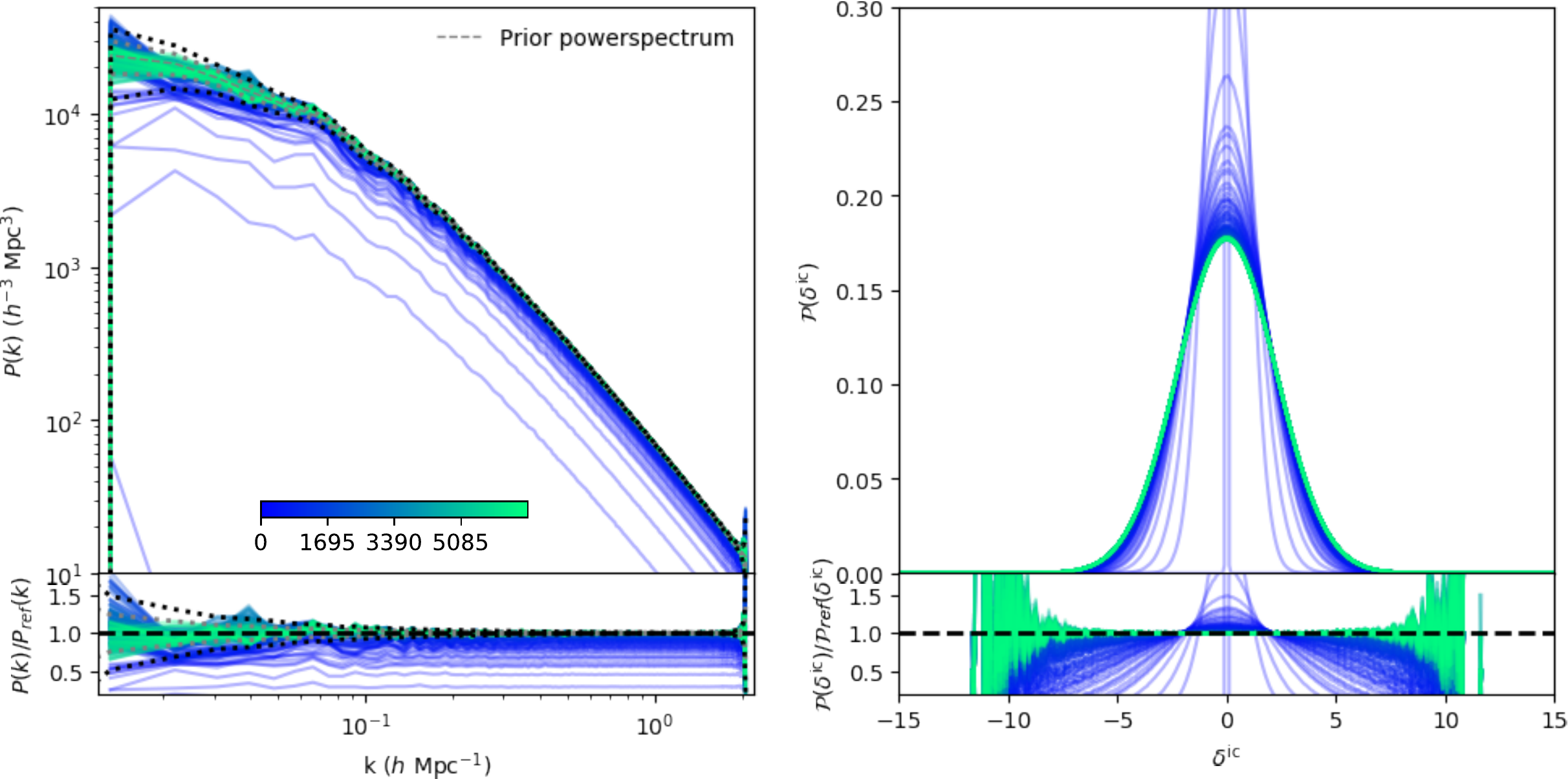}
  \end{center}
  \caption{\label{fig:Pk_burnin} Sequential posterior power-spectrum (left panel) and 1-pt distribution (right panel) of inferred primordial fluctuations measured during the burn-in of the Markov chain with an LPT model. The colour gradient indicates the step number in the chain from zero (random initial condition of small amplitude) to 6783 for which the chain is manifestly stable according to this metric. The top panels give the power spectrum and 1-pt distributions measured a posteriori from the samples, while lower panels show the ratio of these quantities to the expected prior values. Thick dashed lines represent the fiducial prior values. The thin (black respectively) grey dotted line indicates the Gaussian $1\sigma$ limit ($2\sigma$ respectively) in the lower left panel. These results show no sign of any residual systematic artefacts, indicating a healthy burn-in behaviour of the chain. }
\end{figure*}

We intend to go beyond such limitations and to account for the non-linear gravitational dynamics. In this work we update the physics model of our \borg{} algorithm with a numerical particle mesh model \citep[see e.g.][]{1983MNRAS.204..891K,1985ApJS...57..241E,HOCKNEYEASTWOOD1988,1997astro.ph.12217K}.

A particle mesh code solves the gravitational N-body problem by following the dynamical trajectories of a set of simulated dark matter particles including their mutual gravitational interactions. Our implementation of this particle mesh simulator follows closely the description of \citep[][]{1997astro.ph.12217K}. To simulate non-linear gravitational structure formation from some predefined initial conditions to the present state of the cosmic LSS a particle mesh code solves the following equations of motion for positions $\vec{x}$ and momenta $\vec{p}$ of dark matter particles:
\begin{equation}
\frac{\mathrm{d}\vec{x}}{\mathrm{d}a}=\frac{\vec{p}}{\dot{a}a^2} \, ,
\end{equation}
where $a$ is the cosmic scale factor and $\dot{a}$ is its first time derivative.
Corresponding momentum updates are given by:
\begin{equation}
\frac{\mathrm{d}\vec{p}}{\mathrm{d}a}= -\frac{\vec{\nabla}_{\vec x} \Phi}{a H(a)} \,,
\end{equation}
where $\vec{p}=a^2\dot{\vec{x}}$ and the gravitational potential $\Phi$ is given implicitly by the Poisson equation:
\begin{equation}
\nabla^2_{\vec{x}} \Phi = \frac{3}{2} H_0^2 \Omega_{m,0} \frac{\delta_\text{m}(\vec{x})}{a} =
    \frac{H_0}{a} \nabla^2_{\vec{x}} \tilde{\Phi}.
\end{equation}
In the above, we have introduced the reduced gravitational potential $\tilde{\Phi}$. The Poisson relation relating the density of particles to the potential $\tilde{\Phi}$ becomes:
\begin{equation}
	\nabla^2_{\vec{x}} \tilde{\Phi} = \frac{3}{2} H_0 \Omega_\text{m} \delta_\text{m}(\vec{x})\, .
\label{eq:Poisson_eq}
\end{equation}
To estimate densities from simulated particle positions we use the cloud in cell method \citep[see e.g.][]{HOCKNEYEASTWOOD1988}. Then the Poisson equation (\ref{eq:Poisson_eq}) can be solved in Fourier-space by exploiting numerically efficient Fast Fourier Transforms (FFTs). Since our approach requires many model evaluations the numerical implementation of this LSS model has been parallelized via the Message Passing Interface (MPI) \citep[see e.g.][]{BRUCK199719}.
The detailed description of solving the model equations is provided in Appendix \ref{appendix:pm_model}.

To use the non-linear particle mesh model, within the HMC framework, we also need to derive the corresponding gradient of model predictions with respect to changes in the initial conditions. More specifically, the gradient of the particle mesh simulator provides us with the response of the simulation with respect to small changes in the initial conditions. This gradient needs to be evaluated several times within the HMC sampling steps. As discussed above, we typically deal with on the order of ten million parameters, corresponding to the density amplitudes of the primordial fluctuations field. Evaluating such a gradient via finite differencing would be numerically prohibitive. In appendix \ref{appendix:tangent_adjoint_model} we, therefore, derive the tangent-adjoint model of the particle mesh simulator, which encodes the analytic derivative of the numerical forward simulation.

As demonstrated by figure \ref{fig:bench}, both the forward and the tangent adjoint model are fully parallel and exhibit near optimal scaling behaviour as a function of the number of tasks. Also note, that the adjoint model is only a factor two times more expensive than the forward model. Adjoint coding, therefore, provides us with an efficient means to calculate gradients of high dimensional functions.

\subsection{Modelling redshift space distortions}
\label{sec:rsd}
Optical distance estimation via spectroscopic redshift measurements is subject to systematic uncertainties due to the peculiar motions of observed galaxies. Corresponding Doppler effects increase observed redshifts if peculiar velocities are pointing away from the observer and decrease the redshift if velocities are pointing towards the observer. As a consequence exact galaxy positions in three-dimensional space are subject to some uncertainty.

Since the \borg{} algorithm exploits a physical model for LSS formation, predicting also the motion of matter, such redshift space distortions can be taken into account naturally. In this fashion, the \borg{} algorithm will not only exploit positional galaxy information but well also use the dynamic information encoded in the redshift space distortion effect. In principle, there are several different possibilities of implementing a redshift space distortions treatment into the \borg{} algorithm. For the sake of this work we calculate the redshift distorted particle positions as follows:
\begin{align}
\vect{s} & = \vect{r} + \gamma  \frac{\vect{v}\cdot  \vect{r}}{|\vect{r}|^2} \vect{r}\nonumber \\
  &= \vect{r} \left (1 + \gamma  \frac{\vect{v} \cdot \vect{r}}{|\vect{r}|^2}\right)\,,
\end{align}
with $\gamma=a/H(a)$, $\vect{r}=\vect{x}+\vect{x}_{\mathrm{min}}$ being the vector from the observer to a simulation particle and $\vect{v}=\vect{p}/a^2$, where $\vect{p}$ is the momentum vector as discussed in the previous section.
To generate density fields in redshift space we then use the redshift space coordinates $\vect{s}$ rather than the real space coordinates $\vect{x}$ of particles within the cloud in cell approach.

\begin{figure*}
  \begin{center}
    \includegraphics[width=\hsize]{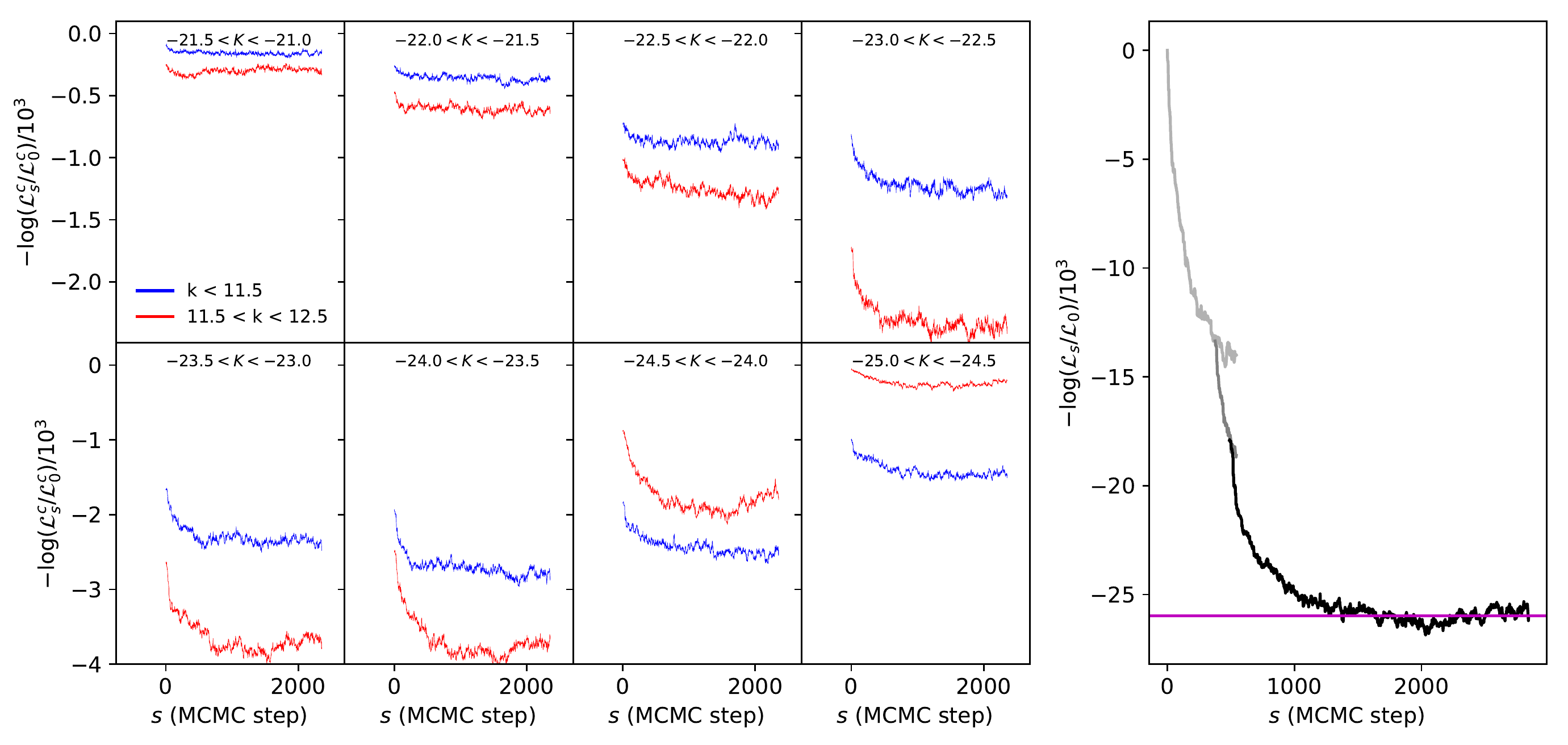}
  \end{center}
  \caption{
  \label{fig:trace_plot_neg_logLH}
  Trace plot of the negative differential logarithmic likelihood as a function of sampling steps $n$. The values represent logarithm of the ratios between the initial likelihood value obtained by the last sample calculated with a LPT model and subsequently evaluated particle mesh models. The right panel shows the trace of the total likelihood, while left panels break down the evolution of logarithmic likelihoods for the respective galaxy sub-catalogues as indicated in the panels.
 It can be seen that the Markov chain starts initially with high values for the negative logarithmic likelihood but successive sampling steps improve the consistency of inferred three-dimensional initial density fields with the observations. After $1200$ steps the trace plot settles at an average value for the negative logarithmic likelihood. In terms of Bayesian odds ratios when comparing the initial guess to a sample at sampling step $2500$ this is an improvement of about five orders of magnitude in logarithmic likelihood. }
\end{figure*}

\subsection{Modelling observed galaxies}
\label{sec:model_observed_galaxies}
One of the most challenging, and yet unsolved, aspects of analysing the galaxy distribution at non-linear regimes is to account for the biased relation between observed galaxies and the underlying distribution of dark matter. For the sake of this work we follow a common approach and approximate the galaxy biasing relation by a local but non-linear bias functions \cite{1988ApJ...333...21S,1995ApJS..101....1M,2000ASPC..201..360S,2011PhRvD..83h3518M,2012JCAP...11..016F,2014MNRAS.441..646N,2015MNRAS.446.4250A,2016arXiv161109787D}. More specifically we model the expected number of galaxies $ n^g$ via the following four parameter function as proposed in  \cite{2014MNRAS.441..646N}:
\begin{equation}
\label{eq:bias_model}
 n^g(\delta,\bar{N},\beta,\rho_g,\epsilon_g)=\bar{N}\,(1+\delta)^{\beta}\,\mathrm{e}^{-\rho_g\,(1+\delta)^{-\epsilon_g}}
\end{equation}
This parametrized bias function is a modification of a power-law bias model to account for suppressed clustering of galaxies in under dense regions by an additional exponential function.

Given this bias model, realizations of galaxy number counts are then assumed to follow a Poisson distribution with the Poisson intensity given as:
\begin{align}
\lambda_i(\delta,\bar{N},\beta,\rho_g,\epsilon_g))
  &= R_i\,n^g_i(\delta,\bar{N},\beta,\rho_g,\epsilon_g)\nonumber \\
&= R_i\bar{N}\,(1+\delta)^{\beta}\,\mathrm{e}^{-\rho_g\,(1+\delta)^{-\epsilon_g}} \, ,
\end{align}
where $R_i$ is the survey response operator consisting of the product of angular and radial selection function \citep[also see][for a discussion on the survey response operator]{JaschePspec2013,JLW15}.
The logarithm of the likelihood part of the posterior distribution of Equation~\eqref{eq:final_posterior} is then:
\begin{multline}
	\mathrm{ln}\left(\Pi(\{N^g_i\}|\delta,\bar{N},\beta,\rho_g,\epsilon_g)\right) =
	- \sum_i\Big(\lambda_i(\delta,\bar{N},\beta,\rho_g,\epsilon_g))\\
	-N^g_i\mathrm{ln}\left(\lambda_i(\delta,\bar{N},\beta,\rho_g,\epsilon_g))\right)+\mathrm{ln}(N^g_i!)\Big) \, ,
\label{eq:poisson_likelihood}
\end{multline}
with the Poisson intensity field $\lambda_i(\delta,\bar{N},\beta,\rho_g,\epsilon_g)$ given by:
\begin{align}
\lambda_i(\delta,\bar{N},\beta,\rho_g,\epsilon_g)&= R_i\,n^g_i(\delta,\bar{N},\beta,\rho_g,\epsilon_g)\nonumber \\
&= R_i\bar{N}\,(1+\delta)^{\beta}\,\mathrm{e}^{-\rho_g\,(1+\delta)^{-\epsilon_g}} \, .
\end{align}
 As can be seen, this is a highly non-linear data model not only due to the bias model but also due to the fact that for a Poisson distribution the noise is signal dependent and is not an additive nuisance.
The four bias parameters $\bar{N}$, $\beta$, $\rho_g$ and $\epsilon_g$ are a priori unknown and have to be inferred jointly together with initial and final density fields.

\begin{table*}[t]
\centering
\begin{tabular}{cccc}
\toprule
sample id & Magnitude range & cut & $-\log(\mathcal{L}_s/\mathcal{L}_0) $   \\
\midrule
\rowcolor{black!20} 1&$-21.5 \leq K \leq -21.0$ &$\ktmpp \le 11.5$     & -159.44  \\
2                    &                          &$11.5 < \ktmpp \le 12.5$                  & -322.79 \\
\rowcolor{black!20} 3&$-22.0 \leq K \leq -21.5$ &$\ktmpp \le 11.5$     & -358.89  \\
4                    &                          &$11.5 < \ktmpp \le 12.5$                  & -644.02 \\
\rowcolor{black!20} 5&$-22.5 \leq K \leq -22.0$ &$\ktmpp \le 11.5$     & -894.60 \\
6                    &      &$11.5 < \ktmpp \le 12.5$                  & -1280.55  \\
\rowcolor{black!20} 7& $-23.0 \leq K \leq -22.5$ &$\ktmpp \le 11.5$    & -1304.67  \\
8                   &                        &$11.5 < \ktmpp \le 12.5$ & -2361.86  \\
\rowcolor{black!20} 9&$-23.5 \leq K \leq -23.0$ &$\ktmpp \le 11.5$     & -2478.00 \\
10                  &                          &$11.5 < \ktmpp \le 12.5$                 & -3777.91 \\
\rowcolor{black!20} 11&$-24.0 \leq K \leq -23.5$ &$\ktmpp \le 11.5$    & -2853.92  \\
12                 &                          &$11.5 < \ktmpp \le 12.5$                 & -3653.12 \\
\rowcolor{black!20} 13&$-24.5 \leq K \leq -24.0$ &$\ktmpp \le 11.5$    & -2472.22  \\
14                 &                          &$11.5 < \ktmpp \le 12.5$                 & -1799.82  \\
\rowcolor{black!20} 15&$-25.0 \leq K \leq -24.5$ &$\ktmpp \le 11.5$    & -1467.34  \\
16                &                           &$11.5 < \ktmpp \le 12.5$& -207.63  \\
\bottomrule
\end{tabular}
\caption{
\label{tbl:model_comparison}
The table provides the logarithmic Bayes factors between a density field generated with the LPT and one with the PM model. It is interesting to note, that the PM model generally outperforms the LPT model in explaining the data. Most improvements are seen for the bright galaxy populations, while fainter galaxies seem to live in regions that can be approximated better by LPT models.}
\end{table*}

As discussed above, the advantage of our Bayesian approach is the possibility to add arbitrarily many parameter sampling procedures to the modular statistical programming approach via sequential block or Gibbs sampling methods. This is relevant since the biasing function as provided in equation~\eqref{eq:bias_model} will not be universally valid, but will require different bias parameters for different populations of galaxies.

In particular, in this work, we will split our galaxy sample into 16 different sub-samples selected by their absolute $K$-band magnitude.
The conditional posterior distribution for bias parameters given a sample of the three-dimensional final density field and corresponding galaxy number counts of the respective sub-samples is given by:
\begin{equation}
  \Pi(\bar{N},\beta,\rho_g,\epsilon_g|\{N^g_i\},\delta) \propto \Pi(\bar{N},\beta,\rho_g,\epsilon_g) \, \Pi(\{N^g_i\}|\delta,\bar{N},\beta,\rho_g,\epsilon_g) \, , \nonumber
\end{equation}
where the first factor on the right-hand side is the prior distribution of bias parameters and the second factor is the Poisson likelihood described in equation (\ref{eq:poisson_likelihood}). We typically follow a maximum agnostic strategy by setting uniform prior distributions for the bias parameters. Since the parameters of the bias model are all required to be positive we choose the following prior distribution:
\begin{equation}
  \Pi(\bar{N},\beta,\rho_g,\epsilon_g) = \Theta(\bar{N}) \, \Theta(\beta) \, \Theta(\rho_g) \, \Theta(\epsilon_g) \, ,
\end{equation}
where $\Theta(x)$ is the Heaviside function.
To explore the space of bias parameters we use a block sampling strategy by iteratively sampling individual parameters conditional on all other parameters. More specifically the algorithm executes the following block sampling scheme:
\begin{align}
\bar{N}^{n+1} & \sim \Pi(\bar{N}|\beta^n,\rho_g^n,\epsilon_g^n,\{N^g_i\},\delta) \nonumber \\
\beta^{n+1} &\sim \Pi(\beta|\bar{N}^{n+1},\rho_g^n,\epsilon_g^n,\{N^g_i\},\delta) \nonumber \\
\rho_g^{n+1} &\sim \Pi(\rho_g |\bar{N}^{n+1},\beta^{n+1},\epsilon_g^{n},\{N^g_i\},\delta) \nonumber \\
\epsilon_g^{n+1}& \sim \Pi(\epsilon_g|\bar{N}^{n+1},\beta^{n+1},\rho_g^{n+1},\{N^g_i\},\delta) \, \nonumber
\end{align}
where the superscript $n$ indicates the sampling step.

Iterating this procedure together with sequential density field updates will yield samples from the joint target distribution. Note, that this approach can easily be extended to account for additional survey systematics, such as foreground contaminations \citep[see e.g.][]{2017A&A...606A..37J}.

A particular challenge arises from the fact, that the specific non-linear shape of the bias function in equation~\eqref{eq:bias_model} does not allow to derive a simple direct sampling approach and we have to resort to standard MCMC techniques to generate bias parameter realizations. In order to have unit acceptance rates for the MCMC bias parameter sampling, we perform a sequence of slice sampling steps \citep{Neal00slicesampling}.

\begin{figure*}
\includegraphics[width=\hsize]{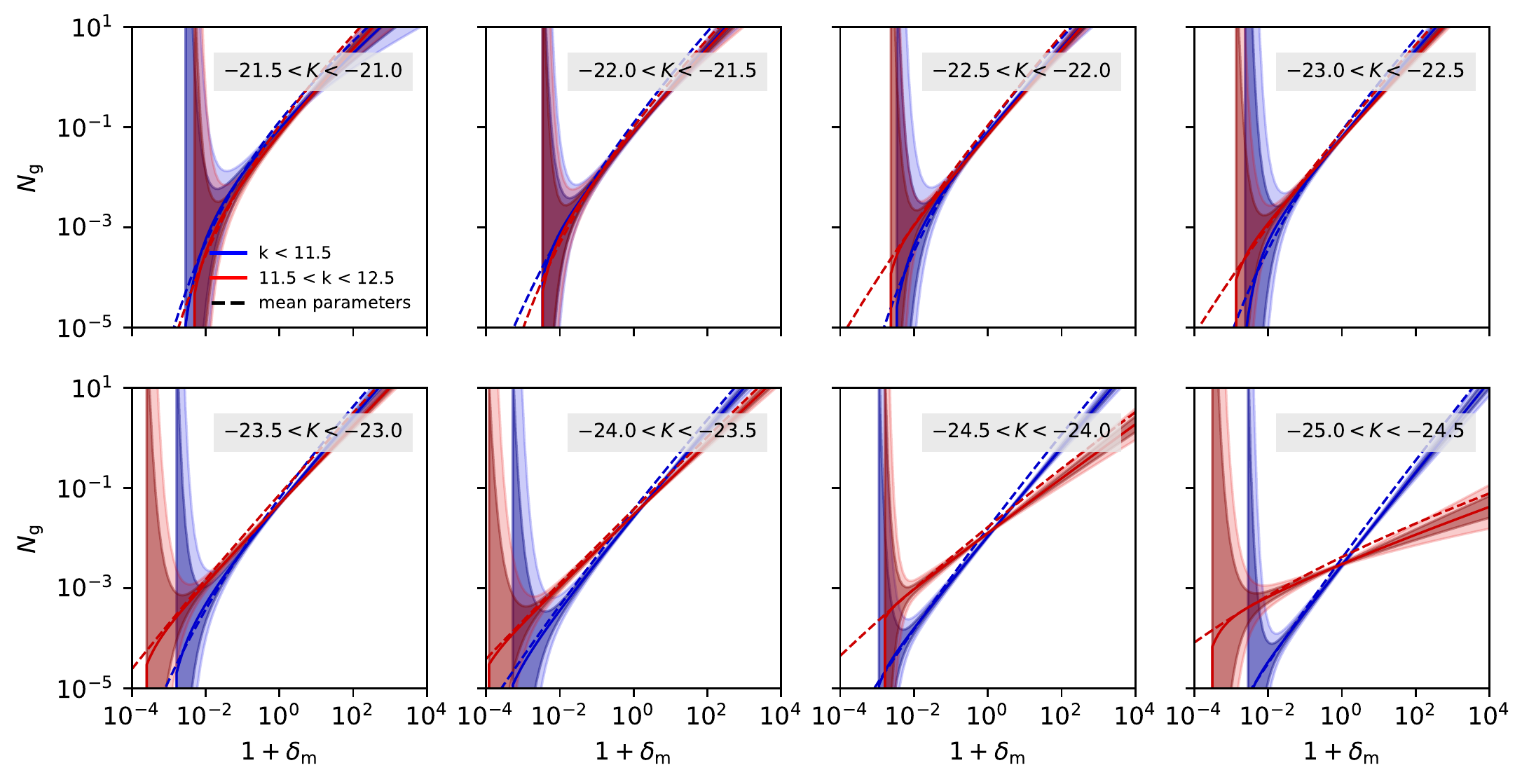}
\caption{\label{fig:mean_bias} Inferred non-linear bias functions for the 16 galaxy subsets of the 2M++ galaxy compilation in 8 absolute K-band magnitude bins. Blue and red lines correspond to ensemble mean bias functions, while shaded regions indicate the $1\sigma$ intervals for the two magnitude cuts as indicated in the upper left panel. Dashed lines correspond to bias functions estimated with the ensemble mean values of the bias parameters.}
\end{figure*}

\subsection{Robust inference with model errors}
\label{sec:robust_inference}
Most often Bayesian inference assumes that the distribution of the data agrees with the chosen class of likelihood models. More specifically it is assumed that the chosen data model is the true and correct explanation for the process that generated the actual observations. Already small deviations from these assumptions may greatly impact the Bayesian procedure.

Currently, several approaches to perform robust Bayesian inference with possible model misspecification have been proposed \citep[see e.g.][]{2014arXiv1412.3730G,2015arXiv150606101M,2016arXiv161101125B,2017arXiv170108515H,2017arXiv170801974F}. Robustness of inferences can be improved by conditioning on a neighbourhood of the empirical likelihood distribution rather than to the data directly \citep{2015arXiv150606101M}. When defining neighbourhoods based on relative entropy estimates it can be shown, that the resulting coarser posterior distribution can be approximated by raising the likelihood to a fractional power \citep{2015arXiv150606101M,2016arXiv161101125B,2017arXiv170108515H}. More specifically this amounts to tempering the likelihood distribution. For a Poisson distribution tempering is equivalent to using only a homogeneous subset of the data. This can be seen by raising the Poisson likelihood to some power $0\leq \beta \leq 1$:
\begin{align}
   \tilde{\Pi}(\{N^g_i\}|\{\lambda^g_i\}) & = \left( \Pi(\{N^g_i\}|\{\lambda^g_i\}) \right)^{\beta} \nonumber \\
  &= \prod_i \mathrm{e}^{-\beta \lambda_i} \frac{\lambda_i^{\beta \, N_i}}{\left(N_i!\right)^{\beta}} \nonumber \\
  & \propto  \prod_i \mathrm{e}^{-\beta \lambda_i} \left(\beta \,\lambda_i\right)^{\beta \, N_i}  \nonumber \\
  & \propto  \prod_i \mathrm{e}^{- \tilde{\lambda_i}} \left( \tilde{\lambda}_i\right)^{\tilde{N}_i}\, .   \label{eq:robust_likelihood}
\end{align}

As can be seen from equation~\eqref{eq:robust_likelihood}, coarsening the posterior distribution amounts to extracting information only from a homogeneous sub-sample of galaxies $\tilde{N}_i= \beta N_i $ while decreasing the expected Poisson intensity $\tilde{\lambda_i}=\beta \lambda_i$. This procedure thus is equivalent to increasing observational uncertainties resulting in conservative interpretations of the data.

The procedure of coarsening the posterior distribution, therefore, does not add spurious information to the inference, quite the contrary it uses only a fraction of the available information provided by the data set. Accessing the full potential of the data would require to develop more accurate data models to compare observations of galaxies to the underlying dark matter distribution at non-linear scales. This is a currently ongoing endeavour in the scientific community. For the sake of this work we choose $\beta=0.3$.

\section{Application to observed galaxy data}
\label{sec:data_application}
This section describes the application of the \borg{} algorithm to galaxy observations provided by the \tmpp galaxy compilation \citep{LH11}. Specifically here we will follow a similar approach as previously discussed in \citep{2016MNRAS.455.3169L}

\subsection{The 2M++ Survey}
\label{sec:data_description}
The 2M++ \citep{LH11} is a combination of the 2MASS Redshift Survey \citep[2MRS,][]{Huchra12}, with a greater depth and a higher sampling rate than the IRAS Point Source Catalogue Redshift Survey \citep[PSCZ,][]{Saunders00}. The photometry is based on the Two-Micron-All-Sky-Survey (2MASS) Extended Source Catalogue
\citep[2MASS-XSC,][]{Skrutskie06}, an all-sky survey in the $J$, $H$ and $K_S$ bands. Redshifts in the $K_S$ band of the 2MASS Redshift Survey
(2MRS) are supplemented by those from the Sloan Digital Sky Survey Data Release Seven \citep[SDSS-DR7, ][]{SDSS7}, and the Six-Degree-Field Galaxy Redshift Survey Data Release Three \citep[6dFGRS,][]{Jones09}.
Data from SDSS was matched to that of 2MASS-XSC using the NYU-VAGC catalogue \citep{BLANTON2005}. As the 2M++ combines multiple surveys, galaxy magnitudes from all sources were first recomputed by measuring the apparent magnitude in the $K_S$ band within a circular isophote at 20 mags arcsec$^{-2}$ . Following a prescription described in \cite{LH11}, magnitudes were corrected for Galactic extinction, cosmological surface brightness dimming and stellar evolution. Then the sample was limited to $\ktmpp \le 11.5$ in regions not covered by the 6dFGRS or the SDSS, and limited to $\ktmpp \le 12.5$ elsewhere. Incompleteness due to fibre-collisions in 6dF and SDSS was accounted for by cloning redshifts of nearby galaxies within each survey region as described in \cite{LH11}.

The galaxy distribution on the sky and the corresponding selection at $\ktmpp \le 11.5$ and $11.5 < \ktmpp \le 12.5$ are given in figure~\ref{fig:data_description}. The top row shows redshift incompleteness, i.e. the number of acquired redshifts versus the number of targets, for the two apparent magnitude bins. The lower row depicts the galaxy distribution as used in this work. We note that the galactic plane clearly stands out and that the incompleteness is evidently inhomogeneous and strongly structured.

\begin{table*}[t]
\centering
\begin{tabular}{ccccccc}
\toprule
sample id & Magnitude range & cut & $\bar{N}$ & $\beta$  & $\rho_g$ & $\epsilon_g$\\
\midrule
\rowcolor{black!20} 1&$-21.5 \leq K \leq -21.0$ &$\ktmpp \le 11.5$     & 0.35 & 0.65 & 0.98 & 0.28 \\
2                    &                          &$11.5 < \ktmpp \le 12.5$                  & 0.31 & 0.74 & 1.06 & 0.26 \\
\rowcolor{black!20} 3&$-22.0 \leq K \leq -21.5$ &$\ktmpp \le 11.5$     & 0.37 & 0.77 & 1.11 & 0.19 \\
4                    &                          &$11.5 < \ktmpp \le 12.5$                  & 0.24 & 0.74 & 0.85 & 0.26 \\
\rowcolor{black!20} 5&$-22.5 \leq K \leq -22.0$ &$\ktmpp \le 11.5$     & 0.25 & 0.75 & 0.91 & 0.27 \\
6                    &      &$11.5 < \ktmpp \le 12.5$                  & 0.40 & 0.79 & 1.30 & 0.12 \\
\rowcolor{black!20} 7& $-23.0 \leq K \leq -22.5$ &$\ktmpp \le 11.5$    & 0.19 & 0.80 & 0.81 & 0.25 \\
8                   &                        &$11.5 < \ktmpp \le 12.5$ & 0.24 & 0.76 & 1.08 & 0.12 \\
\rowcolor{black!20} 9&$-23.5 \leq K \leq -23.0$ &$\ktmpp \le 11.5$     & 0.16 & 0.79 & 0.97 & 0.20\\
10                  &                          &$11.5 < \ktmpp \le 12.5$                 & 0.28 & 0.73 & 1.33 & 0.07\\
\rowcolor{black!20} 11&$-24.0 \leq K \leq -23.5$ &$\ktmpp \le 11.5$    & 0.19 & 0.77 & 1.61 & 0.09 \\
12                 &                          &$11.5 < \ktmpp \le 12.5$                 & 0.14 & 0.67 & 1.31 & 0.05 \\
\rowcolor{black!20} 13&$-24.5 \leq K \leq -24.0$ &$\ktmpp \le 11.5$    & 0.05 & 0.83 & 1.23 & 0.11 \\
14                 &                          &$11.5 < \ktmpp \le 12.5$                 & 0.04 & 0.51 & 0.97 & 0.085 \\
\rowcolor{black!20} 15&$-25.0 \leq K \leq -24.5$ &$\ktmpp \le 11.5$    & 0.01 & 0.88 & 0.98 & 0.12 \\
16                &                           &$11.5 < \ktmpp \le 12.5$& 0.01 & 0.24 & 1.19 & 0.1 \\
\bottomrule
\end{tabular}
\caption{
\label{tbl:bias_params}
The table provides the estimated mean parameter values for the bias functions corresponding to the respective magnitude cuts. Note that for non-linear functions, such as the truncated power-law bias, the means of function parameters do not necessarily agree to the mean of the bias function. As a comparison, in figure~\ref{fig:mean_bias}, we also plotted the bias function corresponding to the means of the parameter values. }
\end{table*}

In addition to the target magnitude incompleteness, and the redshift angular incompleteness, one may also worry about the dependence of the completeness with redshift. This is not a problem for the lower $\ktmpp \le 11.5$ which is essentially 100\% complete. We do not expect much effect in the fainter magnitude bins as the spectroscopic data come from SDSS and 6dFGRS which have both a homogeneous sampling and have fainter magnitude limits as the 2M++.

We account for radial selection functions using a standard luminosity function $\Phi(L)$ proposed by \cite{SCHECHTER1976}. Using this function we can deduce the expected number of galaxies in the absolute magnitude range, observed within the apparent magnitude range of the sample at a given redshift. The $\alpha$ and $M^*$ parameters are given for the K$_S$-band in the line labelled "$|b|>10, K < 11.5$" of the table 2 of \cite{LH11}, i.e. $\alpha=-0.94$, $M^*=-23.28$. The target selection completeness of a voxel, indexed by $p$, is then
\begin{equation}
	c^t_p = \frac{\int_{\mathcal{V}_p} \text{d}^3 \vect{x} \int_{L_\text{app}(|\vect{x}|)}^{L_\text{max}} \Phi(L)\text{d}L}{V_p \int_{L_\text{min}}^{L_\text{max}} \Phi(L)\text{d}L}\,,
\end{equation}
where $\mathcal{V}_p$ the co-moving coordinate set spanned by the voxel, and $V_p = \int_{\mathcal{V}_p} \text{d}^3 \vect{x}$.
The full completeness of the catalogue is derived from the product of $c^t$ and the map corresponding to the considered apparent magnitude cut given in the upper panels of the figure~\ref{fig:data_description} after its extrusion in three dimensions.

Our sampling approach accounts for luminosity dependent galaxy biases. In order to do so the galaxy sample is subdivided into eight bins of same width and without spacing in absolute $K$-band magnitude in the range $-25\le\ktmpp \le-21$. The galaxy sample is further split into two subsets depending on the apparent magnitude: if $\ktmpp \le 11.5$ it belongs to the first set, otherwise, $11.5 < \ktmpp \le 12.5$ it belongs to second set. This yields a total of 16 galaxy subsets.
The bias parameters of each of these subsets are inferred jointly within the multiple block sampling framework as described above. This permits to properly marginalize over these unknown bias parameters within the \borg{} framework. Splitting the galaxy sample permits us to treat each of these sub-samples as an individual data set, with its respective selection effects, biases and noise levels.

\subsection{Non-linear analysis with the \borg{} algorithm}
\label{sec:non_lin_analysis_borg}
The analysis of the 2M++ galaxy sample is conducted on a cubic Cartesian domain of side length of $677.77$\Mpch{} consisting of $256^3$  equidistant grid nodes. This results in a total of $\sim 1.6\times 10^7$ inference parameters, corresponding to primordial density fluctuation amplitudes at respective grid nodes. The inference procedure thus yields data constrained realizations of initial conditions with a Lagrangian grid resolution of about $\sim 2.65$\Mpch.

To integrate the effect of the growth of large scale structure, we assume a fixed standard $\Lambda$CDM cosmology with the following set of cosmological parameters ($\Omega_m=0.307$, $\Omega_{\Lambda}=0.693$, $\Omega_{b}=0.04825$, $h=0.705$, $\sigma_8=0.8288$, $n_s=0.9611$). The cosmological power-spectrum of initial conditions, required by our \borg{} run, was evaluated via the prescription provided by \cite{EH98} and \cite{EH99}.
To guarantee a sufficient resolution of inferred final Eulerian density fields, we oversample the initial density field by a factor of eight, requiring to evaluate the particle mesh model with \(512^3\) particles in every sampling step.

Running the Markov chain with a particle mesh model is numerically expensive. To save some computation time we first ran the Markov Chain for $6783$ transition steps using the numerically less expensive LPT model. This procedure yielded a good starting point for a Markov chain running the full particle mesh model.

\subsection{Testing burn-in behaviour}
\label{sec:sampler_convergence}
To test the burn-in behaviour of the initial LPT sampling procedure we followed a similar approach as described in our previous works \citep[see e.g.][]{JASCHEBORG2012,JaschePspec2013,JLW15,2016MNRAS.455.3169L}. In particular, we initialize the Markov chain with an over-disperse random Gaussian initial density field with amplitudes a factor ten times smaller than expected in a standard $\Lambda$CDM scenario. Starting from such an over-dispersed state the Markov Chain will then follow a persistent drift towards more reasonable regimes in parameter space.

\begin{figure*}
	\begin{center}
		\includegraphics[width=\hsize]{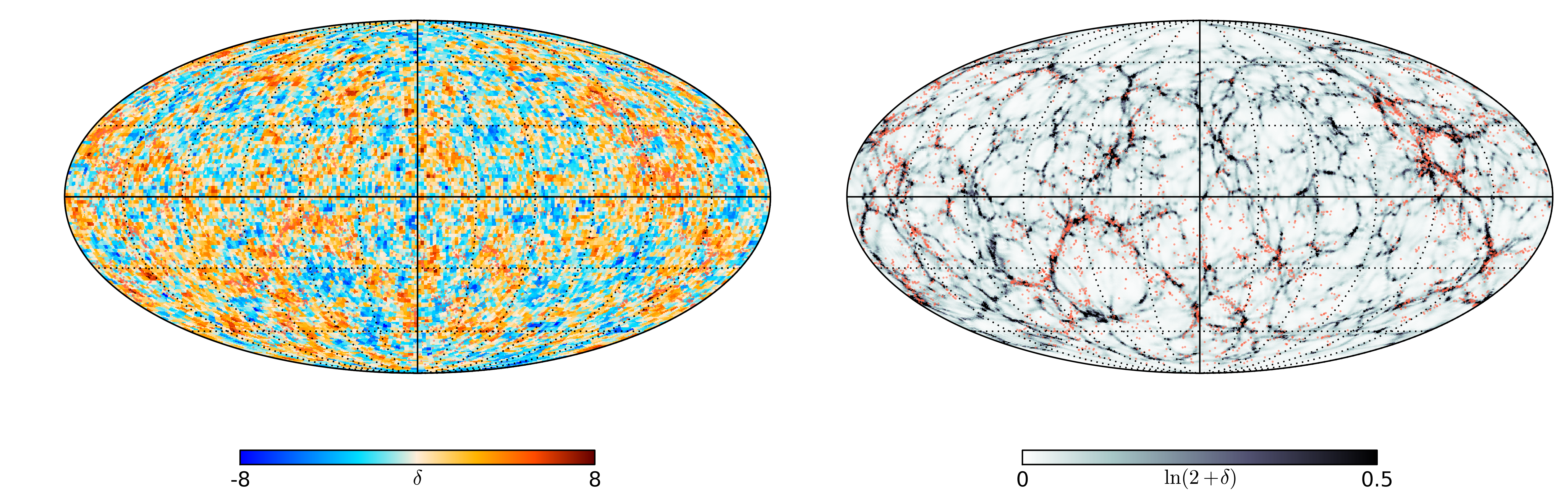}
  	\end{center}
 	\caption{Spherical slices through a data constrained realization of the three-dimensional initial (left panel) and final density field (right panel) at a distance of $R= 100$\Mpch{} from the observer. Initial density fields correspond to the epoch of a cosmic scale factor $a=0.001$ while non-linear final density fields are evaluated at the present epoch ($a=1$). One can see the correspondence of large scale over-densities in the initial conditions and corresponding structures in the gravitationally evolved density field.  Red dots in the right panel denote the observed galaxies in the 2M++ survey. As can be seen observed galaxies trace the inferred dark matter distribution. \label{fig:pm_sky_realization}}
\end{figure*}

To illustrate this initial automatic adjustment of the algorithm in figure~\ref{fig:Pk_burnin}, we illustrate the sequence of posterior power-spectra measured from subsequently inferred three-dimensional initial density fields during the initial burn-in phase. It can be seen that the posterior power-spectra drift towards the expected target power-spectrum. After about 4000 transition steps power-spectra oscillate around the expected values.
In addition, we also trace the evolution of the one point (1-pt) distribution of inferred primordial density fluctuation during the burn-in period. As can be seen in the right panels of \ref{fig:Pk_burnin} the 1-pt distribution of successive density samples approaches the expected normal distribution within about 4000 transitions of the Markov chain.
These results show no sign of any particular systematic artefact and clearly indicate a healthy burn-in behaviour of the chain.

This initial LPT Markov run was stopped after $6783$ transitions and the final result was used as the initial point to start a run with the full particle mesh model. In order to monitor the improvements that the PM model imparts on the previous LPT results, we plot the trace the negative logarithmic likelihood distribution as a function of sample number $n$ in Fig \ref{fig:trace_plot_neg_logLH}.

As can be seen initially the Markov chain starts at high values of the negative logarithmic likelihood. These initial values correspond to the LPT results. During subsequent sampling steps the negative logarithmic likelihood values then drop by more than four orders of magnitude as the particle mesh model method successively improves the inferred non-linear density fields. Finally, it can be seen that the Markov Chain settles at a constant value. At this point we start recording samples of the Markov chain.

It is very interesting to note that the initial starting point of the chain corresponds to a density field inferred with the LPT model, while subsequent samples correspond to density fields inferred with the non-linear particle mesh model. 
Since figure~\ref{fig:trace_plot_neg_logLH} basically shows that the logarithms of the likelihood ratios of the first LPT density fields to all subsequent PM density fields, the plot qualifies as a Bayesian model test in terms of Bayes odds 
ratios. Realizing this fact demonstrates that the data clearly favours density fields inferred with the PM method. On a Jeffreys scale, the statement is far more than decisive. While this statement is true for the combined logarithmic likelihood of all galaxy sub-samples, we may also look at the improvements for the individual catalogues. To show that 
point, we also plot in figure~\ref{fig:trace_plot_neg_logLH} the traces of the negative logarithmic likelihoods for the individual sub-catalogues. As can be seen, especially the fainter galaxies seem to live in regimes of the cosmic LSS that can be acceptably approximated by the LPT method even though PM also provides significant improvements there 
To quantify this effect, we present in table~\ref{tbl:model_comparison} the actual logarithmic likelihood ratios between the initial LPT density model and the last density sample generated with the PM model.
It may be interesting to investigate the details of this effect in future analyses, as it may provide a guideline to optimally select galaxies for cosmological analyses.

To conclude this first diagnostic, the Markov Chain stabilizes after $\sim 1200$ samples the moment from which on we start recording $1500$ samples.
As such the presented \borg{} run does not qualify for a thorough Markov analysis but it provides us with sufficient information on the non-linear dynamics in the Nearby Universe and uncertainty quantification to warrant robust scientific analyses. The exact state of the Markov Chain is stored in a restart file permitting to resume the chain at any later time if the generation of more samples will be required at any point in the future.

\section{Results on cosmological inference}
\label{sec:inference_results}
This section provides an overview of the inference results obtained by applying the \borg{} algorithm to the 2M++ galaxy compilation. In particular, the present work focusses at reconstructing the non-linear LSS and its dynamics in the Nearby Universe.

\begin{figure*}
\centering
 \includegraphics[width=\hsize]{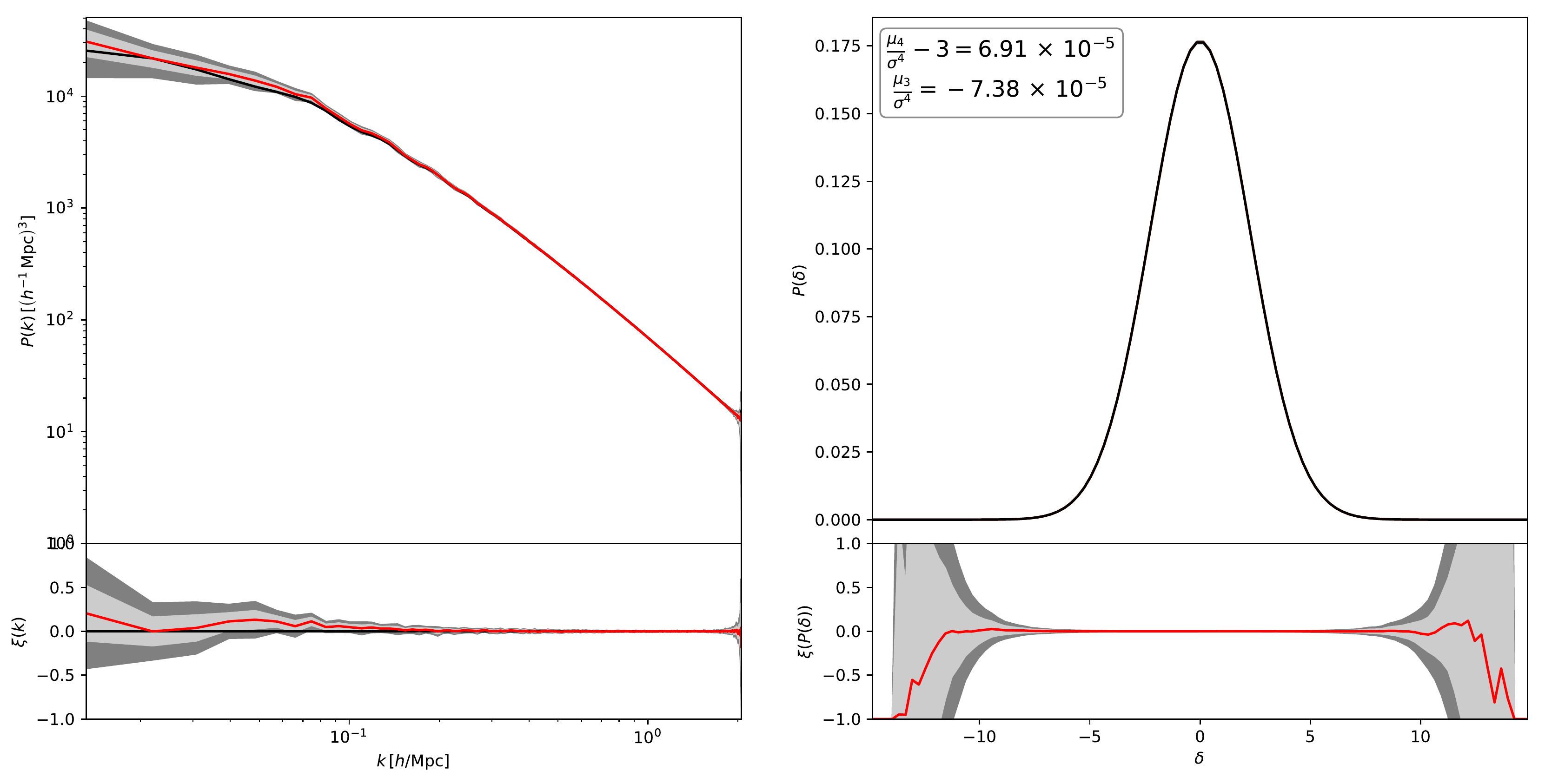}
 \caption{Posterior power-spectra measured from inferred initial density fields (left panel) and the one-point distribution of primordial density fluctuations (right panel) . The plot demonstrates that individual data constrained realizations of the initial density field constitute physically valid quantities. Throughout the entire domain of Fourier modes considered in this work we do not observe any particular bias or attenuation of measured cosmic power-spectra. The measured posterior one-point distribution of primordial fluctuations is compatible with a fiducial normal one-point distribution with variance corresponding to the cosmological parameters as described in section \ref{sec:non_lin_analysis_borg}. Tests of kurtosis and skewness, as indicated in the right panel, confirm inferred initial density fluctuations to follow Gaussian statistics.\label{fig:borg_pm_posterior_density_stats}
 }
\end{figure*}

\subsection{Inferred galaxy biases}
\label{sec:galaxy_biases}
To properly account for the unknown relationship between observed galaxies and the underlying dark matter field, the \borg{} algorithm jointly infers the parameters of a phenomenological, non-linear truncated power-law bias model as discussed in section \ref{sec:model_observed_galaxies}. In particular, the algorithm exploits an iterative block sampling framework to perform a joint Markov Chain over the actual target parameters, the amplitudes of the 3D density field, and the nuisance parameters associated to the employed data model. As a consequence, the \borg{} algorithm also naturally provides measurements of the non-linear galaxy bias.

As described in section~\ref{sec:data_description}, for the sake of this work, we have subdivided the galaxy sample of the 2M++ galaxy compilation into eight bins of same width in absolute $K$-band magnitude in the range $-25<\ktmpp <-21$ respectively for the two selections at $\ktmpp \le 11.5$ and $11.5 < \ktmpp \le 12.5$. This results in a total of 16 sub-samples, for which the \borg{} algorithm infers the respective set of bias parameters. In this fashion, our algorithm can account for the respective systematics in the individual galaxy samples while exploiting their joint information.

Figure~\ref{fig:mean_bias} represents our measurements of the ensemble mean bias functions and corresponding one-sigma uncertainties for the 16 galaxy sub-samples. By comparing inferred bias functions between the two selections at $\ktmpp \le 11.5$ and $11.5 < \ktmpp \le 12.5$, it can be seen that within the absolute $K$-band magnitude in the range $-23<\ktmpp <-21$ the respective bias functions are in agreement. This demonstrates that the galaxies in both selections show the same clustering behaviour for the given absolute mass range. However for  $K$-band magnitudes in the range $-25<\ktmpp <-23$, we observe an increasing difference between the galaxy bias functions of the two selections at $\ktmpp \le 11.5$ and $11.5 < \ktmpp \le 12.5$. In particular, the brighter galaxies in the $\ktmpp \le 11.5$ seem to have a steeper biasing relation as a function of the underlying density field than those in the $11.5 < \ktmpp \le 12.5$ selection.
The true origin of this behaviour is not clear, but it could indicate a contamination or systematic effect of the galaxies selected at $11.5 < \ktmpp \le 12.5$.
These phenomenological bias function shapes agree well with previous findings in numerical simulations \citep[][]{2008ApJ...678..569S}.

In table~\ref{tbl:bias_params} we also report the ensemble mean values for the respective bias parameters. Note, that generally for non-linear functions, the bias function evaluated with the mean parameter values will not correspond to the ensemble mean bias function. This is a simple statement of non-Gaussian and non-linear statistics. To illustrate this fact in \ref{fig:mean_bias} we also plotted the bias functions evaluated at the ensemble mean parameter values.

In section~\ref{sec:mass_reconstruction} we also demonstrate that the masses estimated from our inferred dark matter density fields agree with complementary measurements via X-ray or weak lensing measurements.
This is a strong indication of the fact that our inferred bias functions are a plausible description of the relationship between observed galaxies and the underlying dark matter distribution.

\subsection{The 3D density field in the Nearby Universe}
\label{sec:result_3d_density}
The \borg{} algorithm aims at inferring detailed three-dimensional maps of the matter distribution in the Nearby universe constrained by data of the 2M++ galaxy compilation. In fact, our method simultaneously constrains the present non-linear matter density field and the primordial density fluctuations from which they originate.

We infer the primordial field of matter fluctuations on a  Cartesian equidistant grid of resolution $\sim 2.65$ \Mpch. All primordial matter fluctuations are inferred in the initial Lagrangian space while present structures are determined at their Eulerian coordinates. Since structures collapse under their own gravity, the resolution of the initial Lagrangian grid is sufficiently high to resolve major features in the present Universe, such as the Coma cluster. Corresponding non-linear density fields, as well as positions and velocities of simulation particles, are then estimated by evaluating inferred primordial initial conditions via the PM structure formation model.

\begin{figure*}
\begin{center}
	 \includegraphics[width=\hsize]{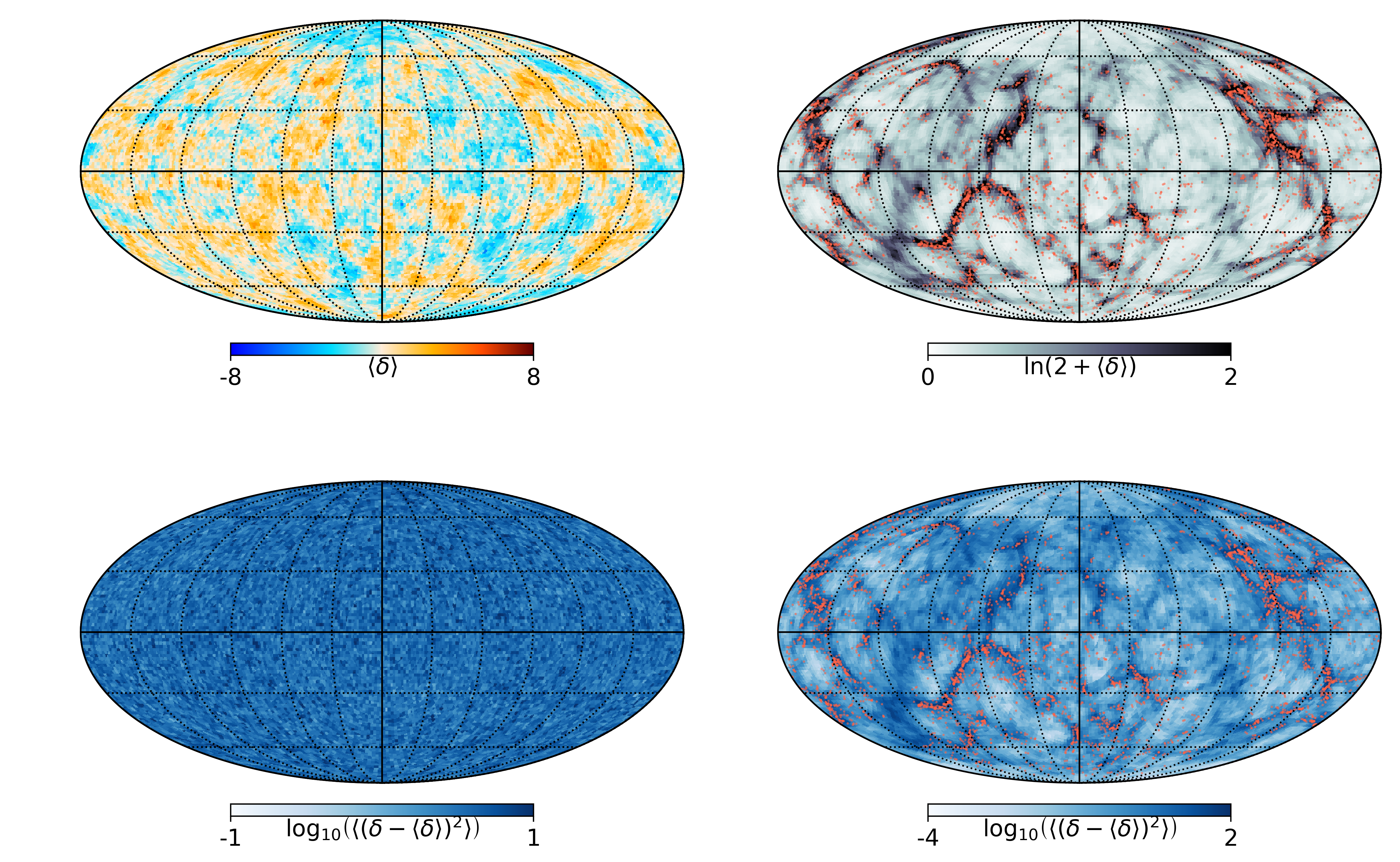}
  \end{center}
 \caption{Spherical slices through the ensemble mean of the three-dimensional initial (left panel) and final density field (right panel) and corresponding pixel-wise variances (lower panels) at a distance of $R= 100$\Mpch{} from the observer. It is interesting to note, that the pixel-wise variance for the final density field imprints the cosmic large scale structure. Correlations between signal and noise are expected for any point process, such as the generation of galaxy observations. The \borg{} algorithm correctly accounts for these effects.   \label{fig:pm_sky_realization_mean}
 }
\end{figure*}

The \borg{} algorithm not only provides simple point estimates, such as mean or maximum a posteriori value but rather provides a numerical approximation to the target posterior distribution in terms of an ensemble of Markov samples. This ensemble of data constrained realizations contains all the information on the three-dimensional density field that can be extracted from the noisy and incomplete data set and at the same time quantifies corresponding observational uncertainties that are necessary in order to not misinterpret the observations.
Unlike point estimates, these posterior realizations constitute physical meaningful quantities which do not suffer from any attenuation or bias due to systematic effects in the data \citep[also see discussions in][]{JASCHEBORG2012,JLW15,2016MNRAS.455.3169L}.

As an illustration of the property, we show in figure~\ref{fig:pm_sky_realization} spherical slices through data constrained realizations of the three-dimensional initial and final density fields, projected onto a HEALPix map  \citep{HEALPIX}. The right panel depicts the non-linear density field at a distance of $R= 100$\Mpch from the observer overlaid by the actually observed galaxies in the 2M++ galaxy compilation.
As can be seen, our algorithm recovered a highly detailed map of the filamentary cosmic web. Observed galaxies in the 2M++ survey trace the recovered spatial distribution of the underlying dark matter. Note that regions that have been traced poorly by galaxies are visually not distinct from those constrained by observations.

This is a crucial feature of the \borg{} algorithm, which augments the information obtained from observations with statistically correct information on the cosmic LSS in unconstrained regions of the galaxy survey. As such, each posterior sample represents a physically meaningful and plausible realization of the actual dark matter distribution in the Universe. The left panel of figure~\ref{fig:pm_sky_realization} shows the corresponding slice through a realization of the initial fluctuations field. This field represents the proto-structures from which the presently observed structures (shown in the right panel) have formed via gravitational collapse. We will further discuss the possibility to follow the structure formation history of objects below in section~\ref{sec:form_hist}.

To further support the qualitative statement that individual posterior realizations represent physically plausible quantities, we test the one- and two-point statistics of inferred primordial density fluctuations realizations. These results are presented in figure~\ref{fig:borg_pm_posterior_density_stats}. As can be seen, the \borg{} algorithm recovers the cosmic LSS over a huge dynamic range covering more than three orders of magnitude in amplitudes of the power-spectrum. In comparison to a fiducial cosmological power-spectrum, corresponding to the set of cosmological parameters as described in section~\ref{sec:non_lin_analysis_borg}, measured power-spectra do not show particular signs of bias or attenuation throughout the entire domain of Fourier modes considered in this work.

We have also tested the one-point probability distribution of inferred primordial density fluctuations. As demonstrated by the right panel of figure~\ref{fig:borg_pm_posterior_density_stats}, inferred primordial density amplitudes are normally distributed. In particular, the inferred one-point distribution is consistent with the expected fiducial Gaussian distribution determined by the cosmological parameters provided in section~\ref{sec:non_lin_analysis_borg}. Residual uncertainties remain only in the tail of the distribution which is dominated by sample variance. To further test the normality of the inferred one-point distribution, we also test the kurtosis $\mu_4 / \sigma^4-3$ and skewness
$\mu_3 / \sigma^4$ as indicated in the right panel of \ref{fig:borg_pm_posterior_density_stats}. Since both values agree numerically with zero, these results demonstrate that the inferred one-point distribution of matter fluctuations is compatible with Gaussian statistics.

\begin{figure*}
 \begin{center}
   \includegraphics[width=\hsize]{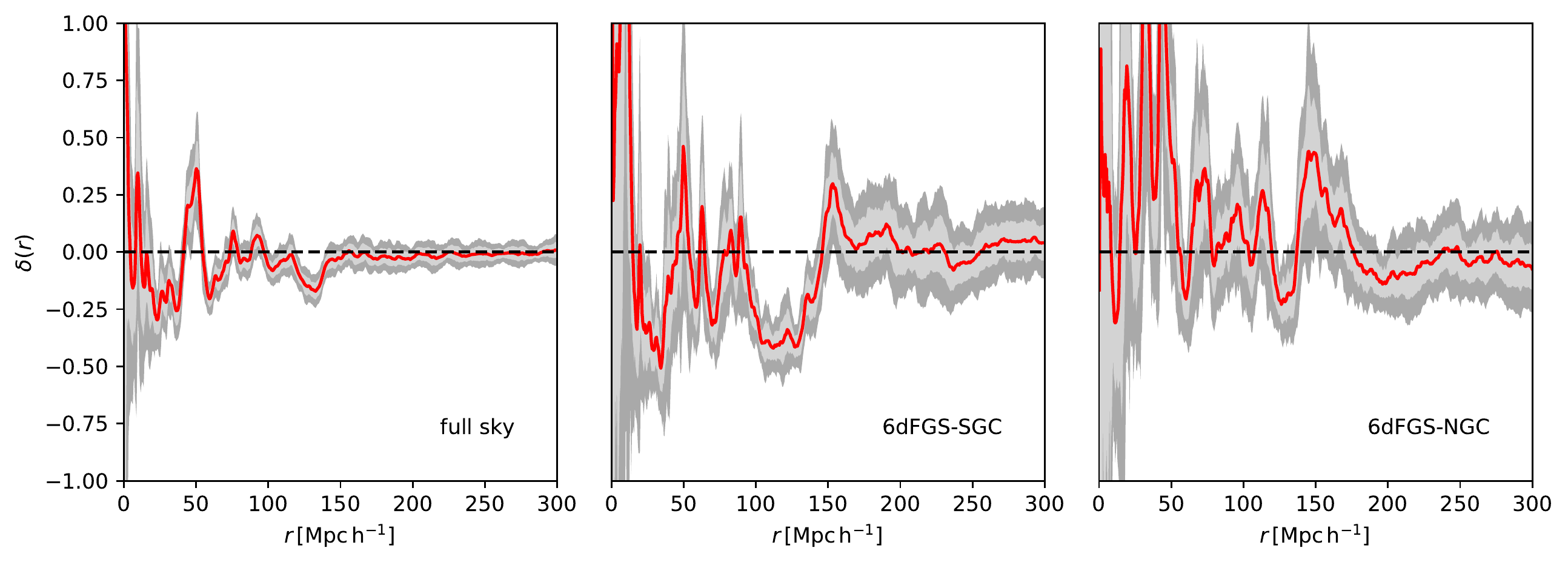}
 \end{center}
 \caption{\label{fig:radial_profile} Radial density profiles of matter fluctuations in shells of radius $r$ around the observer. The left panel corresponds to spherical shells covering the full sky, while middle and right panel show the density fluctuations for the 6dFGS-SGC and 6dFGS-NGC region, as defined in \citet[][]{2014MNRAS.437.2146W}, respectively. Red lines indicate our ensemble mean estimate, while dark and light gray shaded regions indicate the $2\sigma$ and $1\sigma$ limit, respectively. The black dashed line corresponds to cosmic mean density. As can be seen our inference results do not indicate striking evidence for a large scale under-dense region using the 2M++ data.}
\end{figure*}

The unbiased reconstruction of the primordial power-spectrum is also a good indicator that the \borg{} algorithm correctly accounted for various systematic effects. In particular improper treatment of survey geometries, foreground contamination, selection effects, and luminosity-dependent galaxy biases would typically result in excessive erroneous large-scale power \citep[see e.g.][]{TEGMARK_2004,2010JCAP...05..027P,JASCHESPEC2010,2014MNRAS.444....2L,2017A&A...606A..37J}.

As a remark, we have found that the forward modelling approach is particularly sensitive to these effects. Wrong assumptions on galaxy biasing or selection effects would not only introduce erroneous large-scale power to the density field but also affect large-scale matter flows, that are required to translate the initial matter distribution to the present non-linear density field. In particular, the non-linear and non-local nature of the employed particle mesh structure formation model enhances such effects leading to obviously erroneous results. In turn, the high sensitivity of the physical forward approach towards non-linear and luminosity-dependent galaxy biases promises to provide accurate constraints on the relation between observed galaxies and the underlying dark matter distribution, as discussed in the previous and the following sections.

The entire ensemble of physically plausible density field realizations forms a numerical approximation to the target posterior distribution. This permits us to derive any desired statistical summary and quantify corresponding uncertainties. As an example in the figure \ref{fig:pm_sky_realization_mean}, we show the ensemble mean density fields and corresponding pixel-wise standard deviations.
As can be seen, the initial and final ensemble mean density fields both approach cosmic mean density in regions which are poorly constrained by observations. This result is expected. When data does not provide any constraining information, the algorithm will return cosmic mean density on average in unobserved regions. This agrees with the prior assumption of the zero-mean Gaussian distribution of cosmic initial conditions, as described in section \ref{sec:borg_model}.
These results are also in agreement with our previous findings \citep[see e.g.][and discussions therein]{JASCHEBORG2012,JLW15}.

\begin{figure*}
	\begin{center}
    	\includegraphics[width=\hsize]{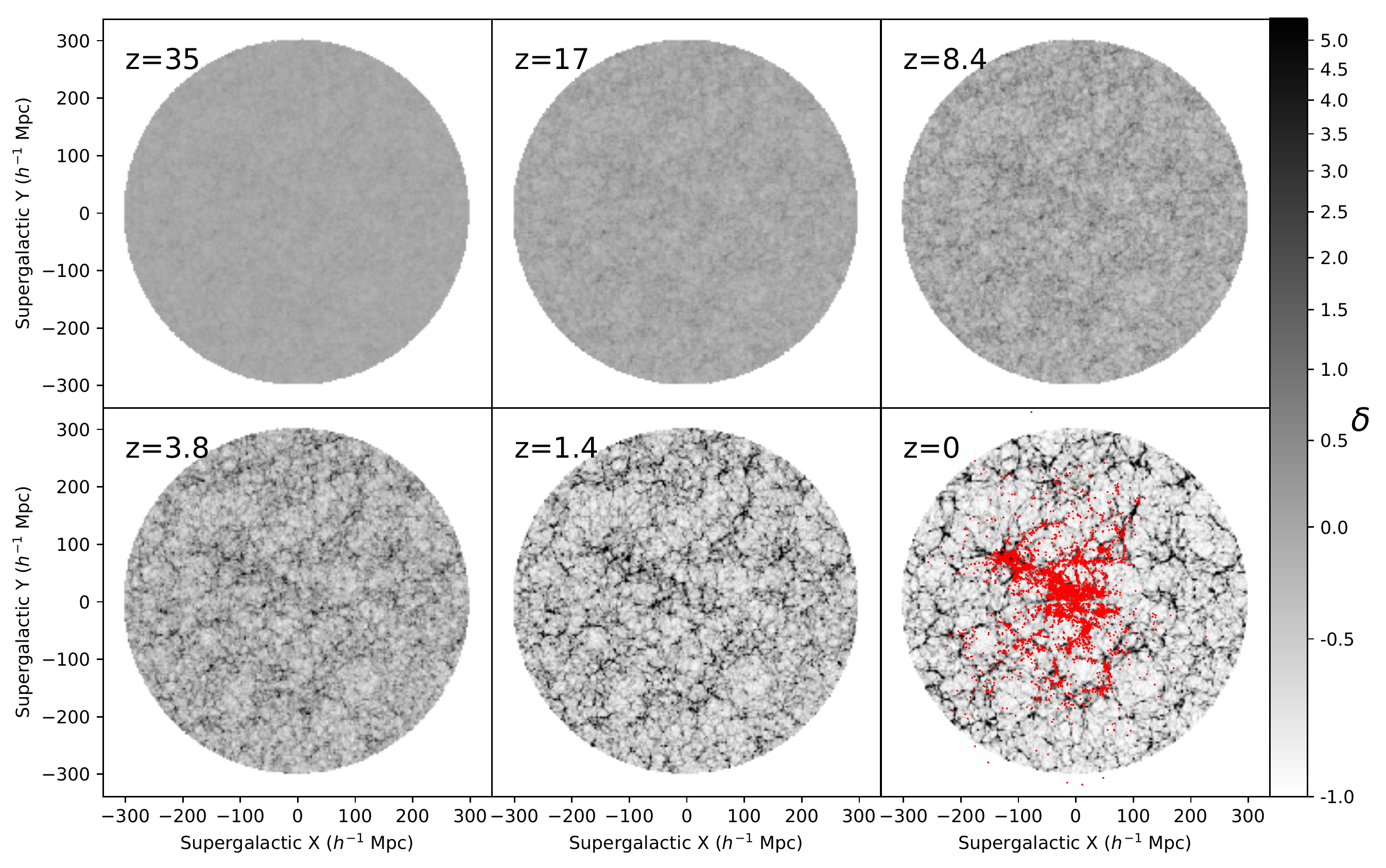}
    \end{center}
    \caption{Slices through the three-dimensional density field of the Supergalactic plane at different cosmic epochs as indicated in the respective panels. The sequence of plots represents a plausible formation history of structures in the Supergalactic plane. Initially, proto-structures arise from almost homogeneous matter distributions forming, through gravitational interaction, the cosmic web of clusters and filaments. To illustrate that this formation history yields actually observed structures we overlay the density field with galaxies of the 2M++ survey in the lower right panel. It can be seen that the galaxies in the Supergalactic plane trace the recovered density field.      \label{fig:chrono_sg}}
\end{figure*}

Figure~\ref{fig:pm_sky_realization_mean} also presents voxel-wise standard deviations of inferred density amplitudes at respective positions inside the analysis domain. It is interesting to note, that estimated standard deviations of the final density amplitudes reflect an imprint of the cosmic large-scale structure. In particular one can recognize the imprinted pattern of filaments and clusters.  This is an immediate consequence of the non-linear noise properties of the galaxy point distribution. In particular, there will be a correlation between signal and noise for any inhomogeneous point process, such as the one generating the galaxy distribution. More explicitly due to the galaxy formation processes, we expect to find more galaxies in high-density regions than in low-density regions. Any such galaxy formation process will, therefore, induce correlations between the underlying dark matter distribution and the noise of the galaxy sample.  As demonstrated by figure \ref{fig:pm_sky_realization_mean} the algorithm correctly accounts for this non-linear relation between noisy observations and the underlying density field.

In contrast, standard deviations of primordial density amplitudes are distributed more homogeneously and show no sign of correlation with the field of primordial fluctuations. This result is anticipated, due to the propagation of information from final to initial density fields as mediated by the physical forward model.

In the course of structure formation, over-densities covering a larger Lagrangian volume in the initial conditions will collapse gravitationally to form higher density objects in the final conditions, covering much smaller volumes. These clusters of matter are then traced by observed galaxies of the 2M++ survey, which provide the data constraints on inferred density fields. While this data constraining information is contained in a smaller Eulerian volume, defined by the cluster at the final conditions, it is distributed over the larger initial Lagrangian volume of the proto-cluster when propagated backward in time.

A similar argument applies to information propagation in void regions. Since voids expand over cosmic times, data constraining information, tracing the outskirts of voids at the final conditions, will be propagated back to a smaller Lagrangian volume in the initial conditions. The process of information propagation through the forward model, therefore, homogenizes the information across inferred initial conditions. This behaviour is correctly reflected by figure \ref{fig:pm_sky_realization_mean}.

In summary, the \borg{} algorithm provides us with highly detailed and physically plausible representation of three-dimensional non-linear cosmic structures and their corresponding initial conditions. The  \borg{} algorithm also provides quantification of uncertainties for all inferred quantities via a sophisticated Markov Chain Monte Carlo approach.

\subsection{No evidence for a local hole}
\label{sec:local hole}
The distribution of matter in our local environment has recently attracted greater interest due to the observed tension between local and CMB measurements of $H_0$ \citep[see e.g.][]{Riess2016,2016A&A...594A..13P}. This has triggered some activity to investigate whether the local cosmic expansion rate is faster than in the remaining Universe due to the existence of a local large scale under density. Indeed several works have claimed growing evidence for the existence of such a local hole. This large-scale void is believed to extend to depth of $r \sim 150$\Mpch{} and beyond with mass deficits of about $\sim 4\% - 40\%$  \citep[see e.g.][]{2003MNRAS.345.1049F,2004MNRAS.354..991B,2012ApJ...754..131K,2013ApJ...775...62K,2014MNRAS.437.2146W,2015A&A...574A..26B,2016MNRAS.459..496W,2018ApJ...854...46H}.

To test the existence of such a large scale under-density we inferred the averaged density contrast in radial shells around the observer.
In particular we determine the ensemble mean profile and corresponding standard deviations from our Markov Chain. The results are presented in figure \ref{fig:radial_profile}.
Averaging over the entire sky, our approach does not provide any indication for a large scale under-density. In fact, at distances of $r \sim 150$\Mpch{} and beyond the averaged density contrast approaches cosmic mean density.

To further compare our results to \citet{2014MNRAS.437.2146W} we also estimate the density contrast profile in the two regions of the northern and southern galactic cap covered by the 6dFGS survey \citep[see e.g.][]{Jones09}. As expected the density field in the 6dFGS-SGC field shows larger voids than the corresponding more massive 6dFGS-NGC field. However, on average we do not find any significant indication for a large-scale under-density on scales of $\sim 150$\Mpch{} or larger sufficiently under-dense to explain the $H_0$ tension.

This result is in agreement with the discussion of \citet{2017MNRAS.471.4946W}, who argue that it would be very unlikely to obtain a sufficiently under-dense large-scale void in a $\Lambda$CDM universe.
Since we fitted a particle mesh model to the data, our results thus indicate that the existence and shapes of nearby cosmic large scale structures can be explained within a standard concordance model of structure formation without invoking a particular violation of the cosmological principle or the scale of homogeneity.
On the contrary, in section~\ref{sec:impact_hubble}, we show that inhomogeneities of the nearby cosmic velocity field can bias local measurements of $H_0$, when not accounted for.

\begin{figure*}
 \begin{center}
   \includegraphics[width=\hsize]{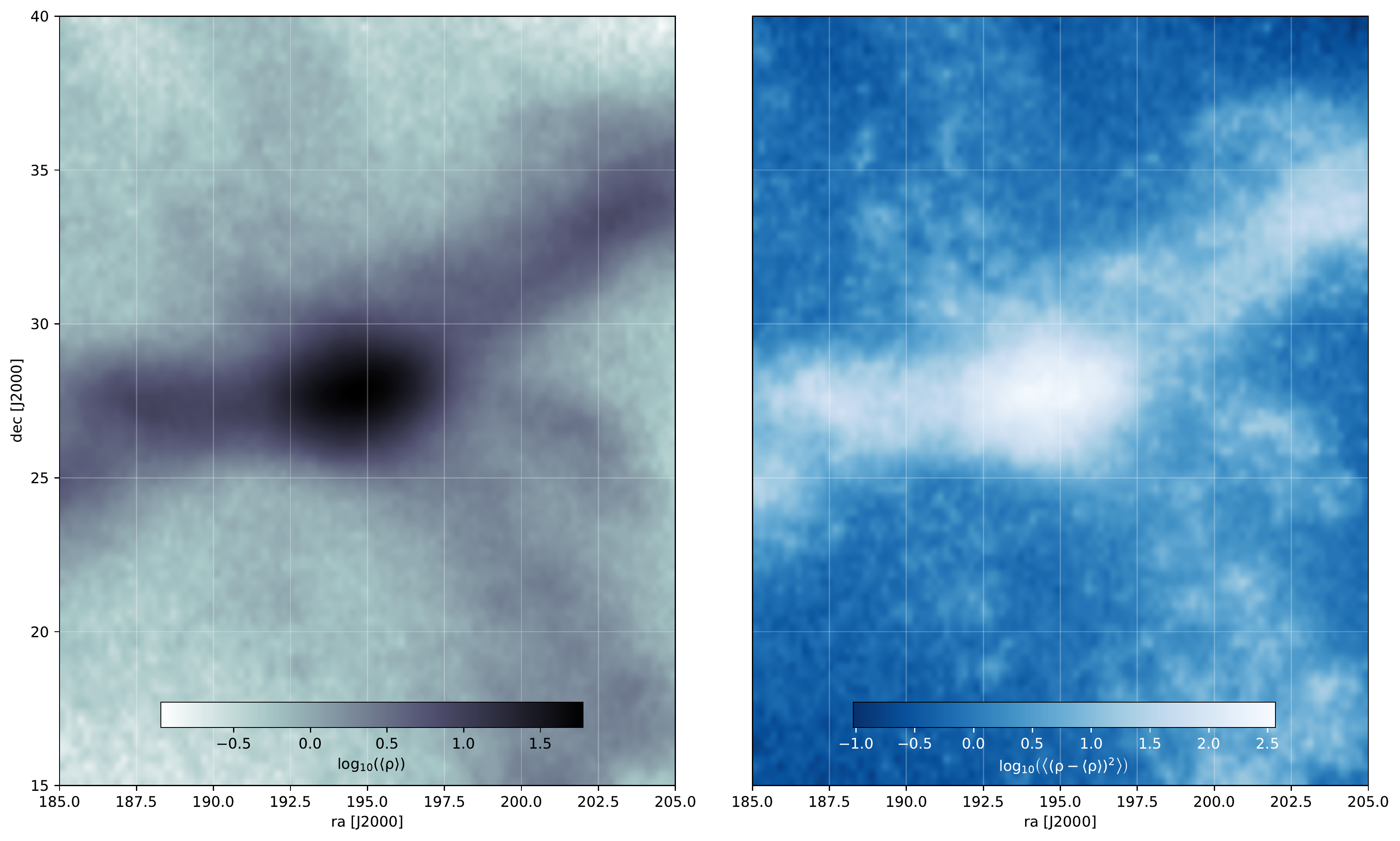}
 \end{center}
 \caption{\label{fig:coma_proj} Projected ensemble mean mass of inferred dark matter particles around the Coma cluster (left panel) and the corresponding ensemble variance field (right panel). One can clearly see the two major filaments along which mass accretes onto the Coma cluster. Additionally one can also notice three fainter filaments. The right panel indicates the ensemble variance estimate for the inferred mean density field. As expected, noise and signal are correlated for a galaxy clustering survey.}
\end{figure*}

\subsection{The formation history of the Supergalactic plane}
\label{sec:form_hist}
As mentioned above, a particularly interesting feature of the \borg{} algorithm is the fact that it links primordial matter fluctuations to actual non-linear structures in the Nearby Universe as traced by observed galaxies. Besides providing information on the state of the cosmic LSS at the initial and final conditions, the physical forward modelling approach of the \borg{} algorithm also traces all intermediate dynamical states. Consequently, our algorithm infers physically plausible structure formation histories of observed objects in the Nearby Universe, permitting to study the formation history of the cosmic web \citep[][]{JLW15,2015JCAP...06..015L}.

As an illustration, here we present the non-linear gravitational formation of cosmic structures in the Supergalactic plane. The Supergalactic plane contains local super-clusters, like the Coma and Pisces-Cetus clusters, the Shapley concentration as well as the southern and northern local super-voids. The Supergalactic plane is of particular interest to study the dynamics in our immediate cosmic neighbourhood. It has been analysed previously with various reconstruction algorithms and data sets \citep[][]{1990ApJ...364..370B,2000MNRAS.312..166L,2007arXiv0707.2607R,2010ApJ...709..483L,2016MNRAS.455.3169L}. In particular the local flows in the Supergalactic plane has been studied with distance and velocity data \citep[][]{1988ApJ...329..519D,1999ApJ...520..413Z,1999ApJ...522....1D,2012ApJ...744...43C,2013AJ....146...69C}.

In figure~\ref{fig:chrono_sg} we show a sequence of slices showing the plausible dynamical formation history of structures in the Supergalactic plane. To demonstrate that this formation history leads to the structures as observed in the super-galactic plane, we overlaid the inferred dark matter density field in the lower right panel of figure \ref{fig:chrono_sg} with the observed galaxies in the 2M++ survey.
Tracing the non-linear formation of cosmic structures provides novel avenues to understand assembly histories and galaxy formation. Details of these effects will be investigated in an upcoming publication.

\subsection{Inferring the mass of the Coma cluster}
\label{sec:mass_reconstruction}
In preceding sections, we demonstrated that the \borg{} algorithm infers detailed three-dimensional matter density fields that are in agreement with the spatial distribution of observed galaxies. We also tested posterior power-spectra to demonstrate that these density fields obey the correct statistical properties and are plausible representations for the dark matter distribution in the Universe.
These obtained density fields also provide a means to estimate the masses of individual large-scale structures in the Nearby Universe.
Mass estimation is feasible, since the \borg{} algorithm uses a physical model to fit redshift space distortions of observed galaxies.
As such the algorithm implicitly performs various dynamical mass estimation approaches that have been proposed in the literature previously \citep[see e.g.][]{1997ApJ...475..421E,2001ApJ...561L..41R,2009arXiv0901.0868D,2010AJ....139..580R,2011MNRAS.412..800S,2014MNRAS.442.1887F,2016ApJ...831..135N}.

For the sake of this work, we will illustrate the feasibility if inferring masses of cosmic structures for the particular case of the Coma cluster.

Besides Virgo and Perseus, Coma is one of the best-studied galaxy clusters in the Nearby Universe and is frequently used as the archetype to compare with clusters at higher redshifts \citep[][]{1998ucb..proc....1B,2014MNRAS.438.3049P}. The Coma cluster is particularly appealing to observers as it is located close to the galactic pole and has almost spherical shape \citep[][]{1998ucb..proc....1B}.

As an illustration in figure~\ref{fig:coma_proj}, we show the inferred ensemble mean of the projected mass density around the Coma cluster in the sky. The plot interestingly shows the two main cosmic filaments along which mass accretes onto the coma cluster. In addition one can also observe three more fainter filaments. For completeness we also present a plot of the corresponding ensemble variance field, reflecting again the expected correlation between signal and uncertainties as discussed above.

First estimates of the mass of Coma date back to \citet{1933AcHPh...6..110Z,1937ApJ....86..217Z}. Since then the mass of the Coma cluster has been estimated via various methods, such as the virial theorem, weak gravitational lensing or X-ray observations \citep[see e.g.][]{1986AJ.....92.1248T,1989ApJ...337...21H,1996ApJ...458..435C,1999ApJ...517L..23G,2007ApJ...671.1466K,2009A&A...498L..33G,2014MNRAS.442.1887F}. Consequently the Coma cluster is an excellent reference to evaluate the mass estimates obtained in this work.

In particular we estimate the cumulative radial  mass profile $M_{\mathrm{Coma}}(R)$ around the Coma cluster given as:
\begin{equation}
\label{eq:coma_mass}
M_{\mathrm{Coma}}(R) = \int_0^{\frac{\pi}{2}} \int_0^{2\,\pi} \int_0^R  \mathrm{sin}(\phi) \rho(\vec{x}(r,\phi,\theta)) \mathrm{d}r\,\mathrm{d}\phi\,\mathrm{d}\theta \, ,
\end{equation}
where $R$ defines the comoving distance from the Coma cluster centre to be found at the coordinates: $z=0.021$, $\mathrm{RA}=195.76\deg$, $\mathrm{DEC}=28.15\deg$.

\begin{figure}
	\begin{center}
    	\includegraphics[width=\hsize]{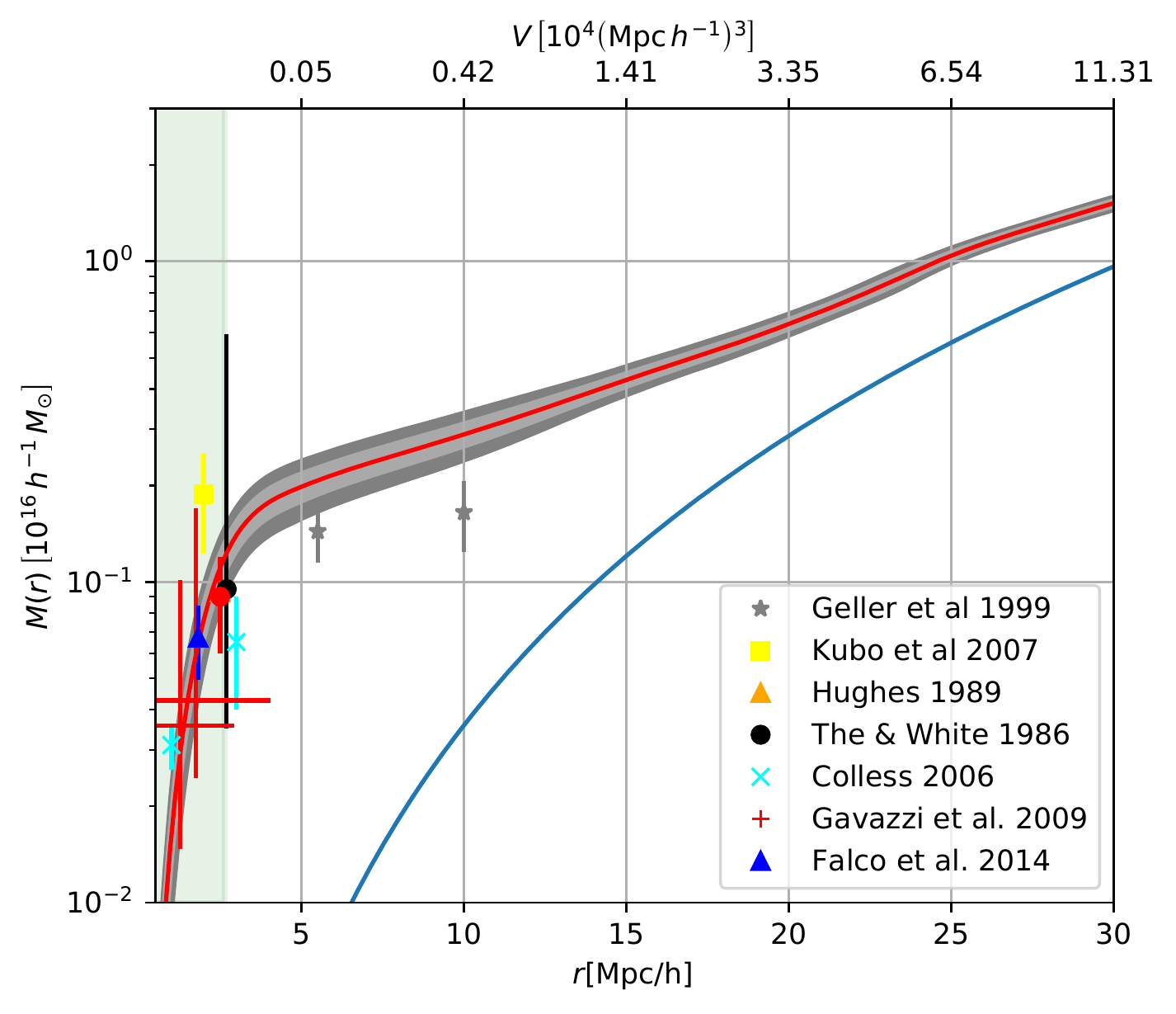}
    \end{center}
    \caption{\label{fig:coma} Coma cumulative mass profile. We show here the relation between the distance $r$ and the mass $M(<r)$ enclosed within that radius. The thick solid red line is the mean relation obtained through density field derived using BORG-PM, while the light grey (dark grey respectively) gives the 68\% (95\%  respectively) limit according to that mean. The thin blue solid line indicates the profile assuming solely the mean density of the universe. We also compare our results with the findings in the literature as indicated in the plot. It can be seen that our mass estimate for Coma agrees particularly well with complementary measurements of weak gravitational lensing or X-ray observations at scales of a few \Mpch{}. }
\end{figure}

We also determine uncertainties in our mass estimates by applying the estimator of equation~\eqref{eq:coma_mass} to the ensemble of inferred density fields. This permits us to estimate the ensemble mean and corresponding variance of the radial cumulative mass profile around the Coma cluster centre. 
In figure~\ref{fig:coma} we present our estimate of the mass profile around the Coma cluster. It can be seen that the algorithm provides a coherent reconstruction over large distances around the Coma cluster.

Literature provides several mass estimates of the Coma cluster at different radii $R$ from its centre \citep[][]{1986AJ.....92.1248T,1989ApJ...337...21H,1996ApJ...458..435C,1999ApJ...517L..23G,2007ApJ...671.1466K,2009A&A...498L..33G,2014MNRAS.442.1887F}. In Fig \ref{fig:coma} we also compare these literature values, as indicated in the figure, to our inferred cumulative mass profile. As demonstrated, our results are in great agreement with mass measurements provided by previous authors.  Most interestingly at radii below a few $h^{-1}$~Mpc we agree well with complementary mass measurements via X-ray and gold-standard gravitational weak lensing observations \citep[see e.g.][]{2009A&A...498L..33G,2014MNRAS.442.1887F}.

These results demonstrate that our inferred density fields provide the correct mass estimates at the correct location in three-dimensional space. Inferred density fields are therefore in agreement with the spatial distribution of observed galaxies, they show the correct physical and statistical features and they reproduce mass estimates at the right locations that agree with complementary gold standard weak lensing or X-ray measurements.

As will be demonstrated in a forthcoming publication, similar results are obtained for all major clusters in the Nearby Universe. In summary, our results indicate that inferred density fields represent coherent and plausible explanations for the expected dark matter distribution in our actual Universe.

\begin{figure*}
\includegraphics[width=\hsize]{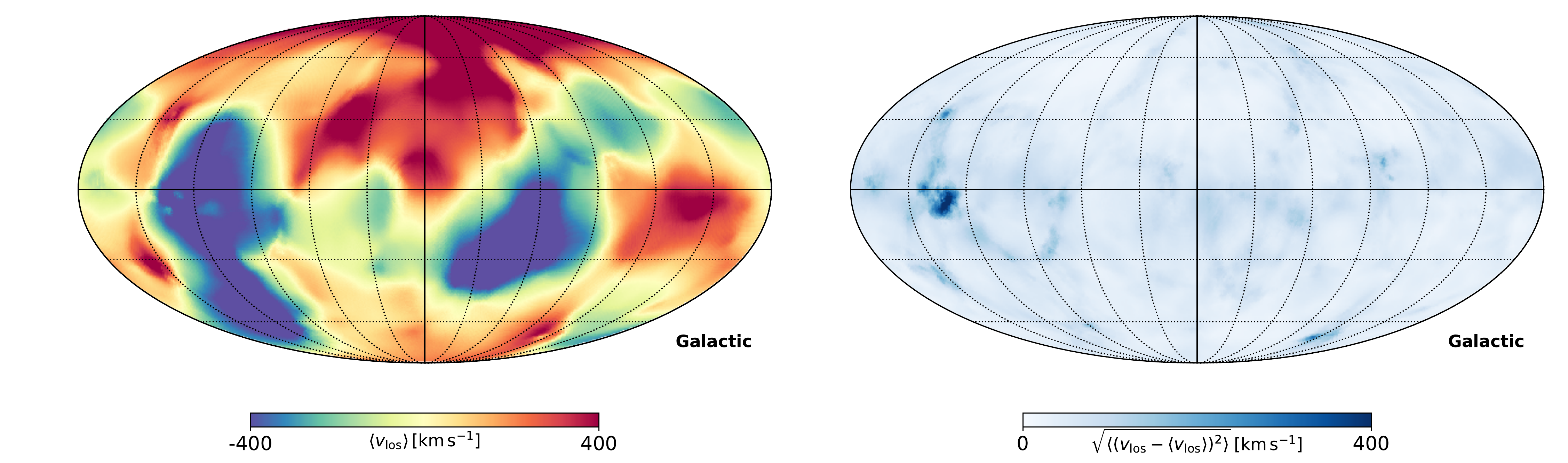}
	\caption{Spherical slice through the inferred three dimensional ensemble mean velocity field (left panel) and corresponding variance field (right panel). Specifically the plot shows the line of sight component of the velocity field. As indicated by the colour bar, in the left panel regions indicated in red are receding from the observer while blue regions are approaching the observer. The plot also shows regions of zero velocity along the line of sight indicating that matter in these regions follows the Hubble flow. The right panel shows the corresponding variance field for the line of sight velocity component. \label{fig:vfield_los}}
\end{figure*}

\subsection{The velocity field in the Nearby Universe}
\label{sec:inf_vel_field}
The \borg{} algorithm provides information on the three-dimensional velocity field, by simultaneously fitting the clustering signal of galaxies and their corresponding redshift space distortions. In particular, in order to explain the three-dimensional distribution of galaxies, the dynamical 
structure formation model has to account correctly for the displacement of matter from its initial Lagrangian to the final Eulerian position. To fit observations, the \borg{} algorithm, therefore, has to reconstruct the non-linear velocity field. Also note, as discussed in 
section~\ref{sec:methodology}, the velocity field derives uniquely from the initial density field. Therefore constraints on the three-dimensional primordial fluctuations will also immediately provide constraints on final non-linear velocity fields. Additional dynamical information is inferred by explicitly fitting redshift space distortions of observed galaxies by the physical forward model. This feature of the \borg{} algorithm permits us to access phase-space information in observations and provide detailed flow models for the Nearby Universe.

More specifically, we estimate that the velocity field in the Nearby Universe from simulated dark matter particle realizations generated by the forward PM model of the \borg{} algorithm. Each particle carries position and velocity information.
Estimating the velocity field from simulation particles is a challenging task. Several optimal estimators have been proposed to solve this problem satisfactorily \citep{Colombi2007,Hahn2015}. In this work, we have opted the adaptive smoothing filter described in \citet{Colombi2007}. This 
algorithm allows projecting the properties carried by particles on any desired grid. The filter is built to preserve summations over particles, notably the mass and momentum. It also guarantees that there are no empty cells by changing the smoothing radius depending on the number of the 
nearest neighbours, which is kept fixed except when the number of particles per cell overflow that number. This last element ensures that the entirety of the information is always used and the momentum in a cell is truly the averaged momentum of the particles in the target cell.

The procedure to generate velocity fields is the following. First, we produce both a mass and a momentum field with the adaptive smoothing filter on a regular grid of the same size as the analysis domain but on a Cartesian grid with $1024^3$ grid nodes, corresponding to a grid resolution of 
$0.67$\Mpch{}. These two fields have the same integrals as the original particle distribution over the same enclosed sub-volumes. Then for each element of the target grids, we divide the momentum components by the mass to obtain the velocity components.

Of course the first application of the obtained 3D velocity field allows for the estimation of bulk flows in the Nearby Universe \citep[][]{2018arXiv180203391H}. But we can generate at least two other interesting scientific products.

The first product is illustrated in figure~\ref{fig:vfield_los}. There we show a spherical slice through the three-dimensional velocity at a distance $R=60$\Mpch{} from the observer. The plot shows the line of sight velocity component of moving cosmic structures. As can be seen, regions coloured in red correspond to structures 
receding from us while regions coloured in blue indicate matter approaching the observer. At the interfaces between these two regions, we can observe a zero-crossing in the line of sight component of the velocity field. Particles close to these zero-crossing regions have almost no radial 
peculiar velocity component and are therefore almost ideal tracers of the Hubble flow. Our results permit to identify critical points of vanishing velocity in the Nearby Universe, as has been proposed to provide unbiased measurements of the Hubble parameter \citep[][]{2016JCAP...06..009L}.

Previous measurements relied on linear perturbation theory and accounted only for the irrotational potential flow of the dark matter \citep[e.g.][]{1995MNRAS.272..885F,1999ApJ...520..413Z,ERDOGDU2004,2008MNRAS.383.1292L,2015MNRAS.450..317C,2017MNRAS.467.3993A,2017MNRAS.468.1812S}.
By exploiting the fully non-linear particle mesh model, the \borg{} algorithm goes beyond such limitations by also inferring the rotational component of the velocity field, which is the second product that is directly derived from our inference framework. This rotational component of the velocity field is particularly associated with non-linear structure formation. Here we use inferred velocity fields to estimate the vorticity field given via the curl of the velocity field:
\begin{equation}
\vec{\omega}(\vec{x})=  \vec{\nabla} \times\vec{v}(\vec{x}) \, .
\end{equation}
We estimate the curl via finite differencing in real space.

\begin{figure*}
  \includegraphics[width=\hsize]{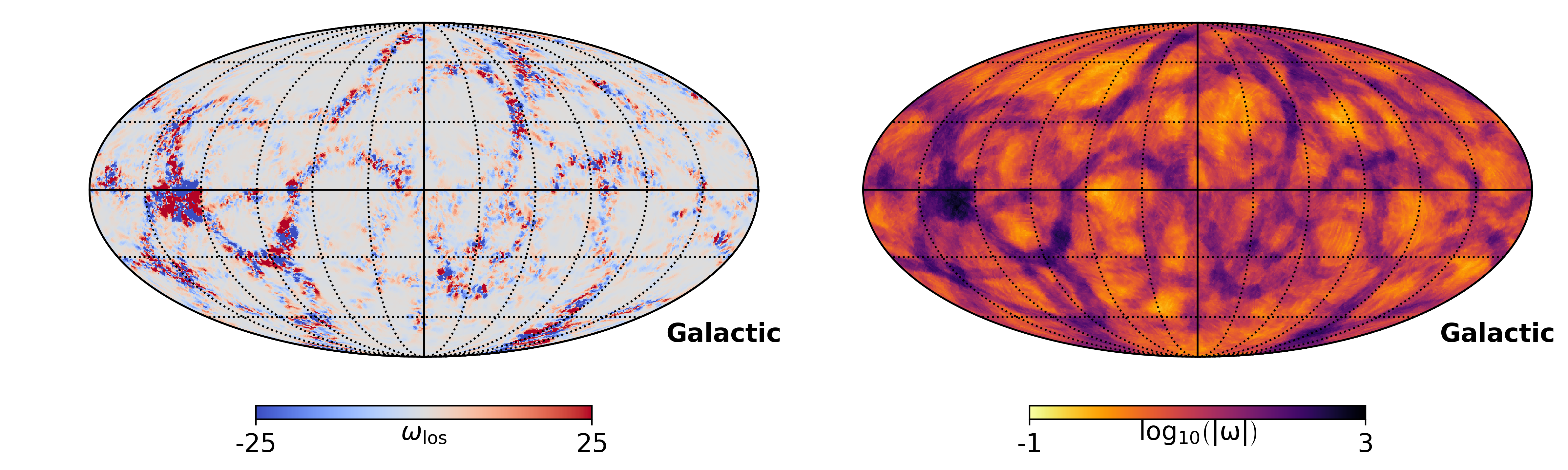}
  \caption{Spherical slices through the three dimensional vorticity of the velocity field at a distance $R=60$\Mpch{} from the observer in galactic coordinates. The left panel shows the projection of the vorticity vector along the line of sight, while the right panel shows the absolute amplitude of the vorticity. As can be seen the vorticity traces the high density filamentary structure of the cosmic web. The left panel also hints towards the quadrupolar structure of the vorticity as found in simulations \citep[see e.g.][]{2015MNRAS.446.2744L}.
    \label{fig:vorticity}
  }
\end{figure*}

In figure~\ref{fig:vorticity} we present the first physical inference of the vorticity field in the Nearby Universe. As can be seen, the absolute amplitude of the vorticity field traces the filamentary large-scale structure. These results are in agreement with the expectations that vorticity contributes to the peculiar motions of observed galaxies 
at scales of  3 - 4\Mpch{} \citep[][]{1999A&A...343..663P}. Vorticity plays an important role in structure formation in the dark matter paradigm as it can explain the generation and alignment of halo angular momentum\citep[][]{2013ApJ...766L..15L,2014MNRAS.441.1974L}. In figure~\ref{fig:vorticity} we also show the line of sight component of the vorticity vector field. The plot shows the characteristic quadrupolar pattern of alternating directions of rotation in the plane of the sky, as expected from simulations \citep[][]{2015MNRAS.446.2744L}.
This specific pattern guarantees the Universe to be irrotational when averaged over sufficiently large scales, such that there is no large-scale net angular momentum build-up.

Inferred vorticity fields could also provide a new step forward in identifying
rotating galaxy clusters. Due to their formation history or recent mergers clusters may have acquired angular momentum. Ignoring such rotations will result in erroneous dynamical mass estimates, affecting the cosmological constraints provided by cluster mass functions \citep[][]{2017MNRAS.465.2584B,2017MNRAS.465.2616M}. Additionally, information on the vorticity may shed new light on the alignment of galaxy angular momentum or shapes with the cosmic large-scale structure \citep[see e.g.][]{2013arXiv1301.0348L,2013ApJ...779..160Z,2014MNRAS.437L..11T,2015MNRAS.450.2727T,2018arXiv180500159C}.

In summary, our results provide new promising paths forward towards studying dynamic structure formation at non-linear scales in observations. Inferred non-linear velocity fields also provide detailed dark matter flow models for the Nearby Universe. These flow models are of particular relevance when attempting to measure the Hubble parameter $H_0$ from observation in the Nearby Universe or calibrating standard candles such as supernov\ae{} \citep[see e.g.][]{2014ApJ...795...45S}.

To provide the community with an accurate flow model of the Nearby Universe, we will make our inferred velocity fields publicly available with a forthcoming publication as well through our web platform\footnote{Available at \url{https://cosmicflows.iap.fr/}}.

\subsection{Hubble variations}
\label{sec:impact_hubble}
As mentioned above, inferred velocity fields permit to build refined flow models for the matter distribution in the Nearby Universe. This may be of relevance when performing measurements of the Hubble parameter $H_0$ in the nearby Universe, where non-vanishing Doppler shifts due to peculiar motions of observed objects may bias results. To quantify this effect for our local environment we estimate fractional variations in the Hubble parameter due to peculiar velocities via:
\begin{equation}
\delta_H(\vec{r}) = \frac{H(\vec{r})-H_0}{H_0} = \frac{\vec{v}(\vec{r}) \cdot \vec{r}}{H_0\, |\vec{r}|^2 } \, .
\end{equation}

In figure~\ref{fig:hubble} we demonstrate the fractional Hubble uncertainty averaged over cumulative spherical shells around the observer. As indicated by the plot, local flows out to about 70\Mpch{} can on average bias local measurements of the Hubble parameter by about three to ten percent. Interestingly that is about the same order of magnitude required to explain the current discrepancy between measurements of the Hubble constant in the Nearby Universe and by measurements of the CMB via the Planck satellite mission \citep[see e.g.][]{2016JCAP...10..019B,2016ApJ...818..132A,2018arXiv180203404F}.

In particular, we indicated the discrepancy between the measurements of the Planck collaboration \citep{2016A&A...594A..13P} and those obtained by \citet{Riess2016} in the figure. Since we have reconstructions of the three-dimensional velocity field we can also estimate the fractional Hubble uncertainty as a function of direction in the sky. These results are presented in figure~\ref{fig:cepheid_sky}. It can be seen, that there exists large coherent bulk flows in the direction towards Perseus-Pisces super-cluster, which may bias measurements of $H_0$.

It is also interesting to note, that we obtain on average a positive bias in the fractional Hubble uncertainty due to peculiar velocities. This seems to be a specific feature of the nearby matter distribution. In general, within a standard $\Lambda$CDM scenario, positive and negative bias should be equally likely on average.

Answering the question, whether or not the discrepancy in measurements of the Hubble parameters can contribute to resolving this issue needs to be investigated further in a future work. By providing new and refined flow models our work will contribute to either identifying the dynamics of the local structure as part of the systematics or ruling it out as a plausible explanation for the discrepancy of measurements of the Hubble parameter in the Nearby Universe.

\section{Summary and Conclusions}
\label{sec:Summary_an_Conclusion}
This work presents an extension of our previously developed \borg{} algorithm to perform analyses of observed galaxy clustering beyond the regime of linear perturbation. To achieve this goal we have implemented a numerical particle mesh algorithm into our previously developed Bayesian inference approach to account for the fully non-linear regime of gravitational structure formation.

As a result, our method fits full numerical structure formation simulations to data and infers the three-dimensional initial conditions from which present observations formed. The method is a fully Bayesian inference algorithm that jointly infers information of the three-dimensional density and velocity fields and unknown observational systematic effects, such as noise, galaxy biasing and selection effects while quantifying their respective and correlated uncertainties via a large scale Markov Chain Monte Carlo framework.

Typically the algorithm explores parameter spaces consisting of the order of ten million dimensions corresponding to the amplitudes of the primordial density field at different positions in the three-dimensional volume.
To perform efficient MCMC sampling with a complex physics model in such high dimensional parameter spaces we rely on a sophisticated implementation of the HMC algorithm. The HMC employs concepts of classical mechanics to reduce the random walk behaviour of standard Metropolis-Hastings algorithms by following a persistent Hamiltonian trajectory in the 
parameter space. In particular, the HMC exploits pseudo energy conservation to guarantee a high acceptance rate of proposed density field realizations and uses gradients of the logarithmic posterior distribution to guide the exploration in high dimensional parameter spaces. This requires providing to the algorithm derivatives of the logarithm of the likelihood distribution.

This likelihood distribution describes the statistical process by which the galaxy observations were generated given a specific realization of the non-linear density field. As described in section~\ref{sec:data_model}, for the sake of this work, the likelihood distribution combines a non-linear galaxy biasing model and the non-linear structure formation model with a Poisson distribution to account for the noise of the observed galaxy distribution.

In order to use the HMC in this scenario, we need to provide derivatives of the logarithm of this likelihood distribution with respect to the three-dimensional field of primordial matter fluctuations, acting as the initial conditions to the employed forward physics model. Since the likelihood incorporates a non-linear numerical structure formation model there exists no analytic gradient with respect to the initial conditions. One, therefore, has to rely on numerical representations of this derivative.

We note that using finite differencing to obtain gradients will not be sufficient. First of all, gradients obtained by finite differencing are numerically too noisy to be useful. Second, evaluating gradients in this fashion would be numerically too expensive. For cases as discussed in this work, finite difference approaches would require more than 10 million model evaluations to calculate a single gradient, which is numerically prohibitive with current computing resources. To obtain gradients of the logarithmic likelihood we, therefore, need to follow a different numerical approach.

As described in section~\ref{sec:pm_model}, any numerical algorithm can be understood as the composition of elementary operations for which there exist analytic derivatives. As described in appendix~\ref{appendix:tangent_adjoint_model}, this permits us to apply the chain rule of differentiation to the physical simulation algorithm to implement an algorithmic gradient of the physics model. As demonstrated in section~\ref{sec:pm_model}, this algorithmic derivative comes at the cost of only two forward physics model evaluations, rendering this approach highly efficient and suitable for high dimensional problems.

To further exploit modern massive parallel computing resources we have also parallelised our algorithm via the MPI message passing systems. As demonstrated in section~\ref{sec:pm_model} and by figure~\ref{fig:bench} the implementation of our algorithm reaches near optimal scaling as a function of the number of used cores. The numerical implementation of our algorithm, therefore, provides an efficient approach to sample the three-dimensional density and velocity fields. This constitutes the core element of our sampling scheme as outlined in figure~\ref{fig:flowchart}.

Employing a forward physics model for the analysis permits us to address observational effects due to non-linear structure formation processes. First of all, our approach accounts for the higher order statistics, associated with the filamentary pattern of the cosmic web. In addition, the dynamical information provided by the physics model permits to account for redshift space distortions effects associated with the peculiar motions of observed objects. As such our method not only extracts information from the clustering signal of the galaxy number counts distribution but also extracts partial dynamic information from redshift space distortions, carrying information on the line of sight projections of peculiar velocities.

Besides accounting for structure formation effects, accessing information at non-linear regimes galaxy data is a non-trivial task. We are confronted with a variety of stochastic and systematic uncertainties, such as unknown noise levels, galaxy biases or incomplete observations. To properly account for these effects we employ a hierarchical Bayes approach in combination with a block sampling approach permitting us to flexibly construct data models to account for individual systematic uncertainties of respective datasets used for the analysis.

\begin{figure}
 \begin{center}
   \includegraphics[width=\hsize]{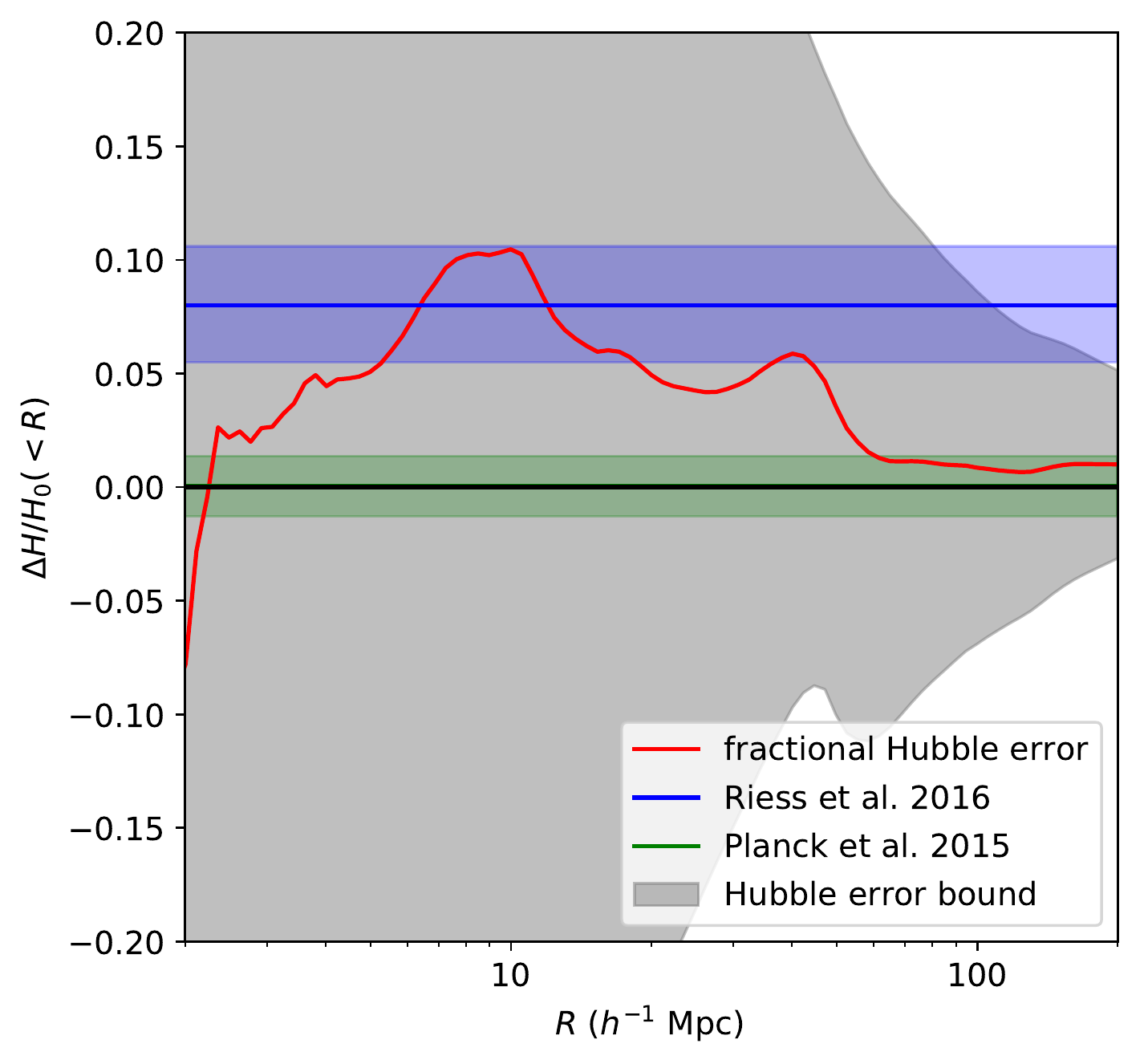}
 \end{center}
 \caption{\label{fig:hubble} This figure illustrates possible biases arising from doing a Hubble measurement with tracers within some volume, neglecting complex cosmic flows effect. We show in red solid line the mean systematic bias for a Hubble measurement per tracer 
 located {\it within} a sphere of radius $R$. The grey area corresponds to the expected $1\sigma$ fluctuation per tracer of that same measurement. We show, for reference, the measurement by \citet{Riess2016} (in blue and shade of blue for the $1\sigma$ limit) of the Hubble constant relatively to the Planck one (centred on zero, and shade of green for the $1\sigma$ limit).
 }
\end{figure}

Specifically, in order to describe the unknown non-linear biasing relation between observed galaxies and the underlying dark matter distribution in this work, we used a phenomenological truncated power-law model as previously proposed by \citet{2014MNRAS.441..646N}. This bias model has three free parameters which are inferred jointly with the three-dimensional density field via a multiple block sampling framework. Similarly, the \borg{} algorithm jointly infers unknown noise levels of the survey, related to the expected number of galaxies. Galaxy biasing can differ as a function of galaxy properties such as luminosity. To account for such luminosity dependent galaxy clustering we typically split our galaxy sample into different subsets according to luminosity or other parameters. The \borg{} algorithm then accounts for the respective uncertainties of individual subsamples while jointly inferring information from the combination of those. Joint and correlated uncertainties between all inference parameters are quantified by performing a thorough Markov Chain Monte Carlo via the block sampling scheme described in section \ref{sec:model_observed_galaxies} and visualized by figure~\ref{fig:flowchart}.

A common issue of analysing the cosmic LSS in galaxy observations is the fact that there exists currently no accurate data model that captures all nuances of unknown galaxy formation processes at non-linear scales. Therefore a necessary requirement for the analyses of present and next-generation surveys is the construction of inference approaches that can cope with unknown systematics and misspecifications of the data model. As discussed in section \ref{sec:robust_inference} we explored the possibility to perform robust inference by not conditioning directly on the likelihood but on some neighbourhood of the specified likelihood distribution. This approach amounts to tempering the likelihood distribution by raising it to some positive power, which is equivalent to using only a homogeneous subset of the data.
 The approach, therefore, provides conservative estimates of the cosmic large-scale structure since it effectively reduces the amount of information that can be used to reliably infer the matter distribution. Exploiting the full potential of observed data requires developing better data models, which is an ongoing activity of the cosmological community.

 \begin{figure*}
 \begin{center}
   \includegraphics[width=0.6\hsize]{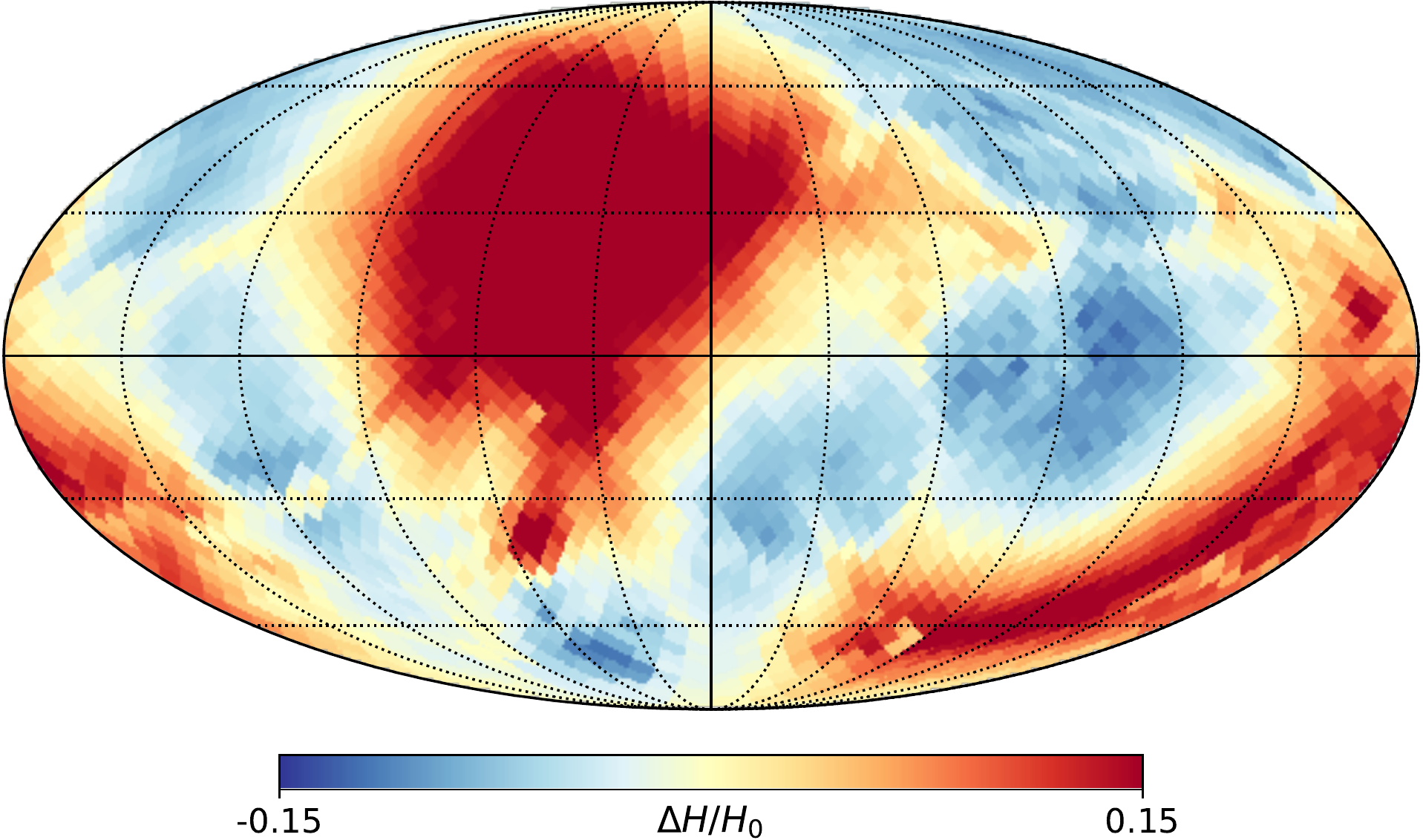}
 \end{center}
 \caption{\label{fig:cepheid_sky} Prediction of the fractional Hubble uncertainty as a function of direction within 60\Mpch{} around the observer. As can be seen, the fractional Hubble uncertainty is highly structured on the sky, with large-scale coherent bulk flows. The dominant red central region points towards the direction of Perseus Pisces. These effects need to be accounted for when inferring the Hubble parameter from data of the Nearby Universe. }

\end{figure*}

In section \ref{sec:data_application} we perform an analysis of the cosmic LSS in the Nearby Universe. This is achieved by applying our \borg{} algorithm to the 2M++ galaxy compilation, covering about 70\% of the sky.  We split the 2M++ galaxy sample into a total of 16 sub-sets as a function of luminosity and the two absolute K-band magnitude cuts at $\ktmpp \le 11.5$ and $11.5 < \ktmpp \le 12.5$. Splitting the galaxy sample into these subsets permits us to treat luminosity dependent galaxy biases as well as respective selection effects due to flux limitations and survey masks.
The \borg{} algorithm infers information jointly from the combination of these galaxy subsets while accounting for their respective systematic and stochastic uncertainties.

As described in section \ref{sec:non_lin_analysis_borg}, we inferred the field of primordial matter fluctuations on a cubic equidistant Cartesian grid of side length 677.77\Mpch{} consisting of $256^3$ volume elements. This amounts to a grid resolution of $\sim 2.6$\Mpch{} in initial Lagrangian space. To guarantee a sufficient resolution of the final Eulerian density field we oversample the initial density field by a factor eight, requiring to evaluate the particle mesh model with a total of $512^3$ simulation particles. Running the particle mesh model for every transition step in the Markov Chain is numerically expensive. To efficiently pass through the initial burn-in period of the Markov chain we initialized the run with a faster but approximate Lagrangian perturbation theory model for about $6700$ Markov transition steps. Then we switched to the full particle mesh model to infer the fully non-linear regime of cosmic structures in the 2M++ survey.

We tested the initial burn-in behaviour of our sampler by initializing the run with a Gaussian random guess for the initial density field and scaled the amplitudes by a factor $0.1$ to start from an over-dispersed state. The initial burn-in behaviour was then tested by following the systematic drift of subsequently measured posterior power-spectra towards the preferred region in parameter space. As discussed in section \ref{sec:sampler_convergence}, this initial burn-in period is completed after about $4000$ sampling steps, when inferred power-spectra start oscillating homogeneously around a fiducial cosmological power-spectrum. Note, that during the initial burn-in period our approach not only adjust the three-dimensional density field but also simultaneously explores the optimal parameter settings for the non-linear galaxy bias model and corresponding unknown noise levels.

Once we switched to running the analysis with the full particle mesh model we follow the initial burn-in behaviour of the non-linear analysis by tracing the logarithmic likelihood across subsequent sampling steps. The observed gains are considerable. With respect to the initial logarithmic likelihood value obtained from the approximate LPT run, we gain five orders of magnitude of improvement in the differential logarithmic likelihood when running the analysis with the non-linear particle mesh model.
The logarithm of the ratio of the likelihood value for the LPT run and the full PM run therefore qualify for a model comparison test for the best representation of the three-dimensional density field able to explain the observations. These results are a clear demonstration that our reconstructions are clearly outperforming any previous results based on Eulerian or Lagrangian perturbation theory.

To further investigate the improvements of the PM over the LPT model, we also studied the traces of the logarithmic likelihood functions for the 16 galaxy sub-samples used for our analysis. We observed that fainter galaxy samples experience fewer improvements than brighter ones. This is expected, since fainter galaxies are believed to live in regions that can be described better by LPT rather than brighter galaxies, living in highly non-linear regimes of the cosmic LSS. It may be interesting to investigate the details of this effect in future analyses, as it may provide a guideline to optimally select galaxies for cosmological analyses.

In section \ref{sec:inference_results} we presented the results of our cosmological analysis of the 2M++ galaxy compilation. We first presented the inference of the 16 non-linear galaxy bias functions for the respective galaxy subsets as split by luminosity. As discussed above the galaxy biasing relation is modelled via a four parameter truncated power-law model.

It is interesting to remark that the inferred shapes of biasing functions are in agreement with the previous findings of \citet{2008ApJ...678..569S}. In general we observe an agreement in the biasing functions for fainter galaxies between galaxies selected in the two K-band magnitude ranges at $\ktmpp \le 11.5$ and $11.5 < \ktmpp \le 12.5$ . For brighter galaxies we observe a difference in the biasing relations between the two apparent magnitude cuts. Whether this indicates a difference in clustering behaviour of galaxies in the respective samples or whether it is due to some contamination or systematic effect needs to be investigated in the future. In any case it can clearly be seen that the galaxy bias functions preferred by the data cannot be described by simple linear biasing relations.

The \borg{} algorithm infers the field of primordial density fluctuations with a Lagrangian resolution of $\sim 2.6$\Mpch{}. This is sufficient to resolve the initial conditions of major features in the Nearby Universe. As demonstrated in section \ref{sec:result_3d_density} the \borg{} algorithm simultaneously infers the present non-linear matter distribution of the Universe together with the three-dimensional initial conditions from which present structures formed. Our algorithm not only provides simple point estimates, such as the mean or maximum a posteriori result but provides a numerical approximation to the actual posterior distribution of the three-dimensional cosmic LSS in terms of an ensemble of density field realizations generated by the Markov Chain.
The ensemble of data constrained Markov samples permits us to quantify the uncertainties of inferred initial and final density fields. To illustrate this fact in figure \ref{fig:pm_sky_realization_mean} we show a plot of the ensemble mean density fields and corresponding standard deviations. The plot nicely demonstrates the feasibility to recover the detailed filamentary pattern of the matter distribution in our Universe.

Unlike simple point estimates, respective Markov samples of the density field represent statistically and physically plausible realizations of the actual matter distribution in our Universe. They are not affected by incomplete observations or selection effects and can be straightforwardly interpreted as physically reasonable quantities.

In particular, figure \ref{fig:pm_sky_realization} demonstrates that regions which are only poorly sampled by observed galaxies are visually similar to regions with much higher signal to noise ratios. Our inferred density samples reveal a highly detailed filamentary cosmic web corresponding to the spatial distribution of actually observed galaxies in the 2M++ survey.
To test whether these inferred density fields are also physically plausible representations of a dark matter density field we measured a posteriori power-spectra from inferred initial conditions.
This test reveals that the \borg{} algorithm is able to reliably recover the dark matter distribution over a huge dynamic range covering three orders of magnitudes of the cosmological power-spectrum. As demonstrated by figure \ref{fig:borg_pm_posterior_density_stats} measured power-spectra agree well with a fiducial cosmological model, demonstrating that our inference results are unbiased throughout the entire ranges of Fourier modes considered in this work. We further tested the one-point distribution of primordial density fluctuations and showed the agreement with the assumption of Gaussian statistics.

The spatial correspondence between inferred density fields and observed galaxies together with the agreement of inferred power-spectra with the fiducial cosmological model indicates that our results are physically plausible representations of the dark matter distribution in our Universe.

To further investigate this fact, in section \ref{sec:mass_reconstruction} we estimated the radial mass profile around the Coma cluster. The Coma cluster is one of the best studied clusters in the Nearby Universe and literature provides a plenitude of measurement results for the Coma mass. In contrast to previous measurements we are able to provide the first continuous measurement of the mass profile around the Coma cluster. We also compared our estimates to previous results obtained via complementary measurements of weak lensing and X-ray observations. As demonstrated by figure \ref{fig:coma} our results agree well with gold standard weak lensing mass estimates at the scales of $\sim 1$\Mpch{}. These results demonstrate that our inferred dark matter density fields provide the correct amount of matter at the correct spatial locations in the Nearby Universe.

In summary we conclude that the obtained density fields are physically plausible representations for the matter distribution in the Nearby Universe. A more detailed analysis and mass estimates for various structures in our nearby neighbourhood will be presented in a coming publication.

The possibility to infer masses of respective cosmological structures is related to the fact that the \borg{} algorithm exploits a dynamical physical model to fit redshift space distortions of observed objects. Thus our algorithm extracts
velocity information from redshift distortions and implicitly applies various dynamical mass estimation techniques that have been presented in the literature \citep[][]{1997ApJ...475..421E,2001ApJ...561L..41R,2009arXiv0901.0868D,2010AJ....139..580R,2011MNRAS.412..800S,2014MNRAS.442.1887F,2016ApJ...831..135N}.

To further illustrate the feasibility to infer the dynamic state of cosmic structures from observations, in section \ref{sec:inf_vel_field} we provide information on the inferred three-dimensional velocity field of our Nearby Universe.

As a complete novelty we are the first to reconstruct the rotational component of the velocity field from observations. As demonstrated by figure \ref{fig:vorticity} this vorticity field traces the non-linear filamentary structures around the Perseus-Pisces and Virgo cluster. When studying the directional components of the vorticity vector field we find a similar quadrupolar structure has been observed in simulations previously. These results therefore provide new avenues to test the alignment of galaxy spin with the cosmic LSS and the generation of angular momentum in the course of structure formation.

Our inferred velocity fields provide a new flow model for the three-dimensional large scale motion of matter in the Nearby Universe. Accounting for the specific realization of the velocity field is of particular relevance to measurements of the Hubble parameter in the Nearby universe. To demonstrate this effect in section \ref{sec:impact_hubble} we estimated the fractional uncertainty in measurements of the Hubble parameter due to the peculiar motion of observed objects. In particular figure \ref{fig:hubble} indicates that there is a risk to bias estimates of the Hubble parameter when not accounting for such peculiar motions. Interestingly for tracer particles at distances between 10 and 
70\Mpch{} our results show a fractional Hubble uncertainty due to peculiar motions that are compatible with the currently debated discrepancy in the measurements of the Hubble parameter from local and CMB observations. As demonstrated by figure \ref{fig:cepheid_sky}, peculiar velocities introduce a highly inhomogeneous and asymmetric distribution of the fractional Hubble uncertainties at different positions in the sky. One needs to investigate further in the future whether these effects can contribute to the observed discrepancy in measurements of $H_0$.

To further investigate the possible impact of nearby cosmic structures on local measurements of the accelerated cosmic expansion, we also investigated the possible existence of a large-scale local under density out to a depth of 150\Mpch{} and beyond, which could mimic the acceleration effects attributed to dark energy. Despite the claim of growing evidence for such a local hole in the literature \citep[see e.g.][]{2016MNRAS.459..496W,2018ApJ...854...46H}, our inferred radial density profiles, shown in figure \ref{fig:radial_profile}, provide no support for the existence of such a large local void. In fact, our results indicate that the existence of local cosmic structures can be explained by the concordance model of structure formation without any violation of the cosmological principle or scale of homogeneity. Our result, therefore, agrees with the discussion of \cite{2017MNRAS.471.4946W}.

In summary, this work presents a modification of our \borg{} algorithm capable of exploring the non-linear regime of the observed galaxy distribution by using physical models of gravitational structure formation. The algorithm provides us with simultaneous reconstructions of present non-linear cosmic structures and the initial conditions from which they formed. We further obtain detailed measurements of the three-dimensional flow of matter and infer plausible structure formation histories for the Nearby Universe. Inferred density and velocity fields represent a detailed and accurate description of the actual matter distribution, resembling correctly the filamentary cosmic web and masses of individual structures in the Nearby Universe.

This work is a clear demonstration that complex analyses of non-linear structures in galaxy surveys subject to several systematic and stochastic uncertainties is feasible and produces significant scientific results.
In consequence, the \borg{} algorithm provides the key technology to study the non-linear regime of three-dimensional cosmic structures in present and coming galaxy surveys.

\begin{acknowledgements}
This research was supported by the DFG cluster of excellence "Origin and Structure
 of the Universe" (www.universe-cluster.de). This work made in the ILP LABEX
 (under reference ANR-10-LABX-63) was supported by French state funds
managed by the ANR within the Investissements d'Avenir programme under reference
 ANR-11-IDEX-0004-02. The Parkes telescope is part of the Australia Telescope
 which is funded by the Commonwealth of Australia for operation as a National
 Facility managed by CSIRO. We acknowledge financial support from "Programme
 National de Cosmologie and Galaxies" (PNCG) of CNRS/INSU, France. This work
 was supported by the ANR grant ANR-16-CE23-0002. This work was granted
 access to the HPC resources of CINES under the allocation A0020410153 made
  by GENCI. We acknowledge several constructive discussions with Benjamin
  Wandelt, St\'ephane Colombi, Franz Elsner, Fabian Schmidt, Licia Verde, Simon White and Carlos Frenk. We also acknowledge reading of the early draft by Doogesh Kodi Ramanah.
  GL acknowledges the hospitality of the University of Helsinki, where part of
  the development of that project took place, notably Peter Johansson, Till Sawala.
This work has made use of the Horizon Cluster hosted by Institut d'Astrophysique de Paris.
This work has been done within the activities of the Domaine d'Int\'er\^et Majeur (DIM) Astrophysique et Conditions d'Apparition de la Vie (ACAV), and received financial support from R\'egion Ile-de-France.

\end{acknowledgements}
\bibliography{paper}

\onecolumn
\appendix

\section{Discrete Fourier transform conventions and properties}
\label{appendix:dft}

In this appendix we summarize some conventions we use for the discrete Fourier transform and some useful properties we make use of.

\subsection{Conventions}

In this manuscript we use the following convention for the discrete Fourier transform $\mathcal{F}$. The matrix element, per dimension, of this transform is set to:
\begin{equation}
  \mathcal{F}_{k,i}= \mathrm{e}^{-\frac{2\pi}{N}\,\sqrt{-1}\, k i}\,,
\end{equation}
which relates the Fourier space representation $\hat{A}_k$ and the real space representation $A_i$ of the same quantity sampled on a regularly spaced mono-dimensional grid of size $N$ through:
\begin{equation}
  \hat{A}_k = \sum_i \mathcal{F}_{k,i} A_i\,.
\end{equation}

The discrete Fourier transform is exactly invertible, and the element of the inverse is:
\begin{equation}
\mathcal{F}^{-1}_{i,k}= \frac{1}{N} \mathrm{e}^{\frac{2\pi}{N}\,\sqrt{-1}\, i\, k}=\frac{1}{N} \mathcal{F}_{i,k}^{\dagger}\,,
\end{equation}
also per dimension.

\subsection{Translation of the discrete Fourier transform}
\label{appendix:translation_dft}
Here we will show a lemma giving the identity between translating the matrix element of a discrete Fourier transform and the translation of the field itself. In this appendix we write $i=(i_0,\ldots,i_d)$ the relation between a matrix index $i$ and the regular grid indices $(i_0,\ldots,i_d)$ in a space of dimension $d$. The discrete Fourier transform is a matrix linear operator $\mathcal{F}$ given as:
\begin{equation}
	\mathcal{F}_{i,k} = \prod_{j=1}^d \omega^{i_j k_j} \,,
\end{equation}
with $\omega=\exp(-2\pi i / N)$ with $i^2 = -1$. Now we express the shifted discrete Fourier transform of a real vector $V_i$ into $\tilde{V}_k$:
\begin{align}
	\tilde{V}_{k,q} = & \sum_{i=0}^{N} \mathcal{F}^*_{\tilde{i}^{+1}_q,k} V_i \\
      = & \sum_i \omega^{-\sum_{\tilde{j}_q} i_j k_j - (i_q+1) k_q} V_i \\
      = & \sum_{\bar{i}_q} \sum_{i_q=1}^{N}
               \omega^{-\sum_{j} i_j k_j} V_{\tilde{i}^{-1}_q} \\
      = & \sum_{i} \omega^{-\sum_{j} i_j k_j} V_{\tilde{i}^{-1}_q} \\
      = &  \sum_i \mathcal{F}^*_{i,k} V_{\tilde{i}^{-1}_q}\,.
\end{align}
with $\tilde{i}^\epsilon_q = (i_0,\ldots,i_{q-1},i_{q}+\epsilon,i_{q+1},\ldots)$ and $\bar{i}_q = (i_0,\ldots,i_{q-1},i_{q+1},\ldots)$. In the above, in the transition from the second to the third line, we have exploited the periodicity of the discrete Fourier transform $\omega^{N k} = \omega^0 = 1$. The above identity stands even if the discrete Fourier transforms have different dimensions along each axis. Additionally, we have similarly:
\begin{align}
	V'_{k,q} = & \sum_{i=0}^{N} \mathcal{F}^*_{\tilde{i}^{-1}_q,k} V_i \\
     = & \sum_{i=0}^{N} \mathcal{F}^*_{i,k} V_{\tilde{i}^{+1}_q,k}
\end{align}

\section{The particle mesh model}
\label{appendix:pm_model}
This work uses a particle mesh (PM) model to evaluate the gravitational formation of cosmic structures from their origins to the present epoch. The General relativity dynamics is approximated using linear perturbations of the background metric. In practice that means we solve for the dynamics of a set of particles interacting via a Newtonian gravitational force. In this appendix we give both a brief overview over the implementation of the particle mesh model and its corresponding derivative required for the HMC framework.

\subsection{PM equation of motions}
As discussed in section \ref{sec:pm_model} our implementation of the particle mesh (PM) algorithm follows closely the description in
\citet[][]{1997astro.ph.12217K}.
In particular the PM algorithm aims at solving the following set of equations of motion for comoving dark matter particle positions $\vec{r}$ and momenta $\vec{p}$ in the simulation domain:
\begin{equation}
\frac{\mathrm{d}\vec{r}}{\mathrm{d}a}=\frac{\vec{p}}{\dot{a}a^2} = \frac{1}{H(a) a^3} \vec{p} = f_r(a) \vec{p} \, ,
\end{equation}
where $a$ is the cosmic scale factor, $\dot{a}$ is its first time derivative and
\begin{equation}
	f_r(a) = \frac{1}{H(a) a^3}\,.
\end{equation}
The corresponding momentum update is given by:
\begin{equation}
  \label{eq:pm_momentum}
\frac{\mathrm{d}\vec{p}}{\mathrm{d}a}
    = -\frac{3}{2} H_0^2 \Omega_\text{m} \frac{\nabla_{\vec{r}} \tilde{\Phi}}{H(a) a^2}
    = - f_v(a) \nabla_{\vec{r}} \tilde{\Phi}\,,
\end{equation}
where the dimensionless gravitational potential is given through the Poisson equation
\begin{equation}
	\nabla^2_{\vec{r}} \tilde{\Phi} = \frac{3}{2} H_0 \Omega_\text{m} \delta_\text{m}(\vec{r})\, ,
\end{equation}
and
\begin{equation}
   f_v(a) = \frac{H_0}{H(a) a^2}\,.
\end{equation}
We will now provide details on the numerical implementation of the ordinary differential equation (ODE) solver.

\subsection{Evaluation of gravitational forces}

We use the standard PM approach, by first estimating the density field from particles via a CIC kernel and then solve equation (\ref{eq:grav_force}) in Fourier space \citep[see e.g.][]{HOCKNEYEASTWOOD1988,1997astro.ph.12217K}. The Fourier kernel is computed from the 5-point stencil, approximating the Laplacian to second order. We thus obtain:
\begin{equation}
	\label{eq:greens_poisson}
	\hat{\tilde{\Phi}}(\vec{k}_{\vec{q}})
      = \left\{\sum_{l=1}^3 \left[\frac{4 N_l}{L_l} \sin\left(\frac{\pi q_l}{N_l}\right)\right]^2\right\}^{-1} \hat{\delta}(\vec{k}_{\vec{q}}) = \mathcal{G}_{\vect{q}} \hat{\delta}(\vect{k}_\vect{q}) \,,
\end{equation}
with $q_l \in \{0,\ldots,N_l-1\}^3$, $N_l$ the number of grid element along the axis $l$, and $L_l$ the comoving length of the simulation box along the axis $l$. We will also write $\mathcal{G}_q$ for the summation over the entire three-dimensional grid of $\vect{q}$ vectors.
Following equation~\eqref{eq:pm_momentum}, the gravitational force acting on the $p$th particle is given as:
\begin{equation}
\label{eq:grav_force}
 \vec{F}_p= -f_v(a) \nabla_{\vec{r}} \tilde{\Phi} (\vec{r}_p) = f_v(a) \vec{\tilde{F}}_p \,,
\end{equation}
with $\vec{r}_p$ the position of the $p$-th particle. Following \citet{HOCKNEYEASTWOOD1988}, to avoid self-interaction the actual value of the gradient must be derived as:
\begin{equation}
 \vec{\tilde{F}}_p = \mathcal{I}[D_{\vec{r}} \tilde{\Phi}](\vec{r}_p)\,
\end{equation}
with $\vec{D}_{\vec{r}}$ the (symmetric) finite difference operator, $\mathcal{I}$ the CIC interpolation kernel.

\subsection{Update of particle positions}

To numerically integrate the equations of motion we use the leap-frog integrator \citep[][and also  related to methods given in Sir Isaac Newton's Dynamica]{HOCKNEYEASTWOOD1988}. Finite differencing then yields the well known update equations for particle momenta and positions \citep[see e.g.][]{1997astro.ph.12217K}:
\begin{align}
\vect{p}^{n+1/2}_p &= \vect{p}^{n-1/2}_p + \left(\int_{a_{n-1/2}}^{a_{n+1/2}} f_v(a)\text{d}a\right) \vect{\tilde{F}}_p \,,  \\
\vect{r}^{n+1}_p &= \vect{r}^{n}_p + \left(\int_{a_{n}}^{a_{n+1}} f_r(a)\text{d}a\right) \vect{p}^{n+1/2}_p\,. \label{eq:update_particles_forward}
\end{align}
By offsetting the initial momentum by half a time step and introducing:
\begin{equation}
\Delta^n_v =  \int_{a_{n-1/2}}^{a_{n+1/2}} f_v(a)\text{d}a\,,
\end{equation}
and
\begin{equation}
\Delta^n_r = \int_{a_{n}}^{a_{n+1}} f_r(a)\text{d}a\, ,
\end{equation}
we can write  the updating scheme local in time:
\begin{align}
\label{eq:update_particles_forward_local}
\vect{p}^{n+1}_q & = \vect{p}^{n}_q +  \vect{\tilde{F}}_q \, \Delta^n_p \nonumber \\
\vect{r}^{n+1}_q & = \vect{r}^{n}_q + \vect{p}^{n+1}_q\, \Delta^n_r \nonumber \\
                 & = \vect{r}^{n}_q + \vect{p}^{n}_q\, \Delta^n_r  +
                      \vect{\tilde{F}}_q \, \Delta^n_p \, \Delta^n_r \, .
\end{align}
Note that at the end of the updating loop one has to move the momenta further by half a time step.
In the rest of this work, notably in appendix~\ref{appendix:tangent_adjoint_model}, we also set $(\vect{y}_q)_\alpha = y_{q,\alpha}$, where
$y$ can be one of $p$, $r$ or $\tilde{F}$.

\section{Tangent adjoint model of the particle mesh code}
\label{appendix:tangent_adjoint_model}

Efficient exploration of high dimensional parameter spaces is facilitated by the use of gradients.
The HMC sampling framework relies on the availability of a gradient of the posterior distribution.
In this appendix we derive in detail such a gradient for the PM model, which is valid for the PM algorithm as described in appendix~\ref{appendix:pm_model}.
Specifically we derive expressions for the following gradient of the negative logarithmic posterior distribution $\psi(\{\delta^\text{init}_l\})$ with respect to a initial density contrast amplitude $\vec{\delta}^\text{init}=\{\delta^{init}_l\}$.
In the following we will describe in detail how to obtain analytic gradients for numerical computer simulations of gravitational structure formation.

\subsection{General framework to derive tangent adjoint model}
Conceptually, any computer model, no matter how complex or non-linear, can be expressed as a succession of elementary algorithmic operations, such as additions and multiplications. A computer algorithm $G(x)$ can therefore be expressed as the composition of several functions. It is simply the nested application of elementary function applications given as:
\begin{equation}
G(x)=(B_N \circ B_{N-1} \circ \ldots \circ B_1 \circ B_0)(x)=B_N(B_{N-1}(\ldots(B_1(B_0(x)))))\, .
\end{equation}
Any derivative of $G(x)$ can then be obtained by use of the chain rule of differentiation:
\begin{equation}
\frac{\partial G}{\partial x}= \frac{\partial B_N}{\partial B_{N-1}} \, \ldots \, \frac{\partial B_{1}}{\partial B_{0}} \frac{\partial B_{0}}{\partial x} \, ,
\end{equation}
As can be seen any derivative of a computer program results just in a long sequential application of linear operations. The same approach applies to any multivariate computer program. In the following we will use this approach to derive the adjoint gradient of our particle mesh computer model.

\subsection{The tangent adjoint model for the LSS posterior distribution}
\label{appendix:adjoint_model_lss}
Having posited the framework, we now proceed with the first step of the derivation
of $\psi(\vec{\delta}^\text{init})$.
The log-likelihood part of the posterior can formally be expressed as follow:
\begin{equation}
    \psi(\vec{\delta}^\text{init}) =
      L \circ \vect{U}^{(N)} \circ \ldots \circ \vect{U}^{(0)}(\vec{\delta}^\text{init})\,.
\end{equation}
Above, $L$ is the log-likelihood function allowing the comparison between the output
of the forward model and the data (i.e. a Poisson distribution in our case as
given in section~\ref{sec:model_observed_galaxies}). $\vect{U}_i$ are the Kick-Drift element of
the PM algorithm, given in equation~\eqref{eq:update_particles_forward}. The
gradient of the total log-likelihood $\psi$ with respect to the initial parameters $\vect{\delta}^{init}$ yields:
\begin{align}
\label{eq:psi_deriv}
\frac{\partial \psi}{\partial\delta^{init}_l}
&= \sum_{q^{(N)}}
    \frac{\partial L}{\partial u_q}
    \frac{\partial \vect{U}_q^{(N)}}{\partial\delta^{init}_l} \nonumber \\
&= \sum_{q^{N},q^{N-1}}
    \left.\frac{\partial \psi}{\partial u_q}\right|_{\vect{u}=\vect{U}^{(N)}(\delta^{init})}
    \left.\frac{\partial \vect{U}^{(N)}_{q^N}}{\partial u_q}\right|_{\vect{u}=\vect{U^{(N-1)}}(\delta^{init})}
    \left.\frac{\partial \vect{U}^{(N-1)}_{q^{N-1}}}{\partial\delta^{init}_l}\right|_{\vect{\delta^{init}}} \nonumber
    \\
&= \sum_{q^{N},q^{N-1},\ldots,q^{0}}
\left.\frac{\partial \psi}{\partial \vect{u}_q}\right|_{\vect{u}=\vect{U}^{(N)}(\delta^{init})}
    \left.\frac{\partial \vect{U}^{(N)}_{q^N}}{\partial \vect{u}_q}\right|_{\vect{u}=\vect{U^{(N-1)}}(\delta^{init})}
    \ldots
    \left.\frac{\partial \vect{U}^{(1)}_{q^1}}{\partial \vect{u}_q}\right|_{\vect{u}=\vect{U^{(0)}}(\delta^{init})}
    \left.\frac{\partial \vect{U}^{(0)}_{q^0}}{\partial\delta^{init}_l}\right|_{\delta^{init}}
\end{align}
where we made frequent use of the chain rule and $\vect{u}=[\vect{r}, \vect{p}]$
is a vector composed of particle positions and momenta. Also we have taken derivatives according to vector, which translates to a derivatives and implicit summations over all elements of the vectors.
Equation~\eqref{eq:psi_deriv} constitutes essentially a sequence of matrix vector
applications permitting to calculate the gradient given by equation~\eqref{eq:psi_deriv}
via the following iterative procedure:
\begin{equation}
\label{eq:gradient_iteration}
a_{p,\beta}^{(m+1)} =\sum_q \vect{a}_{q,\alpha}^{(m)} \mathcal{J}^{(m)}_{(q,\alpha),(p,\beta)} \, ,
\end{equation}
with  $\mathcal{J}^{(m)}_{q,p}$ being the Jacobian matrix between successive time steps.
We note that this operation is exactly an adjoint multiplication by the operator $\mathcal{J}$,
thus the name "tangent adjoint model" given to this whole procedure.
This matrix $\mathcal{J}$ is given by identification in Equation~\eqref{eq:psi_deriv}:
\begin{equation}
\mathcal{J}^{(m)}_{(q,\alpha),(p,\beta)} =
   \left.\frac{\partial U^{(N-m)}_{q,\alpha}}{\partial u_{m,\beta}}\right|_{\vect{u}=\vect{U}^{N-(m+1)}(\delta^{init})} \,,
\end{equation}
for $l < N$. We have also introduced the following notation to indicate components of vectors $(\vect{U}_q)_\alpha = U_{q,\alpha}$. For $l=N$, we have the special case:
\begin{equation}
  \mathcal{J}^{(N)}_{(q,\alpha),\beta} =
     \left.\frac{\partial U^{(0)}_{q,\alpha}}{\partial\delta^{init}_\beta}\right|_{\delta^{init}} \, ,
\end{equation}
and initial conditions with $\vect{a}_p^{0}$ given by:
\begin{equation}
  \vect{a}_q^{(0)} = \left.\frac{\partial \psi}{\partial \vect{u}_q}\right|_{\vect{u}=\vect{U}^{(N)}(\delta^{init})} \, .
\end{equation}
It is important to remark that at no point in the calculation of the gradient
it is required to explicitly store the high dimensional matrix $\mathcal{J}^{(m)}_{q,p}$.
We only need to have a procedure to evaluate exactly the sequence of matrix vector applications.
In the following we will therefore derive the Jacobian $\mathcal{J}^{(m)}_{q,p}$
of successive time steps in standard cosmological particle mesh codes.

\subsection{The Jacobian of particle mesh time stepping}
In this work we use a different implementation of the gradient than described in
\citet{Wang2013}.
The Jacobian between different time steps can then be obtained as:
\begin{equation}
\mathcal{J}^{(n)}_{(q,\alpha),(l,\beta)}
 = \left.\frac{\partial U^{(N-n)}_{q,\alpha}}{\partial u_{l,\beta}}\right|_{\vect{u}=\vect{U}^{(N-(n+1))}(\delta_\text{init})}  = \left[
        \begin{array}{ccccc}
            \displaystyle\frac{\partial r_{q,\alpha}^{(N-n)} }{ \partial r_{l,\beta}^{(N-(n+1))} } &
            \displaystyle\frac{\partial r_{q,\alpha}^{(N-n)}}{ \partial p_{l,\beta}^{(N-(n+1))} } \\[1.5em]
            \displaystyle\frac{\partial p_{q,\alpha}^{(N-n)}}{ \partial r_{l,\beta}^{(N-(n+1))} } &
            \displaystyle\frac{\partial p_{q,\alpha}^{(N-n)}}{ \partial p_{l,\beta}^{(N-(n+1))}}
        \end{array}
      \right]\,.
\end{equation}
Each element of this Jacobian matrix can be derived from the particle mesh
update scheme given in equation~\eqref{eq:update_particles_forward}. We can thus
directly compute the derivatives of a particle position and
velocity of a particle with respect to position and velocity of any other
particle in the simulation at the previous time step:
\begin{align}
  \frac{\partial r^{(m+1)}_{q,\alpha}}{\partial r^{(m)}_{l,\beta}} &
    = \delta^K_{\alpha,\beta} \delta^K_{q,l} +
        \frac{\partial \tilde{F}^{(m)}_{q,\alpha}}{\partial r^{(m)}_{l,\beta}}
          \Delta^m_r\Delta^m_v \\
  \frac{\partial r^{(m+1)}_{q,\alpha}}{\partial p^{(m)}_{l,\beta}} &
    = \delta^K_{\alpha,\beta} \delta^K_{q,l} \Delta^m_r \\
  \frac{\partial p^{(m+1)}_{q,\alpha}}{\partial r^{(m)}_{l,\beta}} &
    = \frac{\partial \tilde{F}^{(m)}_{q,\alpha}}{\partial r^{(m)}_{l,\beta}} \Delta^m_v\\
  \frac{\partial p^{(m+1)}_{q,\alpha}}{\partial p^{(m)}_{l,\beta}} &
    = \delta^K_{\alpha,\beta} \delta^K_{q,l} \,,
\end{align}
where we have used $m=N-n$ in the above to shorten the notation.
Given these calculations $\mathcal{J}^{n}_{(q,\alpha),(l,\beta)}$ can be written as:
\begin{equation}
  \mathcal{J}^{(n)}_{(q,\alpha),(l,\beta)}
    =
      \left[
        \begin{array}{ccccc}
          \displaystyle \delta^K_{\alpha,\beta} \delta^K_{q,l} +
            \frac{\partial F^{(m)}_{q,\alpha}}{\partial r^{(m)}_{l,\beta}} \Delta^m_r\Delta^m_v &
          \displaystyle \delta^K_{\alpha,\beta} \delta^K_{q,l}  \Delta^m_r \\[1.5em]
          \displaystyle \frac{\partial \tilde{F}^{(m)}_{q,\alpha}}{\partial r^{(m)}_{l,\beta}} \Delta^m_v &
          \displaystyle \delta^K_{p,q}
        \end{array}
      \right]
    = \delta^K_{q,l} \delta^K_{\alpha,\beta}
          \left[
            \begin{array}{ccccc}
              \displaystyle 1 &  \displaystyle \Delta^m_r \\
              0 & 1
           \end{array}
          \right] +
          \left[
            \begin{array}{ccccc}
              \displaystyle \frac{\partial \tilde{F}^{m}_{q,\alpha}}{\partial r^{m}_{l,\beta}} \Delta^m_r \Delta^m_v &
              0 \\[1.5em]
              \displaystyle \frac{\partial \tilde{F}^{m}_{q,\alpha}}{\partial r^{m}_{l,\beta}} \Delta^m_v &
              0
            \end{array}
          \right] \, ,
\end{equation}
where the first term describes the Jacobian of the linear equations of motion if there were no forces and the second term accounts for the coupling of the gravitational force and again $m=N-n$.

A single iteration of the gradient calculation step given in \eqref{eq:gradient_iteration} can then be calculated as:
\begin{align}
\left[
   a^{(r),(m)}_{q,\beta} \, , \, a^{(v),(m)}_{q,\beta} \right] =&
      \left[ a^{(r),(m-1)}_{q,\beta} +
      \sum_{p,\alpha} 
          \left(a^{(r),(m-1)}_{p,\alpha} \Delta^m_r\,\Delta^m_v +
                a^{(v),(m-1)}_{q,\alpha} \Delta^m_v \right)
                \frac{\partial \tilde{F}^{(m)}_{p,\alpha}}{\partial r^{(m)}_{q,\beta}}
          \right.
      \,, \nonumber \\
   & \left. \, a^{(v),(m-1)}_{q,\beta} + a^{(r),(m-1)}_{q,\beta} \,\Delta^m_r \right]\, ,
\label{eq:adjoint_update}
\end{align}
where the vector $\vect{a}$ is made of six components decomposed into position and a velocity components as $\vect{a}=[\vect{a}^{(r)} , \vect{a}^{(v)}]$. Each sub-vector having three dimensions indexed by $q$.
The only challenging terms to calculate in equation~\eqref{eq:adjoint_update} are
the terms depending on derivatives of the gravitational force.
In the next subsection we will discuss the evaluation of these terms.

\subsection{Tangent adjoint gradient of the force solver}
\label{sec:gradient_force}

Within a standard particle mesh approach forces at particle positions are
obtained via interpolation of a force field sampled at discrete positions
$\vect{K}^{(m)}_i$ on a grid to continuous particle positions
\citep[][]{HOCKNEYEASTWOOD1988}:
\begin{equation}
  \vect{\tilde{F}}^{(m)}_p =
    		\sum_i \mathcal{W}\left(\vect{x}_i - \vect{r}^{n}_p \, \right) \vect{K}^{(m)}_i\left( \{\vect{r^{(m)}} \}  \right)
            		= \sum_i W_{i,p} \vect{K}^{(m)}_i\, ,
\end{equation}
where $W_{i,p}=W_i(\vect{y}=\vect{r}^{(m)}_p)=\mathcal{W}\left(\vect{x}_i - \vect{r}^{(m)}_p \, \right)$ is the mass
assignment kernel that interpolates between discrete grid $\vect{x}_i$ and
continuous particle positions $\vect{r}^{(m)}_p$, and the discrete force
$\vect{K}^{(m)}_i\left( \{\vect{r}^{(m)} \} \right)$. This force array is a function of all particle positions in the simulation and denotes the force field evaluated
at the grid nodes. For a particle mesh code the force calculation on the grid
can be written as:
\begin{equation}
	\vect{K}^{(m)}_i\left( \{\vect{r}^{(m)}\}\right) =
    \sum_l \vect{M}_{i,l} \left[ \left( \sum_q \frac{1}{\bar{N}}  \mathcal{W}\left(\vect{x}_l - \vect{r}^{(n)}_q\right)\right) -1\right] \, ,
\end{equation}
where we apply the linear operator $\vect{M}_{i,l}$ to the density field as inferred from the particle distribution with the appropriate gridding kernel $\mathcal{W}\left(\vect{y} \right)$. The operator $M_{i,l,\alpha}$ is given as:
\begin{equation}
	M_{i,l,\alpha} =
    	\frac{1}{2 d_a} \sum_k \left(\mathcal{F}^{-1}_{\tilde{i}^{+1}_\alpha,k} -\mathcal{F}^{-1}_{\tilde{i}^{-1}_\alpha,k} \right) \mathcal{G}_k \mathcal{F}_{k,l} \, ,
\end{equation}
where $\mathcal{F}_{i,j}$ and $\mathcal{F}^{-1}_{i,j}$ denotes the forward and backward Fast Fourier transform operators respectively, and $\mathcal{G}_k$ is the Greens operator for the Poisson equation in Fourier space as given in equation~\eqref{eq:greens_poisson}. We have also introduced the notations of appendix~\ref{appendix:translation_dft} to grid indices $i$ and $\tilde{i}$. The gradient of the force with respect to positions is:
\begin{equation}
	\label{eq:grad_force_pm}
	\frac{\partial \tilde{F}^{(m)}_{q,\alpha}}{\partial r^{(m)}_{l,\beta}} =
    	\sum_i \left [ \delta^\text{K}_{q,l} W'^{(m)}_{i,q} K^{(m)}_{i,\alpha} + W_{i,q} \frac{\partial K^{(m)}_{i,\alpha}}{\partial r^{(m)}_{l,\beta}} \right] \,,
\end{equation}
with the introduced kernel derivative
\begin{equation}
  W'^{(m)}_{i,q,\beta} = \left.\frac{\partial W_{i}}{\partial y_\beta}\right|_{\vec{y}=\vect{r}^{(m)}_{q}}\,. \label{eq:define_Wprime}
\end{equation}
We derive the second term in the force derivative given in equation \eqref{eq:grad_force_pm}:
\begin{equation}
  \frac{\partial K^n_{i,\alpha}\left( \{\vect{r}^{(m)}\}\right) }{\partial r^{(m)}_{l,\beta}}
    =  \sum_p M_{i,p,\alpha} \frac{1}{\bar{N}}
    \left.\frac{\partial \mathcal{W}}{\partial y_\beta}\right|_{\vect{y} = \vect{x}_p - \vect{r}^n_l}
    =   \sum_p M_{i,p,\alpha} \frac{1}{\bar{N}} W'^{(m)}_{p,l,\beta}\,.
\end{equation}
We now have to collapse some of these expressions to build an efficient algorithm. We first introduce the updated vector, which is a subcomponent of the vector in \eqref{eq:adjoint_update}:
\begin{equation}
  \vect{b}_p =
    \vect{a}_{p}^{(r),(N-(n+1))} \Delta^n_r\,\Delta^n_v
    + \vect{a}_{p}^{(v),(N-(n+1))} \Delta^n_v \,.
\end{equation}
We now evaluate the force term in the adjoint update given in equation~\eqref{eq:adjoint_update}:
\begin{align}
  \sum_{p,\alpha} b_{p,\alpha} \frac{\partial \tilde{F}^{(m)}_{p,\alpha}}{\partial r^{(m)}_{q,\beta}}
    & =
    \sum_{i,p,\alpha} b_{p,\alpha} \delta^\text{K}_{p,q} W'^{(m)}_{i,p,\beta} K^{(m)}_{i,\alpha} +
    \sum_{i,p,\alpha} b_{p,\alpha} W_{i,p} \sum_p M_{i,p,\alpha} \frac{1}{\bar{N}} W'^{(m)}_{p,q,\beta}  \nonumber \\
    & = \sum_{i,\alpha} b_{q,\alpha} W'^{(m)}_{i,q,\beta}  K^{(m)}_{i,\alpha} + \sum_{p,i,\alpha} B_{i,\alpha}  M_{i,p,\alpha} \frac{1}{\bar{N}} W'^{(m)}_{p,q,\beta}   \\
    & = \sum_{i,\alpha} b_{q,\alpha}  W'^{(m)}_{i,q,\beta} K^{(m)}_{i,\alpha} + \sum_{p} D_{p} \frac{1}{\bar{N}} W'^{(m)}_{p,q,\beta}
    \,,
\end{align}
where we have introduced the vector:
\begin{equation}
\vect{B}_i=\sum_p \vect{b}_p W_{i,p}\, ,
\end{equation}
which is just the vector $\vect{b}_p$ interpolated to the grid with the mass assignment scheme $W_{i,p}$. We now proceed to compute the value of
\begin{equation}
  D_{l} = \sum_{i,a} B_{i,a} M_{i,l,a}\,.
\end{equation}
To achieve this we expand further $\vect{M}_{i,l}$:
\begin{align}
\label{eq:Dl}
  D_{l}
   = & \sum_a \frac{1}{2 d_a} \sum_i B_{i,a}
      \sum_{k}
        \left(\mathcal{F}^{-1}_{\tilde{i}^{+1}_a,k} -\mathcal{F}^{-1}_{\tilde{i}^{-1}_a,k} \right) \mathcal{G}_k \mathcal{F}_{k,l}    \nonumber \\
   = & \sum_{a,k} \frac{1}{2 d_a} C^a_k G_k \mathcal{F}_{k,l} \nonumber \\
   = & \sum_k \mathcal{F}^{*}_{l,k} G_k \left(\sum_a \frac{1}{2 d_a} C^a_k\right)^{*} \nonumber \\
   = & \sum_k \mathcal{F}^{*}_{l,k} G_k \left(\sum_a \frac{1}{2 d_a} C^a_k\right) \,,
\end{align}
with
\begin{equation}
  C^a_k = \sum_{i} B_{i,a} \left(\mathcal{F}^{-1}_{\tilde{i}^{+1}_a,k} -\mathcal{F}^{-1}_{\tilde{i}^{-1}_a,k} \right)\,,
\end{equation}
and the notations of appendix~\ref{appendix:translation_dft}. For the last line of equation~\eqref{eq:Dl}, we have used the hermiticity of the $C^{a}_k$ fields .
We now re-express $C^a_k$ exploiting the periodicity of the discrete Fourier Transform $\mathcal{F}_{i,k}$:
\begin{align}
  C^a_k = & \sum_{i} B_{i,a} \left(\mathcal{F}^{-1}_{\tilde{i}^{+1}_a,k} -\mathcal{F}^{-1}_{\tilde{i}^{-1}_a,k} \right) \\
  = & \frac{1}{N^3} \sum_{i}  \left(\mathcal{F}^{*}_{k,\tilde{i}^{+1}_a} -\mathcal{F}^{*}_{k,\tilde{i}^{-1}_a}\right) B_{i,a} \nonumber \\
  = & \frac{1}{N^3} \sum_{i}  \mathcal{F}^{*}_{k,i} \left( B_{\tilde{i}^{-1}_a,a}-B_{\tilde{i}^{+1}_a,a} \right) \, ,
\end{align}
where  $C^a_k$ is simply the discrete Fourier transform of the differences in the $B^a_i$ field along $a$-th axis, and we have exploited the identity shown in appendix~\ref{appendix:translation_dft}.
The discrete $D_l$ field is now obtained by applying the Greens operator $\mathcal{G}_k$ to the components of the $\vect{C}^a_k$ vector and performing a transposed discrete Fourier transform on the sum of the components (equation~\eqref{eq:Dl}).

If we assume the mass assignment kernel $W_{i,p}$ factorizes along each of the spatial coordinates, as is usually the case in particle mesh codes, then we can write \citep[][]{HOCKNEYEASTWOOD1988}:
\begin{equation}
W_{i,p}=\mathcal{W}(\vect{x}_i - \vect{r}_p)= \prod_{j=0}^2 \omega(x_{i,j} - r_{p,j})
\end{equation}
This yields the gradient of the mass assignment kernel given as:
\begin{equation}
  W'^{(m)}_{i,p,\beta} = -\omega'(x_{i,\beta}-r_{p,\beta}) \prod_{\alpha\neq \beta} \omega(x_{i,\alpha} - r_{p,\alpha})\,,
\end{equation}
with $W'^{(m)}_{i,p,\beta}$ as defined in equation~\eqref{eq:define_Wprime}. Finally we can rewrite the equation~\eqref{eq:adjoint_update}, governing the update of the adjoint gradient vector from time step $(m+1)$  to time step $(m)$ as follow:
\begin{equation}
  \label{eq:adjoint_update_1}
  \left[ 
    {\begin{array}{cc}
    	\vect{a}_{r}^{m} \\ 
    	\vect{a}_{v}^{m} 
    \end{array} }\right] =
       \left[ 
         {\begin{array}{cc}\vect{a}_{r}^{(m+1)}+ \vect{\Theta}^{(m+1)}  \\ 
          \vect{a}_{v}^{(m+1)} +\vect{a}_{r}^{(m+1)} \,\Delta^n_r 
          \end{array} }\right]\, ,
\end{equation}
with
\begin{equation}
 \left[\vect{\Theta}^{(m)}_q\right]_\beta = \Theta^{(m)}_{q,\beta} = \sum_{p,\alpha} b_{p,\alpha}  \frac{\partial \tilde{F}^{m}_{p,\alpha}}{\partial r^{m}_{q,\beta}}\,.
\end{equation}
The total steps involved to compute $\vect{a}^{(m+1)}$ are thus:
\begin{equation}
	\vect{a}^{(m)} \rightarrow  \vect{b} 
    	\underset{\text{CIC}}{\rightarrow }
        	\vect{B}_i  \underset{\mathcal{F}}{\rightarrow} C_{a,k} \underset{\mathcal{F}}{\rightarrow} D_l \underset{\text{CIC}^\dagger}{\rightarrow}\vect{\Theta}^{(m)}_q \rightarrow \vect{a}^{(m+1)}\,,
\end{equation}
with $\mathcal{F}$ denoting the presence of a Fourier transform and $\text{CIC}$ a mass assignment kernel like Cloud-In-Cell.

This demonstrates that the numerical complexity of the tangent adjoint model is the same as for the full forward model evaluation. In fact, as demonstrated by figure~\ref{fig:bench}, the numerical costs of one tangent adjoint model evaluation is equivalent to the costs of two forward model evaluations. A single gradient evaluation requires one full forward model evaluation and a subsequent application of the tangent adjoint model. The numerical complexity of a gradient evaluation and run-times are thus about two times a single forward model evaluation.
It should be remarked that the evaluation of the tangent adjoint model requires to store all
particle positions and velocities at all steps of the forward model evaluation.

\section{Tangent adjoint model of redshift space distortions}

In Section~\ref{sec:rsd}, we have introduced the model we have adopted to introduce redshift space distortions in the analysis. In this appendix we
detail the computation of the tangent adjoint of this model.

We introduce redshift space distortions as an additional displacement of particles compared to their final comoving positions. At first order in $1/c$, we have for a single particle with position $\vect{x}$ and velocity $\vect{v}$
\begin{equation}
	\vect{s} = \vect{x} + \gamma \vect{v}^\text{los}\,,
\end{equation}
where we have set
\begin{equation}
  \vect{v}^\text{los} = \sum_a v_{a} y_{a} \frac{\vect{y}}{|\vect{y}|^2} \,,
\end{equation}
and
\begin{equation}
  \gamma = \frac{a H_0}{H(a)}\,,
\end{equation}
$a$ being the cosmological scale factor. Internally, the particle mesh stores another variant of the velocity, the momentum, which is $\vect{p} = a^2 \vect{v}$. Thus we form $\gamma^p = \frac{H_0}{a H(a)}$ to account for the different scaling
\begin{equation}
	\vect{s} = \vect{x} + \gamma^p \vect{p}^\text{los}\,,
\end{equation}
\begin{equation}
  \vect{p}^\text{los} = \sum_a p_{a} y_{a} \frac{\vect{y}}{|\vect{y}|^2} \,.
\end{equation}

To follow the generic framework indicated in Appendix~\ref{appendix:tangent_adjoint_model}, we introduce the derivative with respect to comoving coordinates $x_i$
\begin{equation}
  \frac{\partial s_i}{\partial x_a}
    =
    \delta^K_{i,a} \left(1 + \gamma^p  \frac{\sum_k p_k x_k}{|\vect{x}|^2} \right)
    + \gamma^p \frac{p_a x_i}{|\vect{x}|^2}
    - 2 \gamma \frac{\sum_k p_k x_k}{|\vect{x}|^4}\, x_i x_a
\end{equation}
Let $\alpha=\sum_k p_k x_k$ and $\beta=|\vect{y}|^2$ then:
\begin{equation}
  \frac{\partial s_i}{\partial x_a} =
    \delta^K_{ir} \left ( 1 + \gamma^p \frac{\alpha}{\beta} \right)
    + \gamma^p  \frac{p_a y_i}{\beta}
    - 2 \gamma^p \frac{\alpha}{\beta^2} y_i y_a
\end{equation}
Similarly we obtain the derivative of $s$ with respect to velocity
\begin{equation}
\frac{\partial s_i}{\partial p_a} = \gamma^p \frac{ y_a y_i}{|\vect{y}|^2} = \gamma^p \frac{ y_a y_i}{\beta}\,.
\end{equation}
Putting back together for we may derive the two adjoint gradient for the position and velocity:
\begin{align}
  x^\text{ag}_a = & \sum_i \frac{\partial s_i}{\partial x_a} s^\text{ag}_i
  = s^\text{ag}_a \left(1 + \gamma^p \frac{\alpha}{\beta}\right) + \frac{\gamma^p \left(\sum_i s^\text{ag}_i x_i\right)}{\beta} \left( p_a - 2 \frac{\alpha}{\beta} x_a \right) \,, \\
  p^\text{ag}_a = & \sum_i \frac{\partial s_i}{\partial p_a} = \gamma^p \left(\sum_i s^\text{ag}_i x_i\right) \frac{y_a}{\beta}\,.
\end{align}
The case for which no redshift space distortions is requested reduces to setting $\gamma=0$. We indeed recover that $\vect{x}^\text{ag}=\vect{s}^\text{ag}$ and $\vect{v}^\text{ag}=0$.

\label{lastpage}

\end{document}